\documentclass[a4paper,11pt]{article}

\usepackage{jheppub} 
\usepackage{etoolbox}
\usepackage[T1]{fontenc} 
\usepackage{latexsym,amsmath,amsfonts,amssymb,amsthm}
\usepackage{amsmath}
\usepackage{mathtools}
\usepackage[dvipsnames]{xcolor}
\usepackage{tikz}
\usetikzlibrary{positioning,arrows.meta}
\usepackage{setspace}
\usepackage{seqsplit}

\allowdisplaybreaks[4]

\title{Exact non-Lagrangian Schur index in closed form}

\author{Yiwen Pan$^1$,}

\affiliation{$^1$ School of Physics, Sun Yat-sen University,\\No. 135 Xingangxi Road, Guangzhou, Guangdong, China}

\author{Peihe Yang$^2$}

\affiliation{$^2$ Center for High Energy Physics,\\Peking University, Yiheyuan Road, Beijing 100871, China}

\emailAdd{panyw5@mail.sysu.edu.cn}
\emailAdd{peiheyang@pku.edu.cn}
\abstract{The Schur index is a powerful tool to probe the spectrum and dualities of 4d $\mathcal{N}=2$ superconformal field theories (SCFTs), deeply related to 2d vertex operator algebras (VOAs). In this paper, we compute the Schur index in closed form for two series of non-Lagrangian theories. We explore and classify the Argyres-Douglas (AD) theories $D_p^b(\mathfrak{sl}_N,[Y])$ realized as the $SU(2)$ gauging of two AD matter theories, where we identify several infinite families with interesting central charge relations analogous to the $a_\text{4d} = c_\text{4d}$ of $\mathcal{N} = 4$ theories. We focus on $D_{N-4}(\mathfrak{sl}(N),[N-4,4])$ and $D_{N-2}(\mathfrak{sl}(N),[N-3,3])$, and compute their flavored and unflavored Schur and Wilson line indices in compact form. We also explore their large-$N$ behavior, and show that they arise as special limits of the $SU(2)$ SQCD flavored index, also analogous to the relation among the $a_\text{4d} = c_\text{4d}$ theories. We also generalize the elliptic function integration formula in the presence of higher order poles to compute in closed form the partially flavored indices of the Minahan-Nemeschansky $E_{6}$ and $E_{7}$ theories. Our results point to a universal structure underlying the residues of elliptic integrands, Wilson loop indices, and non-vacuum modules of the corresponding VOAs.}

\makeatletter
\patchcmd{\maketitle}{\@fpheader}{}{}{}
\makeatother

\begin{document} 
\maketitle

\flushbottom


\section{Introduction}

The superconformal index is a fundamental observable to analyze and characterize 4d $\mathcal{N} = 2$ superconformal field theories (SCFTs). It encodes the representation-theoretic structure of the protected local operator spectrum. The index admits various simplifying limits, among which is the Schur limit \cite{Gadde:2011uv}. The Schur index is particularly interesting because it has been shown to participate prominently in the 4d/2d or the SCFT/VOA correspondence, where every 4d $\mathcal{N} = 2$ SCFT is associated with a 2d vertex operator algebra (VOA), and the Schur index is precisely identified with the vacuum character of the algebra \cite{Beem:2013sza}.

There are several ways of computing the Schur index. For Lagrangian theories, the index is written as a contour integral of a multivariate elliptic function, where the integrand represents the Schur index in the zero-coupling limit, and the contour integral implements the gauging by projecting onto the gauge-invariant states. The integrand can be expanded in $q$-series, and then perform the contour integral term wise, resulting in a $q$-series expansion of the Schur index.

For class-$\mathcal{S}$ theories defined with respect to a puncture Riemann surface, including families of non-Lagrangian $T_{N\ge 3}$ theories and Argyres--Douglas theories, the Schur index can be computed exploiting the topological quantum field theory (TQFT) nature of the index \cite{Gadde:2011uv,Lemos:2014lua,Buican:2015ina,Song:2015wta,Gadde:2011ik,Lemos:2012ph,Mekareeya:2012tn,Buican:2017uka,Buican:2015hsa}. In this approach, the index is typically expressed as a sum over irreducible representations of the gauge group, with each term in the sum being a product of contributions from the punctures and the genus of the Riemann surface.

Some non-Lagrangian 4d $\mathcal{N} = 2$ SCFTs can also be obtained through supersymmetry-enhancing renormalization group flow starting from 4d $\mathcal{N} = 1$ Lagrangian theories \cite{Maruyoshi:2016tqk,Maruyoshi:2016aim,Agarwal:2016pjo,Agarwal:2017roi,Agarwal:2018ejn,Benvenuti:2017bpg,Zafrir:2019hps,Zafrir:2020epd}. In these cases, the Schur index can be computed using the $\mathcal{N} = 1$ index of the UV theory with appropriate modifications to account for decoupled operators along the flow. The usual output of this approach is either the contour integral that comes from the $\mathcal{N} = 1$ Lagrangian theory, or the $q$-series expansion.

Compared with the $q$-series expression or infinite sum over representations, Schur index in some closed-form can be more useful in some circumstances, such as in studying the modular property of the index, representation-theoretic problems of the associated VOA, and the large-$N$ behavior/holography interpretation. There had been series of efforts to perform exact computation of the Schur index. The Fermi-gas formalism was first introduced to compute the unflavored $\mathcal{N} = 4$ $SU(N)$ Schur index in \cite{Bourdier:2015wda}, then applied to circular quiver theories \cite{Bourdier:2015sga}, and more recently generalized to the flavored index and $\mathcal{N} = 4$ theories with other gauge groups \cite{Hatsuda:2022xdv,Du:2023kfu}, and with insertion of line operators. In \cite{Pan:2021mrw}, integration formula of elliptic functions and Eisenstein series were proposed to compute various types of Lagrangian Schur index, and subsequently to include of surface and line defects, where the results are written in terms of quasi-modular or quasi-Jacobi forms \cite{Guo:2023mkn,Zheng:2022zkm,Huang:2022bry,Beemetal}.

There are mainly two known series of closed-form Schur index for non-Lagrangian theories. First is the class of Argyres--Douglas theories whose associated VOAs are given by affine Kac-Moody algebras at boundary admissible levels (and a few non-admissible levels), and W-algebras obtained from quantum Drinfeld-Sokolov reduction of these affine Kac-Moody algebras \cite{Xie:2016evu,Song:2017oew,Xie:2019zlb}. The second class is the $\mathcal{T}_{p, N}$ theories whose Schur index is directly related to that of the 4d $\mathcal{N} = 4$ theories \cite{Buican:2020moo,Kang:2021lic,Jiang:2024baj}. 

In this paper we apply the integration formula to more non-Lagrangian theories and compute the Schur index in closed form. We focus on two classes of theories. The first is the Argyres--Douglas theories $D^b_p(\mathfrak{sl}(N), [Y])$, which has been a subject of intensive study \cite{Wang:2015mra,Xie:2016uqq,Xie:2016evu,Deb:2025ypl,Xie:2017vaf,Song:2017oew,Xie:2017aqx,PhysRevD.100.025001,Xie:2019yds,Xie:2019zlb,Xie:2019vzr,Shan:2023xtw,Cecotti:2013lda,Cecotti:2012jx}. Here $p$ and $N$ are not coprime. These theories can be constructed by going to special points in the Coulomb branch \cite{Argyres:1995jj}, or through geometric engineering on singular Calabi-Yau threefolds \cite{Cecotti:2010fi,Shapere:1999xr}, or in class-$\mathcal{S}$ construction with irregular punctures \cite{Gaiotto:2009we,PhysRevD.100.025001,Xie:2012hs,Wang:2015mra}. As shown in \cite{Beem:2023ofp,Xie:2017vaf}, such theories can be written in a duality frame conformal gauging two $D_{p_i}(\mathfrak{sl}(N_i), [Y_i])$ by an $SU(N)$ vector multiplet, and possibly with some additional fundamental hypermultiplets. In the simplest situation, each pair $p_i, N_i$ are respectively coprime, and the two Argyres--Douglas matter theories do not have exactly marginal deformations. In this case, the Schur index of each component theory is known in closed form in terms of plethystic exponential \cite{Xie:2019zlb,Song:2015wta,Xie:2016evu}, and the gauging gives rise to a multivariate integral of elliptic function. This allows us to analytically compute the Schur index using the integration formula \cite{Pan:2021mrw}. For simplicity, we focus on the cases where the gauge node is an $SU(2)$ gauge node, whose Schur index involves only a single contour integral with integrand $\mathcal{Z}$. Cases with higher-rank gauge node, or where the component theories are not in the coprime class, can be treated similarly, though the computation will be more involved, which we leave for future work.

The requirement of having only one $SU(2)$ gauge node and coprime pairs $(p_i, N_i)$ put significant constraints on the choice of $b, p, N, Y$. This leads to a partial classification for the type I cases, \emph{i.e.}, when $b = N$. Even with this restriction, there are still infinitely many possibilities. Among them, we identify several interesting infinite families, whose central charges satisfy simple relations
\begin{equation}
  a_\text{4d} - c_\text{4d} = \Delta, \quad \Delta = - \frac{1}{12}, - \frac{1}{12} + \frac{1}{4(N - 4)}, - \frac{1}{8}, - \frac{5}{24} \ ,
\end{equation}
reminiscent of the $\mathcal{N} = 4$ and the $\mathcal{T}_{p,N}$ theories. In particular, we study systematically the $D_{N-4}(\mathfrak{sl}(N), [N-4,4])$, $D_{N - 2}(\mathfrak{sl}(N), [N-3,3])$ series. We compute the Schur index for these two families in closed form. Curiously, the Schur index of both series are just special limit of the flavored index of the $SU(2)$ theory with four flavors, in complete analogy to the relation between the $\mathcal{T}_{p,N}$ theories and the $\mathcal{N} = 4$ $SU(N)$ theories \cite{Kang:2021lic,Buican:2020moo,Jiang:2024baj,Pan:2025gzh}. We also explore the large-$N$ behavior of the Schur/Wilson line index for these two families.

BPS non-local operators preserving the supercharges defining the Schur index are believed to give rise to modules of the associated VOA \cite{Cordova:2016uwk,Cordova:2017mhb,Bianchi:2019sxz}. For the two $SU(2)$ Lagrangian theories, the $\mathcal{N} = 4$ $SU(2)$ and the $SU(2)$ SQCD theory, the Wilson line index can be written schematically as
\begin{equation}
  \mathcal{I}_L(b, q) = \sum_{i} R_i(b, q) \operatorname{ch}_i(b,q)
  = R'_0 \mathcal{I} + \sum_{j} R'_j \operatorname{Res}_j  \ ,
\end{equation}
where $R_i, R'_j$ are some rational functions of $q^{1/2}$ and the flavor fugacities $b$, $\operatorname{ch}_i(b,q), \operatorname{Res}_j$ denote the irreducible characters of the associated VOA and the residues of $\mathcal{Z}$ \cite{Pan:2021ulr, Pan:2024epf,Pan:2024hcz,Guo:2023mkn}. In the two examples, all the characters $\operatorname{ch}_i$ and residues $\operatorname{Res}_j$ solve the same set of flavored modular linear differential equations (MLDEs) \cite{Mathur:1988na,zhu1996modular,Gaberdiel:2008pr,Beem:2017ooy,Pan:2023jjw,Pan:2021ulr,Zheng:2022zkm,Pan:2024dod,Kaidi:2022sng}, and the characters $\operatorname{ch}_i$ are given as integral linear combination of the residues (and the Schur index $\mathcal{I}$) \cite{Pan:2023jjw,2023arXiv230409681L}.

The presence of a gauge group allows us to couple the non-Lagrangian theory to BPS non-local operators. We consider the simplest possibility by introducing Wilson line and compute the line index \cite{Gang:2012yr,Hatsuda:2023iwi,Guo:2023mkn}. Although the full analysis of flavor modular differential equations is currently lacking, we find in various examples combinations of residues of $\mathcal{Z}$ that admit unflavoring limit and solve the same unflavored MLDE as the unflavored Schur index. Said differently, the specific combination of residues are linear combination of characters.

We also explore the closed-form Schur index of the Minahan-Nemeschansky $E_{6}$ and $E_7$ theory. Recently the unflavored Schur index of these theories are computed using the generalized partition function method \cite{Deb:2025ypl}, where the index is written as a contour integral of an elliptic function, even though the theory has no Lagrangian description. Nonetheless, the integration formula can be applied. Explicitly, we find that the generalized partition function method can be slightly generalized to include one flavor fugacity, allowing us to extract the partially flavored Schur index in closed-form. In this computation, due to the presence of higher-order poles, we develop new integration formula that incorporates generalized residues. 

Although the contour integral formula does not follow from a Lagrangian description, computationally we can always compute a ``Wilson line index'' by inserting into the integral an $SU(2)$ character, and summing over the generalized residues. A priori, one would not have expected these residues or line index to have any relation with the $(E_6)_{-3}$, $(E_7)_{-4}$ modules. Surprisingly, we do find that the leading partially flavored residues solve the expected partially flavored MLDEs coming from the null states of the VOAs, and hence they are linear combinations of VOA characters. All of these results (together with those from AD theories) suggest universal relation between the residues of an elliptic integrand $\mathcal{Z}$ that integrates to the index, the Wilson line index, and the VOA representation theory.

The rest of the paper is organized as follows. In section \ref{sec:landscape}, we review the landscape of type A Argyres--Douglas theories, and summarize the classification of those admitting an $SU(2)$ gauge theory description. In section \ref{sec:simple-examples}, we compute the Schur index in closed form for several simple examples of $D^b_p(\mathfrak{sl}_N, [Y])$, including both type I and type II. In section \ref{sec:DN-4}, \ref{sec:DN-2} we focus on two infinite families $D_{N-4}(\mathfrak{sl}(N), [N-4,4])$, $D_{N - 2}(\mathfrak{sl}(N), [N-3,3])$. We show that their flavored Schur index is a simple specialization of the $SU(2)$ SQCD Schur index. We compute both the flavored and unflavored Schur/Wilson line index in compact closed form, construct non-vacuum module character from residues, and study the large-$N$ behavior of these indices. In section \ref{sec:MN}, we compute the partially flavored Schur index of the Minahan-Nemeschansky $E_6$ and $E_7$ theories, and show that the first order generalized residues are linear combinations of $(E_6)_{-3}, (E_7)_{-4}$ characters.

\section{\texorpdfstring{Landscape of type $A$ Argyres–Douglas theories}{}}\label{sec:landscape}

\subsection{Generalities on Argyres--Douglas theories}

The Argyres--Douglas (AD) theory was first discovered as the IR fixed point at a special locus on the Coulomb branch of pure $SU(3)$ gauge theory \cite{Argyres:1995jj}. They are intrinsically non-Lagrangian in the $\mathcal{N} = 2$ sense, which can be seen from the presence of local operators with fractional scaling dimension. Alternatively, an AD theory can be constructed by compactifying the $6d$ type-$J=A,D,E$ $\mathcal{N}=(2,0)$ theory on a Riemann surface with one irregular puncture and (optionally) one regular puncture \cite{Xie:2012hs}.

The Coulomb branch of a $4d$ $\mathcal{N}=2$ superconformal field theory on $\mathbb{R}^3\times S^1$ can be described by a Hitchin system \cite{Gaiotto:2009hg}, and hence is characterized by a Higgs field $\Phi(z)$, where $z$ denotes the coordinate on the Riemann sphere.  
The irregular puncture takes the general form
\begin{equation}
	\Phi(z)=\frac{T_k}{z^{2+\frac{k}{n}}}\mathrm{d}z+\sum_{l=-n}^{k-1}\frac{T_l}{z^{2+\frac{l}{n}}}\mathrm{d}z+\cdots,
\end{equation}
where the leading matrix $T_k$ is a semisimple element of $A_{N-1}$.  
A necessary condition for the theory to be superconformal is that the Levi subalgebras for each $T_i$ satisfy \cite{Xie:2017vaf}
\begin{equation}
	\mathfrak{l}_k=\mathfrak{l}_{k-1}=\cdots=\mathfrak{l}_{-n+1}\supset\mathfrak{l}_{-n}.
\end{equation}
The classification of irregular punctures relies on the single-valuedness of the Higgs field under the transformation $z \to e^{2\pi i}z$.  
The allowed leading matrix then takes the form \cite{Xie:2017vaf, Xie:2021ewm}
\begin{equation}
	T_k=\begin{pmatrix}a_1^{(k)}\Sigma_n&&&&\\&a_2^{(k)}\Sigma_n&&&\\&&\ddots&&\\&&&a_m^{(k)}\Sigma_n&\\&&&&0_{N-mn=N-b}\end{pmatrix}
\end{equation}
where
\begin{equation}
	\Sigma_{n}:=\begin{pmatrix}1\\&\omega\\&&\ddots\\&&&\omega^{n-1}\end{pmatrix},\qquad \omega=e^{2\pi i/n}.
\end{equation}

The regular punctures for type-$A$ theories were classified in \cite{Gaiotto:2009we} and take the general form
\begin{equation}
	\Phi(z)=\frac{\Lambda}{z}+\cdots
\end{equation}
where $\Lambda$ is a nilpotent element of $A_{N-1}$.  
Thus, the classification of regular punctures corresponds to the classification of nilpotent orbits, and they can be labeled by Young tableaux $[Y]$.  

In this paper, we require the eigenvalues of each $T_i$ to be distinct.  
In summary, an AD theory can be labeled as
\begin{equation}
	D_p^b(\mathfrak{sl}_N,[Y]),\qquad p=k+b.
\end{equation}
For certain specific values of $b$ and $[Y]$, we follow the convention of \cite{Cecotti:2012jx,Cecotti:2013lda} and simplify the notation as
\begin{equation}
	\begin{aligned}
		&D_p^N\left(\mathfrak{sl}_N,[Y]\right)=:D_p\left(\mathfrak{sl}_N,[Y]\right),\\ &D_p^b\left(\mathfrak{sl}_N,[1^N]\right)=:D_p^b\left(\mathfrak{sl}_N\right),\\ &D_p^N\left(\mathfrak{sl}_N,[1^N]\right)=:D_p\left(\mathfrak{sl}_N\right).
	\end{aligned}
\end{equation}
We also refer to the cases where $b=N$ and $b=N-1$ as type~I and type~II theories, respectively.

Class-$\mathcal{S}$ theories with regular punctures admit different pants decompositions related by S-duality. 
Similarly, the landscape of AD theories also enjoy various dualities that help simplify the overall picture, expressing an AD theory in terms of gauging of other AD matter theories \cite{Xie:2017vaf, Xie:2017aqx, Beem:2023ofp}.

In this paper we make use of the duality discussed in \cite{Beem:2023ofp}, where a generic AD theory $D_p^b(\mathfrak{sl}_N, [Y])$ is expressed in terms of a type~I theory and some additional free hypermultiplets,
\begin{equation}\label{general duality}
	D_p^b\left(\mathfrak{s}\mathfrak{l}_N,[Y]\right)\cong D_p\left(\mathfrak{s}\mathfrak{l}_{(N-b)p+b},[(p-1)^{N-b},Y]\right)\otimes H_{\mathrm{free}}\text{ free hypermultiplets },
\end{equation}
where 
\begin{equation}
	H_{\mathrm{free}}=(N-b)\sum_{i=k+b}^N(i-k-b+1)l_i.
\end{equation}
and $l_i$ denotes the multiplicity of the part $i$ in the Young diagram $[Y] = [N^{l_N},\ldots,1^{l_1}]$.  
The additional hypermultiplets transform in the fundamental representation of each simple flavor factor arising from the regular punctures.

Therefore, the study of generic $D_p^b(\mathfrak{sl}_N, [Y])$ reduces to that  of the type~I theories. 
A type I theory with exactly marginal deformation can be further expressed as a gauge theory with simpler AD matters. Let the regular puncture labeled by the partition $[Y]$,
\begin{equation}
	[Y]=\begin{bmatrix}N^{l_N},\ldots,2^{l_2},1^{l_1}\end{bmatrix},\quad
	L=\sum_{i=1}^Nl_i,
\end{equation}
one can define the ``conjugate'' partition (here the $q \in \mathbb{N}$, not to be confused with the $q = e^{2\pi i \tau}$ in the Jacobi theta functions)
\begin{equation}
	[\widetilde{Y}] \coloneqq \Big[\left(q-1\right)^{l_1},\left(q-2\right)^{l_2},\ldots,\left(q-N\right)^{l_N} \Big],
	\qquad q\coloneqq\frac{p}{m}.
\end{equation}
The entries in the partition $\widetilde{[Y]}$ sum to $qL - N$.  
We separate the entries into three subpartitions based on the sign: positive, negative, and zero.  
Let $[\widetilde{Y}_+]$ be the subpartition of positive entries, and $[\widetilde{Y}_-]$ the subpartition of the absolute values of the negative entries.  
The sums of these two subpartitions are denoted $\widetilde{N}_+$ and $\widetilde{N}_-$, respectively, and the number of zero entries is denoted $\widetilde{N}_0$.

Then, the general type~I theory\footnote{We consider only the case where $k$ is not integer multiple of $N$.} can be expressed as as a gauge theory with AD matters,
\begin{equation}\label{typeIrealization}
	\begin{split}
		D_{p}\left(\mathfrak{sl}_{N},[Y]\right) \cong & \ D_q\left(\mathfrak{sl}_{qL-\frac{N}{m}},\left[\widetilde{Y}_+,1^{N-\frac{N}{m}-\widetilde{N}_-}\right]\right) \\
		&\hspace*{3cm}\Big\uparrow \\
		&\hspace*{1cm} \mathfrak{su}\left(N-\frac{N}{m}-\widetilde N_-\right) \text{\textemdash\textemdash}\; \fbox{$\widetilde N_0$} \\
		&\hspace*{3cm}\Big\downarrow \\
		&D_{(m-1)q}\left(\mathfrak{sl}_{N-\frac{N}{m}},\left[\widetilde{Y}_-,1^{N-\frac{N}{m}-\widetilde{N}_-}\right]\right) 
	\end{split}
\end{equation}
In the gauge theory description, the first AD matter contains no exactly marginal deformation since $\gcd((m-1)q,N-\frac{N}{m})=1$. In the second AD matter, the number of exactly marginal deformations is
\begin{equation}
	\gcd\left((m-1)q,N-\frac{N}{m}\right)-1=m-2 \ ,
\end{equation}
making the total number of exactly marginal deformations of $D_p(\mathfrak{sl}_N, [Y])$ equal to $m - 1 = \gcd(N, p) - 1$. This decomposition can be applied recursively to decompose any $D_p^b(\mathfrak{sl}_N, [Y])$ into a generalized quiver gauge theory with basic AD matter without exactly marginal deformations.

\subsection{Argyres--Douglas theories as \texorpdfstring{$SU(2)$}{SU(2)} gauge theories}

In this paper we are interested in computing the exact Schur index of the simplest class of AD theories, which can be realized as $SU(2)$ gauge theories coupled to two AD matters without exactly marginal deformations. Schematically, they look like
\begin{equation}
	\begin{tikzpicture}[baseline=(current bounding box.center)]
		\node (left) {$D_{p_1}(\mathfrak{sl}_{N_1},[Y_1])$};
		\node (mid) [right=of left, xshift=0.5cm] {$\mathfrak{su}(2)$};
		\node (right) [right=of mid, xshift=0.5cm] {$D_{p_2}(\mathfrak{sl}_{N_2},[Y_2])$};
		\draw[-{Stealth[scale=1.2]}] (mid) -- (right);
		\draw[-{Stealth[scale=1.2]}] (mid) -- (left); 
		\node (box) [draw, rectangle, below=0.5cm of mid] {$\tilde{N}_0$};
		\draw (mid) -- (box);
	\end{tikzpicture}
\end{equation}
where $\operatorname{gcd}(p_1, N_1) = \operatorname{gcd}(p_2, N_2) = 1$. Recall that the number of exactly marginal operators in $D_p(\mathfrak{sl}_N,[Y])$ with $p = N + k$ is given by the formula\footnote{In this paper we focus on the former situation.}
\begin{equation}
	\gcd (N,k)-1\quad \text{for $k\neq m N$}
	, \quad \gcd (N,k)-2\quad \text{for $k=m N$}
	.
\end{equation}
The coprime condition on $p_i, N_i$ implies that both AD matters have no exactly marginal deformations. Imposing that the full theory $D_p(\mathfrak{sl}_N, [Y])$ has only one exactly marginal coupling and the gauge group to be $SU(2)$, the duality \eqref{typeIrealization} puts a set of constraints on the parameters.

The requirement that the full theory has only one exactly marginal deformation and with only one $SU(2)$ gauge node implies
\begin{equation}
	\begin{aligned}\label{eq:constraints-1}
		& m=\gcd(N,k)=2, \quad N-\frac{N}{2}-\tilde{N}_-=2, \quad p=k+N = mq = 2q \ .\\
	\end{aligned}
\end{equation}
Substituting the definition of $\tilde{N}_-$, we have the following constraints
\begin{align}\label{eq:constraints-2}
	l_{q+1} + 2l_{q + 2}+\cdots+(N-q)l_N=\frac{N}{2}-2,\quad
	l_1+2l_2+\cdots Nl_N=N \ .
\end{align}
The condition $\gcd(N,k) = 2$ requires both $N, k$ to be even. Furthermore, if $N = 2n$, $k = 2n'$, then we must have $\gcd(n, n') = 1$. 
\begin{itemize}
	\item If $N = 2$, then $N - \frac{N}{2}- \tilde N_- = 2$ cannot be satisfied. Hence $N \ge 4$.
	\item If $N = 4$, then $k = 2 \mod 4 \ge -2$. Note that $\frac{N}{2} - 2 = 0$ and hence $l_{q+1} = l_{q+2} = \cdots = l_N = 0$. This places no further constraint on $k$, but allows only a subset of the partitions $[Y]$.
	\item If $N > 4$ such that $\frac{N}{2} - 2 > 0$, then $q$ must be strictly less than $N$, otherwise the left hand side of the first equation in (\ref{eq:constraints-2}) would be zero. This in turns places strong constraints on $k$ on which we now elaborate.
\end{itemize}

\vspace{1em}
\underline{Theories with $k > 0$}

First we denote $N = 2n > 4$ and consider $k = 2n' > 0$. The first equation in (\ref{eq:constraints-2}) can be rearranged,
\begin{align}
	l_1 + 2l_2 + \cdots + Nl_N - (l_1 + 2l_2 + \cdots + ql_q) - q(l_{q+1} + l_{q+2} + \cdots + l_N) = & \ \frac{N}{2} - 2 \ ,
\end{align}
giving
\begin{equation}\label{eq:constraint-on-Y}
	l_1 + 2l_2 + \cdots + ql_q + q(l_{q+1} + l_{q+2} + \cdots + l_N) = \frac{N}{2} + 2 \ .
\end{equation}
Recall that $q = n + n' > n = N/2$. The left hand side of (\ref{eq:constraint-on-Y}) as a function of integer partition $Y$ assumes its minimal at $[Y] = [N]$: as $[Y]$ varies away from $[Y] = [N]$, $l_1, \cdots, l_q$ starts picking up non-zero values while $l_{q+1} + \cdots + l_N$ remains 1 for a while. In this process,  the left hand side of (\ref{eq:constraint-on-Y}) increases, until $l_{q+1} = l_{q+2} = \cdots = l_N = 0$ forcing the expression to simply equal $N$: this is the maximal value, since the moment $l_{q + 1} + \cdots + l_N$ drops to zero, $l_1 + 2l_2 + \cdots + ql_q$ increases at least by $q + 1$. To summarize,
\begin{equation}
	n + n' \le l_1 + 2l_2 + \cdots + ql_q + q(l_{q+1} + l_{q+2} + \cdots + l_N) = n + 2  \ .
\end{equation}
Hence,
\begin{equation}
	n' \le 2\ , \qquad k = 2, 4 \ .
\end{equation}
When $n' = 2$, the only partition allowed by the constraint (\ref{eq:constraint-on-Y}) is $[Y] = [N]$. When $n' = 1$, there is only $Y = [N-1,1]$. We summarize the allowed configurations for $k > 0$ below.
$$
\begin{tabular}{c c l l}
	\hline
	even $N$ & $k$ & $[Y]$ & comments \\
	\hline
	$N \ (\neq 4)$ & 4 & $[N]$ & $\cong D_{N + 4}(\mathfrak{sl}_4, [4])$ \\
	&&& $\cong D_{N + 4}(\mathfrak{sl}_{N + 8}, [N+4,4])$\\
	$N \ (\neq 4)$ & 2 & $[N-1,1]$ & $\cong D_{N + 2}(\mathfrak{sl}_{N + 4}, [N + 1, 3])$ \\
	$4$ & $4n'+2 \ge 6$ & $[4]$ & $\cong$ first row \\
	& & $[3,1], [2,2], [1^4]$ \\
	$4$ & 2 & $[2,2], [3,1], [2,1^2], [1^4]$ \\
	\hline
\end{tabular}
$$
In this table, we have made some identification of entries utilizing the dualities among type I Argyres--Douglas theories as discussed in \cite{Beem:2023ofp}. Recall that the duality between general type I Argyres--Douglas theories is encoded in the gauging description \cite{Beem:2023ofp}:
\begin{equation}\label{duality1}
	D_p\left(\mathfrak{sl}_N,[Y]\right)\equiv D_p\big(\mathfrak{sl}_{pl+N},[p^l, Y]\big)
\end{equation}
and
\begin{equation}\label{duality2}
	D_p\left(\mathfrak{sl}_N,[Y]\right)\cong D_p\left(\mathfrak{sl}_{pL-N},[Y^c]\right),
\end{equation}
where $[Y^c]=\begin{bmatrix}(p-1)^{l_1},(p-2)^{l_2},\ldots,(p-N)^{l_N}\end{bmatrix}.$ For instance, one finds the duality
\begin{equation}
	\begin{aligned}
		&	D_{4+N}(\mathfrak{sl}_N,[N])\cong D_{4+N}(\mathfrak{sl}_4,[4]) \cong D_{N + 4}(\mathfrak{sl}_{N + 8}, [N + 4, 4])\\
		&  D_{N+2}(\mathfrak{sl}_N,[N-1,1])\cong D_{2+N}(\mathfrak{sl}_{N+4},[N+1,3]).
	\end{aligned}
\end{equation}
This gives the ``comments'' column in the table above.

\vspace{1em}
\underline{Theories with $k < 0$}

For $2-N \leq k = 2n' < 0$, the analysis is similar, where the left hand side of (\ref{eq:constraint-on-Y}) assumes its maximum $N = 2n$ and minimum $q = n + n'$ at $[Y] = [1^N]$ and $[Y] = [N]$, respectively. Given that $n + n' \le n + 2 \le N = 2n$, and the fact that the left hand side of (\ref{eq:constraint-on-Y}) can take all integer values between $n + n'$ and $N$, there are always some partitions $[Y]$ satisfying (\ref{eq:constraint-on-Y}).

However, we should remove unphysical solutions where the partition $[Y] = [r^{l_r > 0}, Y']$ with $r > p$. From the gauge theory description \eqref{typeIrealization}, when $r > p = 2q$, the conjugate partition $[\widetilde{Y}]$ must contain at least one negative entry, $\widetilde{Y} = [\dots, (q - r)^{l_r}, \dots]$. Hence $\widetilde{Y}_- = [\dots, (r - q)^{l_r}, \dots]$ while $(r - q) > (m - 1)q = q$, making the second AD matter $D_{(m - 1)q}(\mathfrak{sl}_{N- \frac{N}{m}}, [\widetilde{Y}_-, 1^{N - \frac{N}{m} - \widetilde{N}_-}])$ in \eqref{typeIrealization} unphysical. We also identify and omit cases that have been listed in the previous table for $k > 0$ using the dualities \eqref{duality1} and \eqref{duality2}. Let us spell out the physical solutions in detail.

Consider $n' = -1$, then $q = n - 1$. The sum $l_{q + 1} + \cdots + l_{N} \in \{0,1,2\}$, where $2$ is assumed when $l_{q + 1} = 1, l_{i \ne q+1} = 0$. It is then straightforward to see that the only possibility of satisfying (\ref{eq:constraint-on-Y}) is $Y = [N-3, \text{partitions of }3]$, except for the special case $n = 4$: in this case, $2q = 2n - 2 = n + 2$, and there is additional solution $Y = [4^2]$.

Consider $n' = -2$, then $q = n - 2$, $p = 2n - 4$. The sum $l_{q + 1} + \cdots + l_{N} \in \{0,1,2\}$. When the sum equals $0$, the left hand side of (\ref{eq:constraint-on-Y}) gives $N = 2n > n + 2$. When the sum equals $2$, $l_{q + 1} + l_{q + 2} = 2$, and the left hand side of (\ref{eq:constraint-on-Y}) gives three possible values,
\begin{equation}
	2q + 0 = 2n - 4, \quad 2q + 1 = 2n - 3, \quad 2q + 2 = 2n - 2 \ .
\end{equation}
Only $n = 5$ is allowed such that $2n - 3 = n + 2$, since $\gcd(n, n') = 1$. Finally, when the sum equals $1$, since $n + 2 = q + 4$, only the partitions $Y = [N - 4, \text{partitions of }4]$ satisfy the constraint (\ref{eq:constraint-on-Y}). However, these are dual to the cases with $n' = 2$ in the previous table, and hence are omitted.

The general cases with $1 - n\le n' < -2$ is a lot trickier. However, we can show that for given $n'$ and relative large $N$ such that $n > |3n'|$, there are no new solutions. The large-$N$ condition implies the following useful inequalities,
\begin{equation}
	3q = 3n + 3n' > 2n, \qquad
	2 - n' < n + n' = q, \qquad
	2n - 2|n'| > n + 2 \ .
\end{equation}
Due to the first inequality, the sum $l_{q + 1} + \cdots + l_N \in \{0,1,2\}$. When the sum equals $0$, the left hand side of (\ref{eq:constraint-on-Y}) gives $N = 2n > n + 2$. When the sum equals $2$, the multiplicities must have $l_{q + 1} + l_{q + 2} + \cdots + l_{q + |n'|} = 2$ and $l_{i > q + |n'|} = 0$. The left hand side of (\ref{eq:constraint-on-Y}) gives a set of possible values,
\begin{equation}
	= 2q + (2n - (2q + i + j)) = 2n - (i + j) \ge 2n - 2|n'|, \qquad i, j \in \{1, 2, \ldots, |n'|\} \ .
\end{equation}
But $2n - 2|n'| > n + 2$, hence no solution. The only remaining choice is to have the sum equals $1$. Since $n + 2 = q \cdot 1 + (2 - n')$, only the partitions $Y = [N - (2 - n'), \text{partitions of }(2 - n')]$ satisfy the constraint (\ref{eq:constraint-on-Y}). However, these are unphysical since $n' < -2 \Rightarrow N + n' - 2 > p$. For $n < |3n'|$, there are solutions which do not exhibit illuminating patterns, some of which are listed in Table \ref{tab:negative-k}.
\begin{table}
	\centering
	\begin{tabular}{c c l l}
		\hline
		even $N$ & $k$ & $[Y]$ \\
		\hline
		4 & $-2$ & $[1^4]$ \\
		6 & $-2$ & $[3,2,1],\ [3,1^3]$ \\
		8 & $-2$ & $[4,4],\ [5,3],\ [5,2,1],\ [5,1^3]$ \\
		10 & $-4$ & $[5,4,1]$ \\
		10 & $-6$ & $[3^3,1]$ \\
		14 & $-6$ & $[7,6,1]$ \\
		16 & $-6$ & $[9,7], [8,8]$ \\
		$N\geq 10$ & $-2$ & $[N-3,1^3]$ & $\cong D^{b=2}_{N-2}(\mathfrak{sl}_3, [1^3])$ \\
		& & $[N-3,3]$ & $\cong D^{b=2}_{N-2}(\mathfrak{sl}_3, [3])$ \\
		& & $[N-3,2,1]$ & $\sim D^{b=2}_{N-2}(\mathfrak{sl}_3, [2,1])$ \\
		\hline
	\end{tabular}
	\caption{Cases with $k < 0$ and $N \le 10$.\label{tab:negative-k}}
\end{table}
Note that the three infinite families at the bottom of the table can be identified as type II Argyres--Douglas theories with $b = 2$ through the duality \eqref{general duality}, up to some free hypermultiplets (as indicated by the $\sim$ sign).



	

In the above we have identified a few infinite series of AD theories that can be realized as $SU(2)$ gauge theories coupled to two AD matters without exactly marginal deformations. Their central charges satisfy interesting relations, analogous to the relation $a_\text{4d} - c_\text{4d} = 0$ of the $\mathcal{N} = 4$ theories and $\mathcal{T}_{p, N}$ theories \cite{Kang:2021lic,Jiang:2024baj,Pan:2025gzh}.
{\renewcommand{\arraystretch}{1.8}
	\begin{center}
	\begin{tabular}{c c}
	Theories & $a_\text{4d} - c_\text{4d}$\\
	\hline
	$D_{N -4} (\mathfrak{sl}_N, [N-4,4])$ & $\displaystyle - \frac{1}{12} + \frac{1}{4(N - 4)}$ \\
	$D_{N - 2}(\mathfrak{sl}_N, [N - 3, 3])$ & $\displaystyle - \frac{1}{12}$\\
	$D_{N - 2}(\mathfrak{sl}_N, [N -3, 2, 1])$ & $\displaystyle- \frac{1}{8}$\\
	$D_{N - 2}(\mathfrak{sl}_N, [N -3, 1, 1, 1])$ & $\displaystyle- \frac{5}{24}$\\
	\end{tabular}
\end{center}}
In the following discussions we will compute the large-$N$ limit of the Schur index and Wilson line index of the first two infinite families. It would be very interesting to explore the holography interpretation.


\section{Schur index and non-vacuum characters}\label{sec:simple-examples}

The Schur index can be defined for any 4d $\mathcal{N} = 2$ SCFT, given by the trace formula
\begin{equation}
  \mathcal{I}(q,b) = q^{\frac{c_\text{4d}}{2}} \operatorname{tr}_\mathcal{H} q^{E - R} b^f \ .
\end{equation}
Here $E$ denotes the scaling dimension, $R$ the $SU(2)_\mathcal{R}$ charge and $b$ the flavor fugacity associated to the flavor charge $f$. It is also the Schur limit $t \to q$ of the full $\mathcal{N} = 2$ superconformal index $\mathcal{I}(p,q,t; b)$. The index can be viewed as an $S^3 \times S^1$ partition function with suitable background $U(1)_r, SU(2)_\mathcal{R}$ gauge field \cite{Pan:2019bor,Dedushenko:2019yiw}. In the 4d/2d correspondence \cite{Beem:2013sza}, the Schur index is identified with the vacuum character of the associated VOA, and also viewed as a $T^2$ partition function where the complex moduli of $T^2$ is denoted by $\tau$, with $q = e^{2\pi i \tau}$ in the Schur index. In this section we explicitly compute a few examples of the Schur index for the type A Argyres-Douglas theories using the integration formula proposed in \cite{Pan:2021mrw}.

\subsection{Index of \texorpdfstring{$D_p(\mathfrak{sl}_N,[Y])$}{} with \texorpdfstring{$\gcd(p, N) = 1$}{}}

From the duality \eqref{typeIrealization}, the basic building blocks in the decomposition of type I AD theories are those with $\gcd(p, N) = \gcd(k,N) = 1$. These theories have been studied extensively \cite{Xie:2019zlb,Song:2017oew,Xie:2016evu,Song:2015wta}. When $[Y] = [1^N]$ is the maximal regular puncture, the associated VOA is given by the affine Kac-Moody algebra $\widehat{\mathfrak{su}}(N)_{k_\text{2d}}$, where the level
\begin{equation}
  k_\text{2d} = - N + \frac{N}{p} 
\end{equation}
is boundary admissible. For general regular puncture $[Y]$, the associated VOA is given by the simple $\mathcal{W}$-algebra $W_{k_\text{2d}}(\mathfrak{su}(N), Y)$ obtained from the quantum Drinfeld-Sokolov reduction of $\widehat{\mathfrak{su}}(N)_{k_\text{2d}}$ specified by $[Y]$. The Schur index of $D_p(\mathfrak{sl}_N, [Y])$ or the vacuum character of the associated VOA is well-known \cite{2016arXiv161207423K,Xie:2019zlb}, and can be written in the form of plethystic exponential.



For a generic regular puncture labeled by $[Y]$ the corresponding flavor symmetry is given by the commutant $G_F$ of the $SU(2)$ embedding specified by the partition $[Y]$. Explicitly, writing $[Y] = [1^{l_1}, \dots, N^{l_N}]$, the flavor symmetry is given by $G_F = (\prod_{i = 1}^N U(l_i))/U(1)$. The adjoint representation of $SU(N)$ decomposes under $SU(2) \times G_F$ as  
\begin{equation}
	\mathfrak{g}=\sum_jV_j\otimes R_j,
\end{equation}
where $V_j$ denotes the spin-$j$ representation of $SU(2)$ and $R_j$ is the corresponding representation of $G_F$. Note that $j \in \mathbb{Z}$. The Schur index of $D_p(\mathfrak{sl}_N, [Y])$ then takes the form \cite{Xie:2019zlb} 
\begin{equation}
	\operatorname{PE}\left[\frac{\sum_jq^{1+j}\chi_{R_j}(z)-q^{N+k}\sum_jq^{-j}\chi_{R_j}(z)}{(1-q)(1-q^{N+k})}\right].
\end{equation}

\textbf{Example:~$D_p(\mathfrak{sl}_N,[1^N])$} 

The associated flavor symmetry $G_F = SU(N)$, and the decomposition is trivial.  
The only non-trivial contribution corresponds to $j=0$ with $\chi_{R_j} = \chi_{\mathrm{adj}}$, leading to the Schur index  
\begin{equation}
	\operatorname{PE}\left[\frac{q-q^{N+k}}{(1-q)(1-q^{N+k})}\chi_{\text{adj}}(z)\right].
\end{equation}

\textbf{Example:~$D_p(\mathfrak{sl}_N,[r^m,1^{N-rm}])$}

The associated flavor symmetry $G_F = SU(N - rm) \times SU(m) \times U(1)$.  
The decomposition of the adjoint representation of $\mathfrak{sl}_N$ under this symmetry reads 
\begin{align}
	\chi_\text{adj}^{\mathfrak{sl}_N}
  = & \ \chi_\text{adj}^{SU(N-rm)}\\
  & \ +\sum_{j=0}^{r-1}\chi_{V_j}(\chi_\text{adj}^{SU(m)}+1)+(a\chi_{\square}^{SU(m)}\chi_{\bar{\square}}^{SU(N-rm)}+a^{-1}\chi_{\bar{\square}}^{SU(m)}\chi_{\square}^{SU(N-rm)})\chi_{V_{\frac{r-1}{2}}}, \nonumber
\end{align}
where $a$ is the fugacity associated with the $U(1)$ symmetry, and $\square, \bar square$ denote respectively the fundamental and anti-fundamental representations of the corresponding $SU$ group. The Schur index takes the form  
\begin{align}\label{schurrm}
		& \operatorname{PE}\Bigg[ 
		\frac{1}{(1 - q)(1 - q^{N + k})} \Big( 
		(q - q^{N + k}) \, \chi_{\text{adj}}^{SU(N - rm)} + (\chi_{\text{adj}}^{SU(m)} + 1) \sum_{i = 1}^r (q^i - q^{N + k - i + 1}) \nonumber \\
		& \qquad + \left(q^{\frac{r+1}{2}} - q^{N + k - \frac{r - 1}{2}}\right) 
		(a\chi_{\square}^{SU(m)}\chi_{\bar{\square}}^{SU(N-rm)}+a^{-1}\chi_{\bar{\square}}^{SU(m)}\chi_{\square}^{SU(N-rm)})
		\Big) 
		\Bigg].
\end{align}

\paragraph{Vector and hypermultiplet contributions.}  
In addition to the above building blocks, we must also include the contributions from vector multiplets and hypermultiplets.  

The Schur index of a hypermultiplet in the representation $R$ of the flavor symmetry includes contributions from a pair of half-hypermultiplets transforming in $R$ and its conjugate $\bar{R}$: 
\begin{equation}
	\mathrm{PE}\left[\frac{q^{\frac{1}{2}}}{(1-q)}(a\chi_R(\mathbf{x})+a^{-1}\chi_{\bar{R}}(\mathbf{x}))\right],
\end{equation}
where $a$ denotes the $U(1)$ fugacity which rotates the scalars $q, \bar q$ oppositely, and $\chi_R(\mathbf{x})$ is the character of $R$ under the flavor symmetry.  
The Schur index of a vector multiplet is  
\begin{equation}
	\mathrm{PE}\left[-\frac{2q}{1-q}\chi_{\text{adj}}(\mathbf{z})\right].
\end{equation}

\paragraph{Unphysical theories from the Schur index}

There is a physical consistency requirement arising from the structure of the Schur index. 
Recall that the Schur index involves the plethystic exponential of 
\begin{equation}
	\frac{\sum_jq^{1+j}\chi_{R_j}(z)-q^{h^\vee+k}\sum_jq^{-j}\chi_{R_j}(z)}{(1-q)(1-q^{h^\vee+k})} \ .
\end{equation}
The sum over spin-$j$ can be split into two parts,
\begin{equation}
  \sum_{j} \frac{q^{1 + j} - q^{p - j}}{1-q} \chi_{\mathcal{R}_j}(z)
  = \sum_{j \ge p} \frac{q^{1 + j} - q^{p - j}}{1-q} \chi_{\mathcal{R}_j}(z)
  + \sum_{j < p} \frac{q^{1 + j} - q^{p - j}}{1-q} \chi_{\mathcal{R}_j}(z) \ .
\end{equation}
The latter sum can be expanded as
\begin{equation}
  = \sum_{j < p}(q^{1 + j} + q^{2 + j} + \cdots + q^{p - 2j -2}) \chi_{\mathcal{R}_j}(z) \ ,
\end{equation}
while the former
\begin{equation}
  = - \sum_{j \ge p} q^{p - j}(q^0 + q^1 + \cdots + q^{2j - p}) \chi_{\mathcal{R}_j}(z) \ .
\end{equation}
Each integer spin $j \ge p$ contributes one $\frac{-1}{1 - q^p}$ in the plethystic exponential, leading to a factor
\begin{equation}
  \operatorname{PE}\Big[\sum_{\substack{j \in \mathbb{N}\\ j \ge p}}^{n} \frac{-1}{1-q^p}\Big] = (1;q^p)^{\# \text{integer spin $j \ge p$}} = 0 \ ,
\end{equation}
which is unphysical.


Now consider a general regular puncture specified by $Y=[r^m,h^n,\ldots],$ with $ r\geq h \geq \cdots .$ The fundamental representation of $SU(N)$ then decomposes as 
\begin{equation}
	\chi_{\square}^{SU(N)}=\chi_{\frac{r-1}{2}}^{SU(2)}\chi_{\square}^{SU(m)}+\cdots.
\end{equation}
Consequently, the adjoint representation necessarily contains a factor with maximal spin-$j$:
\begin{equation}
	\chi_{\frac{r-1}{2}}^{SU(2)}\chi_{\frac{r-1}{2}}^{SU(2)}=\chi_{r-1}^{SU(2)}+\cdots.
\end{equation}
Therefore, the consistency requirement becomes 
\begin{equation}
	p>r-1.
\end{equation}

\subsection{\texorpdfstring{$D_{6}(\mathfrak{sl}_4, [2,2])$}{D6(sl(4), [2,2])}}

We consider the $D_{6}(\mathfrak{sl}_4, [2,2])$ theory as our first non-trivial example of an type I AD theory with a duality frame as an $SU(2)$ gauge theory. In this and the following simple examples, we will compute analytically the flavored Schur index using the integration formula \cite{Pan:2021mrw}
\begin{equation}
  \oint \frac{da}{2\pi i a}\mathcal{Z}(\mathfrak{a})
  = \mathcal{Z}(\mathfrak{a}_0) + \sum_j R_j E_1 \begin{bmatrix}
    -1 \\ a_j/a_0 q^{\pm \frac{1}{2}}
  \end{bmatrix}(\tilde \tau) \ .
\end{equation}
Here we follow the convention $a = e^{2\pi i \mathfrak{a}}$, $q = e^{2\pi i \tau}$. The integrand $\mathcal{Z}(\mathfrak{a})$ is an elliptic function with double periodicity $\mathcal{Z}(\mathfrak{a}) = \mathcal{Z}(\mathfrak{a} + 1) = \mathcal{Z}(\mathfrak{a} + \tilde \tau)$; note that $\tilde \tau$ is not necessarily the same $\tau$ that defines the Schur index. For the formula to work, $\mathcal{Z}(\mathfrak{a})$ has only simple poles $\mathfrak{a}_j$ within the fundamental parallelogram spanned by $1$ and $\tilde \tau$. The residues $R_j$ are given by
\begin{equation}
  R_j = 2\pi i \operatorname{Res}_{\mathfrak{a} \to \mathfrak{a}_j} \mathcal{Z}(\mathfrak{a}) = \operatorname{Res}_{a \to a_j} \frac{1}{a}\mathcal{Z}(\mathfrak{a}_j) \ .
\end{equation}
Finally, poles $\mathfrak{a}_j = \operatorname{real} + \lambda \tau$ with a positive $\lambda$ is called an imaginary pole, and a real pole otherwise. The factor $q^{\pm \frac{1}{2}} = q^{\frac{1}{2}}$ or $q^{- \frac{1}{2}}$ when $\mathfrak{a}_j$ is real or imaginary, respectively.

Once the flavored index is obtained, we take the unflavored limit to obtain the unflavored Schur index in closed form, and find the unflavored modular differential equation that it satisfies. Although the residues $R_j$ that appear in the computation individually do not have unflavoring limit, we will show in all examples that there exist non-trivial combinations that admit unflavoring limit, which give rise to additional solutions to the unflavored modular differential equation. This phenomenon suggest close relation between the residues and the module character of the associated VOA, which we leave for future investigation.

The gauge theory description of $D_{6}(\mathfrak{sl}_4, [2,2])$ is given by
\begin{equation}
	\begin{tikzpicture}[baseline=(current bounding box.center)]
		\node (left) {$D_{3}(\mathfrak{sl}_2,[1,1])$};
		\node (mid) [right=of left, xshift=0.5cm] {$\mathfrak{su}(2)$};
		\node (right) [right=of mid, xshift=0.5cm] {$D_{3}(\mathfrak{sl}_4,[1,1,1,1])$};
		\draw[-{Stealth[scale=1.2]}] (mid) -- (right);
		\draw[-{Stealth[scale=1.2]}] (mid) -- (left); 
		\node (box) [draw, rectangle, below=0.5cm of mid] {$0$};
		\draw (mid) -- (box);
	\end{tikzpicture}
\end{equation}
Concretely, we read off the following data
\begin{align}
  & p = 6, N = 4, m = 2, k = 3, q = \frac{p}{m} = 3, Y = [2,2],  L = 2\\
  & \widetilde Y = [1,1], \quad \widetilde Y_+ = [1,1], \quad \widetilde Y = \emptyset,\\
  & N - \frac{N}{m} - \tilde N_- = 2, \quad qL - \frac{N}{m} = 4, \quad N - \frac{N}{m} = 2
\end{align}

The building blocks are $D_3(\mathfrak{sl}_2, [1^2])$ and $D_3(\mathfrak{sl}_4, [1^4])$, with the Schur index given by
\begin{align}
	\mathcal{I}_{D_3(\mathfrak{sl}_2, [1^2])} = & \ \operatorname{PE}\left[\frac{q-q^{2+1}}{(1-q)(1-q^{2 + 1})}(a^2 + a^{-2} + 1)\right] \ ,\\
  \mathcal{I}_{D_3(\mathfrak{sl}_4, [1^4])} = & \ \operatorname{PE}\left[\frac{q-q^{4+(-1)}}{(1-q)(1-q^{4+(-1)})}\Bigg( - 1 + \sum_{A, B = 1}^{4}\frac{a_A}{a_B}\Bigg) \right]_{a_1 a_2 a_3 a_4 = 1} \ .
\end{align}
Upon gauging the $SU(2)$ subgroup, we identify the fugacities
\begin{equation}
  a_1 = a b_1, \qquad
  a_2 = a^{-1}b_1, \qquad
  a_3 = b_1 b_2, \qquad
  a_4 = b_1^{-3}b_2^{-1} \ .
\end{equation}

The Schur index of $D_6(\mathfrak{sl}_4, [2,2])$ is given by the contour integral
\begin{equation}
  \oint \frac{da}{2\pi i a} \mathcal{I}_{D_3(\mathfrak{sl}_2, [1^2])}(\mathfrak{a}) \mathcal{I}_{D_3(\mathfrak{sl}_4, [1^4])}(\mathfrak{a}, \mathfrak{b}) \mathcal{I}_\text{VM}(\mathfrak{a})
  = \oint \frac{da}{2\pi i a}\mathcal{Z}(\mathfrak{a}) \ ,
\end{equation}
where the integrand can be rewritten in terms of $\vartheta_1$ functions,
\begin{align}
  \mathcal{Z}(\mathfrak{a})
  = & \ \frac{\eta(3\tau)^{15}}{q^{\frac{15}{4}}\eta(\tau)^3}
  \frac{\vartheta_1(2 \mathfrak{a})^2 \vartheta_1(4 \mathfrak{b}_1 + 2 \mathfrak{b}_2|3\tau)}{
    \prod_{\alpha=\pm} \vartheta_1(2\alpha \mathfrak{a} + \tau|3\tau)^2
  } \nonumber\\
  & \times \frac{1}{\prod_{\alpha, \beta = \pm}
    \vartheta_1(\alpha \tau + \beta \mathfrak{a} + \mathfrak{b}_2|3\tau)
    \vartheta_1(\alpha \tau + \beta \mathfrak{a} + 4 \mathfrak{b}_1 + \mathfrak{b}_2|3\tau)}\ ,
\end{align}
The integrand contains apparent double poles.
Howeveer, using the identity
\begin{equation}
  \prod_{\alpha = -1,0,1}\vartheta_1(\alpha \tau + \mathfrak{z}|3\tau)
  = \frac{q^{-1/3} \eta(3\tau)^3}{\eta(\tau)} \vartheta_1(\mathfrak{z})\ ,
\end{equation}
the integrand can be further simplified to
\begin{align}
  \mathcal{Z}(\mathfrak{a}) = -\frac{1}{2}\frac{\eta(\tau)^3}{q^{7/4}\eta(3\tau)^3}
  & \ 
  \frac{\vartheta_1(4\mathfrak{b}_1 + 2 \mathfrak{b}_2|3\tau)}{\vartheta_1(4\mathfrak{b}_1 + 2 \mathfrak{b}_2)}
  \nonumber \\
  & \times \prod_{\alpha = \pm} \frac{
    \vartheta_1(2 \alpha \mathfrak{a}|3\tau)
    \vartheta_1(\alpha \mathfrak{a} + \mathfrak{b}_2 |3\tau)
    \vartheta_1(\alpha \mathfrak{a} + 4 \mathfrak{b}_1 + \mathfrak{b}_2|3\tau)
  }{
    \vartheta_1(\alpha \mathfrak{a} + \mathfrak{b}_2)
    \vartheta_1(\alpha \mathfrak{a} + 4 \mathfrak{b}_1 + \mathfrak{b}_2)
  } \ .
\end{align}
After the rewriting, the double poles are removed by canceling with zeroes. This integrand is elliptic with respect to $\mathfrak{a}$ with period $3\tau$,
\begin{equation}
  \mathcal{Z}(\mathfrak{a} + 3\tau) = \mathcal{Z}(\mathfrak{a} + 1) = \mathcal{Z}(\mathfrak{a}) \ .
\end{equation}

To apply the integration formula, we identify the relevant poles,
\begin{align}
  \pm \mathfrak{b}_2 + \ell \tau, \qquad \pm (4\mathfrak{b}_1 + \mathfrak{b}_2) + \ell \tau, \qquad \ell = 0, 1, 2\ .
\end{align}
Note that $\pm \mathfrak{b}_{1}$ and $\pm (4 \mathfrak{b}_1 + \mathfrak{b}_2)$ are cancelled by a zero in the numerator, hence there are only eight non-zero residues, which sum to zero since the integrand is elliptic. The integration formula leads to the closed form,
\begin{align}
  & \ \mathcal{I}_{D_6(\mathfrak{sl}_4, [2,2])}(q, b)\nonumber \\
  = & \ - \frac{\eta(3\tau)^6}{
    q^{23/12} \eta(\tau)^2 \vartheta_1(4 \mathfrak{b}_1|3\tau) \vartheta_1(4\mathfrak{b}_1 + 2\mathfrak{b}_2)} \nonumber \\
    & \ \times \sum_{\pm} \mp  \bigg(
    \frac{ \vartheta_1(\pm \tau + 2 \mathfrak{b}_2|3\tau)}{
      \vartheta_1(\mp \tau + 4 \mathfrak{b}_1|3\tau)
      \vartheta_1(\pm \tau + 4 \mathfrak{b}_1 + 2\mathfrak{b}_2|3\tau)
      \vartheta_1(2\mathfrak{b}_2|3\tau)
    }E_1 \begin{bmatrix}
      -1 \\ b_2 q^{\pm \frac{1}{2}}
    \end{bmatrix}(3\tau)\\
    & \ \qquad + \frac{\vartheta_1(\pm \tau + 8 \mathfrak{b}_1 + 2 \mathfrak{b}_2|3\tau)}{
      \vartheta_1(\pm \tau + 4 \mathfrak{b}_1|3\tau)
      \vartheta_1(\pm \tau + 4 \mathfrak{b}_1 + 2 \mathfrak{b}_2|3\tau)
      \vartheta_1(8 \mathfrak{b}_1 + 2 \mathfrak{b}_2|3\tau)
    } E_1 \begin{bmatrix}
      -1 \\ b_1^4 b_2 q^{\pm \frac{1}{2}}
    \end{bmatrix}(3\tau)
    \bigg) \ . \nonumber
\end{align}
The expression is a sum of four terms, where the coefficients of the Eisenstein series are the residues from the poles. Let us denote these residues as $R_i$,
\begin{align}
  R_1 = & \ \Lambda 
    \frac{\vartheta_1(2 \mathfrak{b}_2 + \tau|3\tau)}{\vartheta_1(2 \mathfrak{b}_2|3\tau) \vartheta_1(4 \mathfrak{b}_1 - \tau|3\tau) \vartheta_1(4 \mathfrak{b}_1 + 2 \mathfrak{b}_2 + \tau|3\tau)} \ ,\\
  R_2 = & \ - \Lambda
    \frac{\vartheta_1(2 \mathfrak{b}_2 - \tau|3\tau)}{\vartheta_1(2 \mathfrak{b}_2|3\tau) \vartheta_1(4 \mathfrak{b}_1 + \tau|3\tau) \vartheta_1(4 \mathfrak{b}_1 + 2 \mathfrak{b}_2 - \tau|3\tau)} \ ,\\
  R_3 = & \ \Lambda \frac{\vartheta_1(8 \mathfrak{b}_1 + 2 \mathfrak{b}_2 + \tau|3\tau)}{\vartheta_1(8 \mathfrak{b}_1 + 2 \mathfrak{b}_2|3\tau) \vartheta_1(4 \mathfrak{b}_1 + \tau|3\tau) \vartheta_1(4 \mathfrak{b}_1 + 2 \mathfrak{b}_2 + \tau|3\tau)} \ , \\
  R_4 = & \ - \Lambda \frac{\vartheta_1(8 \mathfrak{b}_1 + 2 \mathfrak{b}_2 - \tau|3\tau)}{\vartheta_1(8 \mathfrak{b}_1 + 2 \mathfrak{b}_2|3\tau) \vartheta_1(4 \mathfrak{b}_1 - \tau|3\tau) \vartheta_1(4 \mathfrak{b}_1 + 2 \mathfrak{b}_2 - \tau|3\tau)} \ ,
\end{align}
where
\begin{equation}
  \Lambda = \frac{\eta(3\tau)^6}{q^{23/12} \eta(\tau)^2 \vartheta_1(4 \mathfrak{b}_1|3\tau) \vartheta_1(4\mathfrak{b}_1 + 2\mathfrak{b}_2)}\ .
\end{equation}
As residues of an elliptic function with only simple poles, these four residues satisfy
\begin{equation}
  R_1 - R_2 - R_3 + R_4 = 0 \ .
\end{equation}

It is straightforward to take the unflavored limit given in a closed-form,
\begin{align}
  & \ \mathcal{I}_{D_{6}(\mathfrak{sl}_4, [2,2])}(q) \nonumber\\
  = & \ -\frac{i \vartheta_1^{(3)}(0|3\tau) \vartheta_1^{(1)}(\tau |3\tau)^2}{48 \pi ^5 \eta (3\tau) \vartheta_1(\tau |3\tau)^3}+\frac{i \vartheta_1^{(3)}(0|3\tau) \vartheta_1^{(2)}(\tau |3\tau)}{48 \pi ^5 \eta (3\tau) \vartheta_1(\tau |3\tau)^2}+\frac{3 i \vartheta_1^{(1)}(\tau |3\tau)^4}{16 \pi ^4 \vartheta_1(\tau |3\tau)^5} \nonumber\\
  & \ -\frac{i \vartheta_1^{(2)}(\tau |3\tau) \vartheta_1^{(1)}(\tau |3\tau)^2}{4 \pi ^4 \vartheta_1(\tau |3\tau)^4}
  +\frac{3 \vartheta_1^{(2)}(\tau |3\tau) \vartheta_1^{(1)}(\tau |3\tau)}{8 \pi ^3 \vartheta_1(\tau |3\tau)^3}+\frac{i \vartheta_1^{(3)}(\tau |3\tau) \vartheta_1^{(1)}(\tau |3\tau)}{6 \pi ^4 \vartheta_1(\tau |3\tau)^3} \nonumber\\
  & \ -\frac{\vartheta_1^{(3)}(\tau |3\tau)}{8 \pi ^3 (\vartheta_1(\tau |3\tau))^2}
  -\frac{3 i \vartheta_1^{(2)}(\tau |3\tau)^2}{32 \pi ^4 \vartheta_1(\tau |3\tau)^3}
  -\frac{\vartheta_1^{(1)}(\tau |3\tau)^3}{4 \pi ^3 \vartheta_1(\tau |3\tau)^4}
  -\frac{i \vartheta_1^{(4)}(\tau |3\tau)}{96 \pi ^4 \vartheta_1(\tau |3\tau)^2} \ .
\end{align}
Expanded in $q$-series, the Schur index reads
\begin{align}
  \mathcal{I}_{D_{6}(\mathfrak{sl}_4, [2,2])}(q)
  = & \  q^{\frac{7}{4}}(1 + 4 q + 20 q^2 + 76 q^3 \nonumber \\
  & \ + 263 q^4 + 816 q^5 + 2381 q^6 + 6482 q^7 + 16841 q^8 + 41776 q^9 + 99863 q^{10} \nonumber \\
  & \ + 230608 q^{11} + 517077 q^{12} + 1128130 q^{13} + 2402649 q^{14 \nonumber}\\
  & \ + 5004024 q^{15} + 10213532 q^{16} + 20459252 q^{17} + 40283830 q^{18} \nonumber \\
  & \ + 78055088 q^{19} + 149007235 q^{20} + 280514896 q^{21}  + 521243939 q^{22}\nonumber \\
  & \ + 956743326 q^{23} + 1735941489 q^{24} + 3115550718 q^{25} + \cdots
  ) \ . 
\end{align}
The unflavored index satisfies a 12-th order unflavored monic MLDE,
\begin{align}
  0 = \Big[& \ D^{(12)}_q - 1490 E_4 D^{(10)}_q - 68460 E_6 D^{(9)}_q - 644625 E_4^2 D^{(8)}_q \nonumber \\
 & \ - 8022000 E_6 E_4 D^{(7)}_q + 56202500 E_4^3 D^{(6)}_q - 212170000 E_6^2 D^{(6)}_q \nonumber \\
 & \ - 954765000 E_6 E_4^2 D^{(5)}_q - 9022740625 E_4^4 D^{(4)}_q \nonumber \\
 & \ - 17200960000 E_6^2 E_4 D^{(4)}_q 
  - 441015750000 E_6 E_4^3 D^{(3)}_q - 174724200000 E_6^3 D^{(3)}_q \nonumber \\
 & \ - 1100917031250 E_4^5 D^{(2)}_q - 9954509250000 E_6^2 E_4^2 D^{(2)}_q \nonumber \\
 & \ - 55495407187500 E_6 E_4^4 D^{(1)}_q - 75831126000000 E_6^3 E_4 D^{(1)}_q \nonumber \\
 & \ - 49750337109375 E_4^6 - 587402077500000 E_6^2 E_4^3 \nonumber \\
 & \ - 137044278000000 E_6^4 \Big] \mathcal{I}_{D_{6}(\mathfrak{sl}_4, [2,2])}  (q)= 0 \ .
\end{align}
Here $E_n$ denotes $E_n(\tau)$ for brevity. The equation has ten rational indicial roots and two irrational indicial roots,
\begin{equation}
  \alpha = - \frac{1}{4}, \quad \bigg[\frac{1}{12}\bigg]_2, \quad \bigg[\frac{5}{12}\bigg]_2, \quad \bigg[\frac{3}{4}\bigg]_2, \quad \frac{13}{12}, \quad \frac{17}{12}, \frac{7}{4}, \frac{1}{12}(27 \pm 2 \sqrt{5}) \ ,
\end{equation}
where the subscripts denote the multiplicity. The presence of irrational indicial roots suggests the existence of another unflavored non-monic MLDE, which is of 10-th order but at modular-weight 32,
\begin{align}
0 = & \ \Big[\left( 20 E_4^3 + 147 E_6^2 \right) D^{(10)}_q 
- 3360 E_6 E_4^2 D^{(9)}_q 
- \left( 19500 E_4^4 + 84525 E_6^2 E_4 \right) D^{(8)}_q \nonumber\\
& \ + \left( 1932000 E_4^3 E_6 - 6379800 E_6^3 \right) D^{(7)}_q 
- \left( 6195000 E_4^5 - 12090750 E_4^2 E_6^2 \right) D^{(6)}_q \nonumber\\
& \ + \left( 508620000 E_4^4 E_6 - 1289427000 E_4 E_6^3 \right) D^{(5)}_q \nonumber\\
& \ + \left( 75625000 E_4^6 - 1073896250 E_6^2 E_4^3 + 172872000 E_6^4 \right) D^{(4)}_q \\
& \ - \left( 6457500000 E_6 E_4^5 + 10109925000 E_6^3 E_4^2 \right) D^{(3)}_q \nonumber\\
& \ - \left( 36360937500 E_4^7 - 154931109375 E_6^2 E_4^4 + 817900650000 E_6^4 E_4 \right) D^{(2)}_q \nonumber\\
& \ + 525000E_6\left( 2267000 E_4^6-15161335 E_6^2 E_4^3+2247336 E_6^4 \right) D^{(1)}_q \nonumber\\
& \ - \left( 1009687500000 E_4^8 - 6180571289062.5 E_6^2 E_4^5 + 51391350000000 E_6^4 E_4^2 \right) \Big] \mathcal{I}_{D_6(\mathfrak{sl}_4, [2,2])} .\nonumber
\end{align}
The indicial roots are precisely the previous ten rational indicial roots. The largest root $\frac{7}{4}$ corresponds to the Schur index, while the minimal root $\alpha_\text{min}$ is related to the 4d central charges,
\begin{equation}
  6 = -24 \alpha_\text{min} = c_\text{eff} = 48 (c_\text{4d} - a_\text{4d}), \qquad c_\text{4d} = \frac{11}{4}, a_\text{4d} = \frac{21}{8} \ .
\end{equation}
The four residues $R_i$ have precisely one linear combination that admits smooth limit $b_i \to 1$, which leads to an additional solution to both of the unflavored equations:
\begin{align}
  &\ \frac{1}{2}(R_1 + R_2 + R_3 + R_4)|_{\mathfrak{b}_i \to 0} \nonumber\\
  = & \ - \frac{2}{\eta(\tau)^6} \bigg(
    E_1 \begin{bmatrix}
      1 \\ q
    \end{bmatrix}(3\tau)^3
    + 3 E_1 \begin{bmatrix}
      1 \\ q
    \end{bmatrix}(3\tau)E_2 \begin{bmatrix}
      1 \\ q
    \end{bmatrix}(3\tau)
    + 3 E_3 \begin{bmatrix}
      1 \\ q
    \end{bmatrix}(3\tau)
  \bigg) \nonumber \\
  = & \ q^{\frac{7}{4}}(\frac{1}{q} + 9 + 54 q + 246 q^2 + 954q^3 + 3267q^4 + 10215q^5
  + 29637q^6  + \cdots)  \ .
\end{align}

\subsection{\texorpdfstring{$D_{6}(\mathfrak{sl}_4, [3,1])$}{D6(sl(4), [3,1])}{}}

According to the duality~\eqref{typeIrealization}, the theory can be decomposed as  
\begin{equation}
	\begin{tikzpicture}[baseline=(current bounding box.center)]
		\node (left) {$D_{3}(\mathfrak{sl}_4,[2,1,1])$};
		\node (mid) [right=of left, xshift=0.5cm] {$\mathfrak{su}(2)$};
		\node (right) [right=of mid, xshift=0.5cm] {$D_{3}(\mathfrak{sl}_2,[1,1])$};
		\draw[-{Stealth[scale=1.2]}] (mid) -- (right);
		\draw[-{Stealth[scale=1.2]}] (mid) -- (left); 
		\node (box) [draw, rectangle, below=0.5cm of mid] {$1$};
		\draw (mid) -- (box);
	\end{tikzpicture}
\end{equation}
The Schur index of $D_3(\mathfrak{sl}_2,[1,1])$ is given by 
\begin{equation}
	\operatorname{PE}\left[\frac{q-q^{3}}{(1-q)(1-q^{3})}\chi_{\mathrm{adj}}(z)\right] \ ,
\end{equation}
and the schur index of $D_{3}(\mathfrak{sl}_4,[2,1,1])$ takes the form
\begin{equation}
	\begin{aligned}
		& \operatorname{PE}\Bigg[ 
		\frac{1}{(1 - q)(1 - q^{3})} \Big( 
		(q - q^{3}) \, \chi_{\text{adj}}^{SU(2)} +  \sum_{i = 1}^r (q^i - q^{4- i }) \\
		& \qquad \qquad\qquad\qquad \qquad + \left(q^{\frac{3}{2}} - q^{\frac{5}{2}}\right) 
		(a\chi_{\bar{\square}}^{SU(2)}+a^{-1}\chi_{\square}^{SU(2)})
		\Big) 
		\Bigg].
	\end{aligned}
\end{equation}
Collecting all building blocks, the Schur index of $D_{6}(\mathfrak{sl}_4, [3,1])$ is given by the contour integral
\begin{equation}
  \mathcal{I}_{D_6(\mathfrak{sl}_4, [3,1])}
  = \oint \frac{da}{2\pi i a} \frac{\eta(\tau)\eta(3\tau)^3\vartheta_1(2 \mathfrak{a}|3\tau)^2}{2 q^{\frac{5}{6}}\prod_\pm\vartheta_4(\pm \mathfrak{a}+ \mathfrak{b}_1|3\tau) \vartheta_4(\pm \mathfrak{a} + \mathfrak{b}_2)}
   = \oint \frac{da}{2\pi i a} \mathcal{Z}(\mathfrak{a})\ .
\end{equation}
The integrand $\mathcal{Z}(\mathfrak{a})$ is elliptic with period $3\tau$, hence there are eight imaginary poles within the fundamental parallelogram,
\begin{equation}
  \pm \mathfrak{b}_1+\frac{3 \tau }{2},\qquad
  \pm \mathfrak{b}_2+\frac{\tau }{2}, \qquad
  \pm\mathfrak{b}_2+\frac{3 \tau }{2},\qquad
  \pm \mathfrak{b}_2+\frac{5 \tau }{2} \ .
\end{equation}
The sum of residues of these poles vanish since the integrand is elliptic. Using the integration formula, the Schur index reads
\begin{align}
  \mathcal{I}_{D_6(\mathfrak{sl}_4, [3,1])}
  = & \  \frac{i\eta(\tau)\vartheta_1(2 \mathfrak{b}_1|3\tau)}{
      2q^{\frac{5}{6}}\prod_\pm\vartheta_1(\mathfrak{b}_1 \pm \mathfrak{b}_2)} \sum_{\alpha= \pm}\alpha E_1 \begin{bmatrix}
        -1 \\ b_1^\alpha
      \end{bmatrix} (3\tau) \nonumber\\
    & \ - \frac{i \eta(3\tau)^3 \vartheta_1(2 \mathfrak{b}_2|3\tau)^2}{
      2q^{\frac{5}{6}} \eta(\tau)^2 \vartheta_1(2\mathfrak{b}_2)
      \prod_\pm\vartheta_1(\mathfrak{b}_1 \pm \mathfrak{b}_2)
    }\sum_{\alpha = \pm}\alpha E_1 \begin{bmatrix}
      -1 \\ b_2^\alpha
    \end{bmatrix} (3\tau)\\
    & \ - \frac{i q^{\frac{1}{6}} \eta(3\tau)^3 }{
      2\vartheta_1(2 \mathfrak{b}_2)
      \eta(\tau)^2
    }\sum_{\alpha = \pm} \frac{\alpha \vartheta_1(- 2 \tau + 2 \alpha \mathfrak{b}_2|3\tau)^2}{b_2^{2\alpha}
    \prod_{\beta = \pm}\vartheta_1(\beta\tau + \alpha \mathfrak{b}_1 - \beta\alpha \mathfrak{b}_2|3\tau)
    } E_1 \begin{bmatrix}
      -1 \\ b_2^\alpha/q
    \end{bmatrix}(3\tau) \nonumber\\
    & \ - \frac{i  \eta(3\tau)^3}{
      2q^{\frac{11}{6}}\vartheta_1(2 \mathfrak{b}_2)
      \eta(\tau)^2
    }\sum_{\alpha = \pm} \frac{
      \alpha \vartheta_1(+ \tau + 2 \alpha \mathfrak{b}_2|3\tau)^2}{
        \prod_{\beta = \pm}\vartheta_1(2 \beta \tau + \alpha \mathfrak{b}_1 + \alpha \beta \mathfrak{b}_2|3\tau)} E_1 \begin{bmatrix}
      -1 \\ b_2^\alpha/q
    \end{bmatrix}(3\tau) \ . \nonumber
\end{align}
The six coefficients in front of the Eisenstein series are again residues of the simple poles of the integrand $\mathcal{Z}(\mathfrak{a})$. For later convenience, let us denote these coefficients by $R_i$,
\begin{align}
  R_1 = & \ \frac{i \eta (\tau) \vartheta_1(2 \mathfrak{b}_1, q^3)}{2 q^{5/6} \vartheta_1(\mathfrak{b}_1-\mathfrak{b}_2|\tau) \vartheta_1(\mathfrak{b}_1+\mathfrak{b}_2|\tau)} \ , \\
  R_2 = & \ \frac{i \eta (3\tau)^3 \vartheta_1(2 \mathfrak{b}_2, q^3)^2}{2 q^{5/6} \eta (\tau)^2 \vartheta_1(2 \mathfrak{b}_2|\tau) \vartheta_1(\mathfrak{b}_1-\mathfrak{b}_2|3\tau) \vartheta_1(\mathfrak{b}_1+\mathfrak{b}_2|3\tau)} \ ,
\end{align}
\begin{align}
  R_3 = & \ -\frac{i b_2^2 q^{\frac{1}{6}} \eta (3\tau)^3 \vartheta_1(-2 \mathfrak{b}_2-2 \tau |3\tau)^2}{2 \eta (\tau)^2 \vartheta_1(2 \mathfrak{b}_2|\tau) \vartheta_1(-\mathfrak{b}_1-\mathfrak{b}_2-\tau |3\tau) \vartheta_1(-\mathfrak{b}_1+\mathfrak{b}_2+\tau |3\tau)}\ ,\\
  R_4 = & \ \frac{i b_2^{-2} q^{\frac{1}{6}} \eta (3\tau)^3 \vartheta_1(2 \mathfrak{b}_2-2 \tau |3\tau)^2}{2 \eta (\tau)^2 \vartheta_1(2 \mathfrak{b}_2|\tau) \vartheta_1(\mathfrak{b}_1-\mathfrak{b}_2+\tau |3\tau) \vartheta_1(\mathfrak{b}_1+\mathfrak{b}_2-\tau |3\tau)}\ ,\\
  R_5 = & \ -\frac{i \eta (3\tau)^3 \vartheta_1(\tau -2 \mathfrak{b}_2|3\tau)^2}{2 q^{\frac{11}{6}} \eta (\tau)^2 \vartheta_1(2 \mathfrak{b}_2|\tau) \vartheta_1(-\mathfrak{b}_1-\mathfrak{b}_2+2 \tau |3\tau) \vartheta_1(-\mathfrak{b}_1+\mathfrak{b}_2-2 \tau |3\tau)}\ ,\\
  R_6 = & \ \frac{i \eta (3\tau)^3 \vartheta_1(2 \mathfrak{b}_2+\tau |3\tau)^2}{2 q^{\frac{11}{6}} \eta (\tau)^2 \vartheta_1(2 \mathfrak{b}_2|\tau) \vartheta_1(\mathfrak{b}_1-\mathfrak{b}_2-2 \tau |3\tau) \vartheta_1(\mathfrak{b}_1+\mathfrak{b}_2+2 \tau |3\tau)} \ .
\end{align}
By inspection, $R_3 + R_5 = R_4 + R_6 = 0$, hence only four residues $R_1, R_2, R_3, R_4$ are independent. 

The unflavored limit of the Schur index is given by the close form
\begin{align}
  \mathcal{I}_{D_6(\mathfrak{sl}_4, [3,1])}(q)
  =& \  \frac{\eta(3\tau)^3}{\eta(\tau)^5}
  \bigg(
    + E_1 \begin{bmatrix}
      -1 \\ q
    \end{bmatrix}(3\tau)
    + 3E_2(3\tau)
    + E_1 \begin{bmatrix}
      -1 \\ q
    \end{bmatrix}(3\tau)^2\\
    & \ + 6 E_1 \begin{bmatrix}
      -1 \\ q
    \end{bmatrix}(3\tau)E_1 \begin{bmatrix}
      1 \\ q
    \end{bmatrix}(3\tau)
    + 4 E_2 \begin{bmatrix}
      -1 \\ 1
    \end{bmatrix}(3\tau)
    + 2 E_2 \begin{bmatrix}
      -1 \\ q
    \end{bmatrix}(3\tau)
  \bigg) \ . \nonumber
\end{align}
It satisfies a 7-th order unflavored monic MLDE,
\begin{align}
  0 = & \ \Big[D_q^{(7)} - 200 E_4 D_q^{(5)} - 7000 E_6 D_q^{(4)} - 62000 E_4 D_q^{(3)}
  - 924000 E_4 E_6 D_q^{(2)} \\ 
  & \ \qquad\qquad - 30000 (80 E_4^3 + 147 E_6^2) D_q^{(1)}
  - 29400000E_4^2 E_6 \Big]\mathcal{I}_{D_6(\mathfrak{sl}_4, [3,1])}(q)\ . \nonumber
\end{align}
The equation has seven rational indicial roots,
\begin{equation}
  \alpha = - \frac{1}{6}, \frac{1}{6}, \frac{1}{6}, \frac{1}{2}, \frac{5}{6}, \frac{5}{6}, \frac{7}{6} \ .
\end{equation}
The $\alpha = \frac{7}{6}$ solution corresponds to the Schur index. The minimal indicial root $\alpha_\text{min}$ is related to the 4d central charges
\begin{equation}
-24\alpha_\text{min} = 4 = c_\text{eff} = 48 (c_\text{4d} - a_\text{4d}), \qquad c_\text{4d} = \frac{5}{3}, a_\text{4d} = \frac{19}{12} \ .
\end{equation}

The four independent residues $R_i$ has one unique linear combination that admits smooth limit $b_i \to 1$, which leads to an additional solution to the unflavored equation:
\begin{align}
  - 2(R_3 + R_4)\Big|_{b_i \to 1} = & \ \frac{2\eta(3\tau)^3}{\eta(\tau)^5}
  \bigg(1 + 3 E_1 \begin{bmatrix}
    +1 \\ q
  \end{bmatrix}(3 \tau)\bigg)\nonumber \\
  = & \ q^{\frac{1}{6}} + 11 q^{\frac{7}{6}}
  + 50 q^{\frac{13}{6}}
  + 188 q^{\frac{19}{6}}
  + 583 q^{\frac{25}{6}}
  + 1646 q^{\frac{31}{6}} + \cdots \ .
\end{align}

\subsection{\texorpdfstring{$D_{10}(\mathfrak{sl}_4, [4])$}{D10(sl(4), [4])}{}}

The theory $D_{10}(\mathfrak{sl}_4, [4])$ is also known as $(A_{3}, A_{5})$ in the notation of \cite{Cecotti:2010fi}. From the duality (\ref{typeIrealization}), the theory has the following duality frame,
\begin{equation}
	\begin{tikzpicture}[baseline=(current bounding box.center)]
		\node (left) {$D_{5}(\mathfrak{sl}_3,[1^3])$};
		\node (mid) [right=of left, xshift=0.5cm] {$\mathfrak{su}(2)$};
		\node (right) [right=of mid, xshift=0.5cm] {$D_{5}(\mathfrak{sl}_2,[1^2])$};
		\draw[-{Stealth[scale=1.2]}] (mid) -- (right);
		\draw[-{Stealth[scale=1.2]}] (mid) -- (left); 
		\node (box) [draw, rectangle, below=0.5cm of mid] {$0$};
		\draw (mid) -- (box);
	\end{tikzpicture}
\end{equation}
We can also read off the following data
\begin{align}
  & p = 10, N = 4, m = 2, k = 3, q = \frac{p}{m} = 5, Y = [4], L = 1 \\ 
  & \widetilde Y = [1], \quad \widetilde{Y}_+ = [1], \quad \widetilde{Y} = \emptyset, \\
  & N - \frac{N}{m} - \tilde N_- = 2, qL - \frac{N}{m} = 3, N - \frac{N}{m} = 2 \ .
\end{align}
The basic building blocks of the Schur index for this theory are given by
\begin{align}
  \mathcal{I}_{D_5(\mathfrak{sl}_3, [1,1,1])} = & \ \operatorname{PE}\bigg[
    \frac{q - q^{3 + 3}}{(1 - q)(1 - q^{3 + 3})} \Big(-1 + {\sum_{A, B = 1}^{3} \frac{a_A}{a_B}}\Big)
    \bigg] \ ,\\
  \mathcal{I}_{D_5(\mathfrak{sl}_2, [1,1])} = & \ \operatorname{PE}\bigg[
    \frac{q - q^{2 + 3}}{(1 - q)(1 - q^{2 + 3})} \Big(a^2 + a^{-2} + 1\Big)
    \bigg]\ .
\end{align}
We gauge the diagonal of the flavor subgroup $SU(2) \subset SU(3)$ from $D_5(\mathfrak{sl}_3, [1,1,1])$ and the $SU(2)$ flavor symmetry of $D_5(\mathfrak{sl}_2, [1,1])$. The fugacities are identifieds by
\begin{equation}
  a_1 = ab, \qquad a_2 = \frac{b}{a}, \qquad a_3 = \frac{1}{b^2} \ ,
\end{equation}
which leads to the following Schur index,
\begin{equation}
  \mathcal{I}_{D_{10}(\mathfrak{sl}_4, [4])}
  = \oint \frac{da}{2\pi i a} \mathcal{I}_{D_5(\mathfrak{sl}_3, [1,1,1])}(\mathfrak{a}, \mathfrak{b})\mathcal{I}_{D_5(\mathfrak{sl}_2, [1,1])}(\mathfrak{a}) \mathcal{I}_\text{VM}(\mathfrak{a}) = \oint \frac{da}{2\pi i a}\mathcal{Z}(a) \ ,
\end{equation}
where
\begin{align}
  \mathcal{Z}(\mathfrak{a}) = 
  & \ \frac{a^{12} \eta(5\tau)^{19} \vartheta_1(2\mathfrak{a})^2}{2q^{\frac{169}{12}}\eta(\tau)^3}
  \frac{1}{\prod_\pm \prod_{\ell = 1}^4 \vartheta_1(\pm 3 \mathfrak{b} - \ell \tau + \mathfrak{a}|5\tau)}\frac{1}{\prod_{\ell = 1}^4  \vartheta_1(2\mathfrak{a} - \ell \tau|5\tau)^2}\ . \nonumber 
\end{align}
The apparent double poles can be removed by using the identity
\begin{equation}
  \prod_{\ell = 0}^4 \vartheta_1(2 \mathfrak{a} - \ell \tau | 5 \tau)
  = a^4 q^{-3} \vartheta_1(2 \mathfrak{a}) \frac{\eta(5\tau)^5}{\eta(\tau)} \ ,
\end{equation}
which implies
\begin{equation}
  \mathcal{Z}(\mathfrak{a}) = \frac{\eta(\tau)}{2 q^{\frac{25}{12}}\eta(5\tau)}
  \frac{\vartheta_1(2 \mathfrak{a}|5\tau)^2 \vartheta_1(\pm 3 \mathfrak{b} + \mathfrak{a}|5\tau)}{\vartheta_1(\pm 3 \mathfrak{b} + \mathfrak{a})} \ .
\end{equation}

The integrand $\mathcal{Z}(\mathfrak{a})$ is elliptic with period $5\tau$, hence the relevant poles are
\begin{equation}
  \pm 3 \mathfrak{b} + \ell \tau, \qquad \ell = 0, \cdots, 4 \ .
\end{equation}
The corresponding residues are, 
\begin{align}
  0,\qquad
  \frac{i b^6 \vartheta_1\left(\tau |5\tau\right) \vartheta_1\left(6 \mathfrak{b}+\tau |5\tau\right) \vartheta_1\left(6 \mathfrak{b}+2 \tau |5\tau\right)^2}{2 q^{13/12} \eta (\tau)^2 \eta (5\tau) \vartheta_1(6 \mathfrak{b})},\\
  \frac{i \vartheta_1\left(2 \tau |5\tau\right) \vartheta_1\left(6 \mathfrak{b}-\tau |5\tau\right)^2 \vartheta_1\left(6 \mathfrak{b}+2 \tau |5\tau\right)}{2 q^{13/12} \eta (\tau)^2 \eta (5\tau) \vartheta_1(6 \mathfrak{b})},\\
  -\frac{i \vartheta_1\left(2 \tau |5\tau\right) \vartheta_1\left(6 \mathfrak{b}-2 \tau |5\tau\right) \vartheta_1\left(6 \mathfrak{b}+\tau |5\tau\right)^2}{2 q^{13/12} \eta (\tau)^2 \eta (5\tau) \vartheta_1(6 \mathfrak{b})},\\
  -\frac{i \vartheta_1\left(\tau |5\tau\right) \left(\vartheta_1\left(6 \mathfrak{b}-2 \tau |5\tau\right)\right)^2 \vartheta_1\left(6 \mathfrak{b}-\tau |5\tau\right)}{2 b^6 q^{13/12} \eta (\tau)^2 \eta (5\tau) \vartheta_1(6 \mathfrak{b})},\\
  0,\qquad
  \frac{i \vartheta_1\left(\tau |5\tau\right) \left(\vartheta_1\left(6 \mathfrak{b}-2 \tau |5\tau\right)\right)^2 \vartheta_1\left(6 \mathfrak{b}-\tau |5\tau\right)}{2 b^6 q^{13/12} \eta (\tau)^2 \eta (5\tau) \vartheta_1(6 \mathfrak{b})},\\
  \frac{i \vartheta_1\left(2 \tau |5\tau\right) \vartheta_1\left(6 \mathfrak{b}-2 \tau |5\tau\right) \vartheta_1\left(6 \mathfrak{b}+\tau |5\tau\right)^2}{2 q^{13/12} \eta (\tau)^2 \eta (5\tau) \vartheta_1(6 \mathfrak{b})},\\
  -\frac{i \vartheta_1\left(2 \tau |5\tau\right) \vartheta_1\left(6 \mathfrak{b}-\tau |5\tau\right)^2 \vartheta_1\left(6 \mathfrak{b}+2 \tau |5\tau\right)}{2 q^{13/12} \eta (\tau)^2 \eta (5\tau) \vartheta_1(6 \mathfrak{b})},\\
  -\frac{i b^6 \vartheta_1\left(\tau |5\tau\right) \vartheta_1\left(6 \mathfrak{b}+\tau |5\tau\right) \vartheta_1\left(6 \mathfrak{b}+2 \tau |5\tau\right)^2}{2 q^{13/12} \eta (\tau)^2 \eta (5\tau) \vartheta_1(6 \mathfrak{b})} \ .
\end{align}

The Schur index in closed form reads
\begin{align}
  \mathcal{I} = & \ \frac{i}{q^{13/12}\eta(\tau)^2 \eta(5\tau)\vartheta_1(6 \mathfrak{b})} \bigg( \\
  & \ \vartheta_1(2 \tau|5\tau)
    \sum_{\alpha = \pm} \alpha
    \vartheta_1(-6 \alpha \mathfrak{b} + 2 \tau|5\tau)
    \vartheta_1(6 \alpha \mathfrak{b}|5\tau)^2 
    E_1 \begin{bmatrix}
      -1 \\ b^{3\alpha}q^{\frac{1}{2}}
    \end{bmatrix}(5\tau)\\
  & \ - \vartheta_1(\tau|5\tau) \sum_{\alpha = \pm} \alpha b^{6\alpha} \vartheta_1(6 \alpha \mathfrak{b} + \tau|5\tau) 
  \vartheta_1(6 \alpha \mathfrak{b} + 2 \tau|5\tau)^2
  E_1 \begin{bmatrix}
    -1 \\ b^{-3 \alpha} q^{\frac{3}{2}}
  \end{bmatrix}(5\tau)
  \bigg) \ .
\end{align}
The coefficients of the Eisenstein series come from the residues of integrand $\mathcal{Z}(\mathfrak{a})$, where only three are actually linear independent. The unflavored limit gives
\begin{align}
  \mathcal{I}_{D_{10}(\mathfrak{sl}_4, [4])}(q) & \
  = - \frac{\eta(5\tau)}{\eta(\tau)^3}
  \Big(
    1 + E_1 \begin{bmatrix}
      +1 \\ q
    \end{bmatrix}(5\tau)
    + 3 E_1 \begin{bmatrix}
      + 1 \\ q^2
    \end{bmatrix}(5\tau)
    - 2E_2(5\tau)\\
    & \ + E_1 \begin{bmatrix}
      + 1 \\ q
    \end{bmatrix}(5\tau)^2
    + E_1 \begin{bmatrix}
      +1 \\ q^2
    \end{bmatrix}(5\tau)^2
    - 2 E_2 \begin{bmatrix}
      + 1 \\ q
    \end{bmatrix}(5\tau)
     - 2 E_2 \begin{bmatrix}
       1 \\ q^2
     \end{bmatrix}(5\tau)
  \Big) \ . \nonumber
\end{align}

The unflavored Schur index satisfies a 16-th order unflavored monic MLDE,
{\fontsize{8pt}{8pt}\begin{align}
  0 = \bigg[ & \ D^{(16)}_q - \frac{11171050208504}{2189278165} E_4 D^{(14)}_q - \frac{464197483598576}{2189278165} E_6 D^{(13)}_q \nonumber\\
  & + \frac{9860330680149964}{2189278165} E_4^2 D^{(12)}_q + \frac{779235372995814112}{2189278165} E_6 E_4 D^{(11)}_q \nonumber\\
  & + \left( \frac{32456611659695917304}{54731954125} E_4^3 + \frac{200995901927582792800}{54731954125} E_6^2 \right) D^{(10)}_q
  - \frac{1187733051446491930128}{54731954125} E_6 E_4^2 D^{(9)}_q \nonumber\\
  & - \left( \frac{148712814577023697558778}{273659770625} E_4^4 - \frac{155652147396466305175360}{273659770625} E_6^2 E_4 \right) D^{(8)}_q \nonumber\\
  & - \left( \frac{3283674181201442502482624}{273659770625} E_6 E_4^3 - \frac{578783920798494958080000}{273659770625} E_6^3 \right) D^{(7)}_q \nonumber\\
  & + \left( \frac{438912948272493815363138136}{273659770625} E_4^5 - \frac{132702672164353531952051648}{273659770625} E_6^2 E_4^2 \right) D^{(6)}_q \nonumber\\
  & + \left( \frac{652908814152742315130442976}{273659770625} E_6 E_4^4 - \frac{1166821289023326402246437888}{273659770625} E_6^3 E_4 \right) D^{(5)}_q \nonumber\\
  & - \left( \frac{16835730318835206451559968884}{1368298853125} E_4^6 + \frac{16890224717433561754340326400}{1368298853125} E_6^2 E_4^3 \right. \nonumber\\ & \left. \qquad + \frac{29935674500039554254749440000}{1368298853125} E_6^4 \right) D^{(4)}_q \nonumber\\
  & - \left( \frac{443891989782486916804297504032}{1368298853125} E_6 E_4^5 + \frac{85848527366234686773233272576}{1368298853125} E_6^3 E_4^2 \right) D^{(3)}_q \nonumber\\
  & - \left( \frac{384426650637641472733259675640}{1368298853125} E_4^7 + \frac{13182902853285529377609168626592}{1368298853125} E_6^2 E_4^4 \right. \nonumber\\ & \left. \qquad + \frac{7332153876129784543672065337600}{1368298853125} E_6^4 E_4 \right) D^{(2)}_q \nonumber\\
  & - \left( \frac{3673534225094072904022225764800}{1368298853125} E_6 E_4^6 + \frac{153792329386800366550880928720384}{1368298853125} E_6^3 E_4^3 \right. \nonumber\\ & \left. \qquad + \frac{26099008472373459377401584640000}{1368298853125} E_6^5 \right) D^{(1)}_q \nonumber\\
  & - \left( \frac{17740578315181492500104332125}{1368298853125} E_4^8 + \frac{489994112866540076845404221440}{1368298853125} E_6^2 E_4^5 \right. \nonumber\\ & \left. \qquad + \frac{608157430757144976639991429353984}{1368298853125} E_6^4 E_4^2 \right)\bigg] \mathcal{I}_{D_{10}(\mathfrak{sl}_4, [4])}(q) \ .
\end{align}
}
The equation has 12 rational indicial roots
\begin{equation}
  - \frac{7}{60}, \quad \bigg[\frac{1}{12}\bigg]_2, \quad \bigg[\frac{17}{60}\bigg]_2, \quad \frac{29}{60}, \quad \frac{41}{60}, \quad \bigg[\frac{53}{60}\bigg]_2, \quad \frac{13}{12}, \quad \frac{77}{60}, \quad \frac{25}{12} \ ,
\end{equation}
and 4 additional irrational ones. The largest indicial root $\alpha = \frac{25}{12}$ corresponds to the Schur index, while the minimal indicial root $\alpha_\text{min}$ is related to the 4d central charges
\begin{equation}
-24\alpha_\text{min} = \frac{14}{5} = c_\text{eff} = 48 (c_\text{4d} - a_\text{4d}), \qquad c_\text{4d} = \frac{25}{6}, a_\text{4d} = \frac{493}{120} \ .
\end{equation}

Among the eight non-zero residues list earlier, there are only three independent ones, from which we can construct two combinations $\operatorname{ch}_{1,2}$ that admit smooth $b \to 1$ limit,
\begin{align}
  \operatorname{ch}_1 = &\ \frac{i \vartheta_1\left(6 \mathfrak{b}+\tau |5\tau\right)}{2 q^{13/12} \eta(\tau)^2 \eta(5\tau) \vartheta_1(6 \mathfrak{b}|\tau)}  \bigg(
    b^6 \vartheta_1\left(\tau | 5\tau\right) \vartheta_1\left(6 \mathfrak{b}+2 \tau | 5\tau\right)^2 \nonumber\\
    & \ +\vartheta_1\left(2 \tau | 5\tau\right) \vartheta_1\left(6 \mathfrak{b}+\tau | 5\tau\right) \vartheta_1\left(6 \mathfrak{b}-2 \tau | 5\tau\right)
  \bigg) \ ,\\
  \operatorname{ch}_2 = & \ \frac{i \vartheta_1\left(2 \tau |5\tau\right)}{2 q^{13/12}\eta (\tau)^2\eta(5\tau) \vartheta_1(6 \mathfrak{b}|\tau)} \nonumber \\
  & \ \bigg(
    \vartheta_1\left(6 \mathfrak{b}+2 \tau |5\tau\right) \vartheta_1\left(6 \mathfrak{b}-\tau |5\tau\right)^2+\vartheta_1\left(6 \mathfrak{b}+\tau |5\tau\right)^2 \vartheta_1\left(6 \mathfrak{b}-2 \tau |5\tau\right)
  \bigg) \ .
\end{align}
Sending $b \to 1$, the two expressions reduces to
\begin{align}
  \operatorname{ch}_1 = & \ \frac{\eta(5\tau)}{\eta(\tau)^3} \Big(
    1 - E_1 \begin{bmatrix}
      +1 \\ q
    \end{bmatrix}(5\tau)
    + 3 E_1 \begin{bmatrix}
      + 1 \\ q^2
    \end{bmatrix}
  \Big) \nonumber\\
  = & \ q^{\frac{25}{12}}(\frac{1}{q} + 1 + 7q + 13 q^2 + 35q^3 + 71 q^4 + 155 q^5 + 293 q^6 + \cdots) \ , \\
  \operatorname{ch}_2 = & \ - \frac{2\eta(5\tau)}{\eta(\tau)^3}\Big(
    2E_1 \begin{bmatrix}
      + 1 \\ q
    \end{bmatrix}(5\tau)
    - E_1 \begin{bmatrix}
      1 \\ q^2
    \end{bmatrix}(5 \tau)
  \Big) \nonumber \\
  = & \ q^{\frac{25}{12}}(\frac{1}{q^2} + \frac{7}{q} + 23 + 70 q + 173q^2 + 407 q^3 + 882 q^4 + 1830 q^5 + 3610 q^6 + \cdots) \ .
\end{align}
Both of these expressions solves the same 16-th order unflavored MLDE as the unflavored Schur index, corresponding to the indicial roots $\frac{13}{12}$, $\frac{1}{12}$.

\subsection{\texorpdfstring{$D_{8}(\mathfrak{sl}_6, [5,1])$}{}}

As a last standalone example, we consider the $D_8(\mathfrak{sl}_6, [5,1])$ theory. The theory admits the gauge theory description
\begin{equation}
	\begin{tikzpicture}[baseline=(current bounding box.center)]
		\node (left) {$D_{4}(\mathfrak{sl}_5,[3,1,1])$};
		\node (mid) [right=of left, xshift=0.5cm] {$\mathfrak{su}(2)$};
		\node (right) [right=of mid, xshift=0.5cm] {$D_{4}(\mathfrak{sl}_3,[1,1,1])$};
		\draw[-{Stealth[scale=1.2]}] (mid) -- (right);
		\draw[-{Stealth[scale=1.2]}] (mid) -- (left); 
		\node (box) [draw, rectangle, below=0.5cm of mid] {$0$};
		\draw (mid) -- (box);
	\end{tikzpicture}
\end{equation}
The theory has central charges
\begin{equation}
  a_\text{4d} = \frac{49}{12} , \qquad c_\text{4d} = \frac{25}{6} \ ,
\end{equation}
and it belongs to the infinite series $D_{N + 2}(\mathfrak{sl}(N), [N-1, 1]) $ which is also dual to $ D_{N + 2}(\mathfrak{sl}(N + 4), [N+1, 3])$. We will discuss this series later.

The $SU(2)\subset SU(3)$ flavor subgroup from $D_4(\mathfrak{sl}(3), [1,1,1])$ is gauged with the $SU(2)$ flavor symmetry from $D_4(\mathfrak{sl}_5), [3,1,1]$. The leftover $U(1)^2$ flavor fugacities are labled by $b_{1,2}$. The full Schur index is given by the contour integral
\begin{align}
  \mathcal{I}_{D_8(\mathfrak{sl}_6, [5,1])}
  = - \oint \frac{da}{2\pi i a}
  \frac{\eta(4\tau)^2}{2 q^{17/6}} \frac{\vartheta_1(2 \mathfrak{a}|4\tau)^2 \prod_\pm\vartheta_1(\mathfrak{b}_1 \pm \mathfrak{b}_2|4\tau)}{
    \prod_\pm \vartheta_1(\mathfrak{a} \pm \mathfrak{b}_1|\tau)
    \vartheta_1(-2 \tau + \mathfrak{a}_1 \pm \mathfrak{b}_2|4\tau)
  } = \oint \frac{da}{2\pi i a}\mathcal{Z}(\mathfrak{a}) \ . \nonumber
\end{align}

The integrand $\mathcal{Z}(\mathfrak{a})$ is periodic with respect to $4\tau$,
\begin{equation}
  \mathcal{Z}(\mathfrak{a} + 4\tau) = \mathcal{Z}(\mathfrak{a} + 1) = \mathcal{Z}(\mathfrak{a}) \ .
\end{equation}
There are ten simple imaginary and real poles $\pm \mathfrak{b}_2 + 2 \tau$ and $\pm \mathfrak{b}_1 + \ell \tau|_{\ell = 0, 1, 2, 3}$. We then compute the index using the integration formula, which gives
\begin{align}
  = & -\frac{\eta(4\tau)^6 \vartheta_1(\tau|4\tau)  \vartheta_1(2\mathfrak{b}_1 - 2\tau|4\tau)}{2 q^{\frac{19}{12}} \eta(\tau)^4 \vartheta_1(2\mathfrak{b}_1|4\tau) \vartheta_1(2\mathfrak{b}_1 - \tau|4\tau) \vartheta_1(\mathfrak{b}_1 - \mathfrak{b}_2 - \tau|4\tau) \vartheta_1(\mathfrak{b}_1 + \mathfrak{b}_2 - \tau|4\tau)} E_1 \begin{bmatrix}
    -1 \\ b_1/q
  \end{bmatrix} \nonumber \\
  & +\frac{i \eta(4\tau)^2 \vartheta_1(2\mathfrak{b}_1|4\tau)^2 \vartheta_1(2\tau|4\tau)  \vartheta_1(2\mathfrak{b}_1 - 2\tau|4\tau)}{b_1 q^{\frac{5}{6}} \eta(\tau)^3 \vartheta_1(2\mathfrak{b}_1|\tau) \vartheta_1(\mathfrak{b}_1 - \mathfrak{b}_2|4\tau) \vartheta_1(\mathfrak{b}_1 + \mathfrak{b}_2|4\tau)} E_1 \begin{bmatrix} -1 \\ b_1 \end{bmatrix}(4\tau) \nonumber\\
  & +\frac{i \eta(4\tau)^2 \vartheta_1(\tau|4\tau) \vartheta_1(2\mathfrak{b}_1 - 2\tau|4\tau)^2 \vartheta_1(2\mathfrak{b}_1 - \tau|4\tau)}{b_1^3 q^{\frac{5}{6}} \eta(\tau)^3 \vartheta_1(2\mathfrak{b}_1|\tau) \vartheta_1(\mathfrak{b}_1 - \mathfrak{b}_2 + \tau|4\tau) \vartheta_1(\mathfrak{b}_1 + \mathfrak{b}_2 + \tau|4\tau)} E_1 \begin{bmatrix} -1 \\ b_1 q \end{bmatrix}(4\tau)\nonumber \\
  & -\frac{i \eta(4\tau)^2 \vartheta_1(\tau|4\tau) \vartheta_1(2\mathfrak{b}_1 - 2\tau|4\tau)^2 \vartheta_1(2\mathfrak{b}_1 + \tau|4\tau)}{2 b_1 q^{\frac{5}{6}} \eta(\tau)^3 \vartheta_1(2\mathfrak{b}_1|\tau) \vartheta_1(\mathfrak{b}_1 - \mathfrak{b}_2 - \tau|4\tau) \vartheta_1(\mathfrak{b}_1 + \mathfrak{b}_2 - \tau|4\tau)}E_1 \begin{bmatrix} -1 \\ b_1q^{-1} \end{bmatrix}(4\tau) \nonumber \\
  & +\frac{i \vartheta_1(2\mathfrak{b}_2|4\tau)  \vartheta_1(\mathfrak{b}_1 - \mathfrak{b}_2 - 2\tau|4\tau) \vartheta_1(\mathfrak{b}_1 + \mathfrak{b}_2 - 2\tau|4\tau)}{b_1 q^{\frac{5}{6}} \eta(4\tau) \vartheta_1(\mathfrak{b}_1 - \mathfrak{b}_2|\tau) \vartheta_1(\mathfrak{b}_1 + \mathfrak{b}_2|\tau)}E_1 \begin{bmatrix} -1 \\ b_2 \end{bmatrix}(4\tau) \ .
\end{align}
The coefficients of the Eisenstein series are simply the residues of $\mathcal{Z}(\mathfrak{a})$.

The unflavoring limit is given by the simple formula
\begin{align}
  \mathcal{I}_{D_8(\mathfrak{sl}_{10},[7,3])}(q)
  = & \ \frac{\eta(4\tau)^4}{\eta(\tau)^2\vartheta_1(\tau|4\tau)}\bigg(
    3E_2(4\tau)
    + 4 E_2 \begin{bmatrix}
      -1 \\ 1
    \end{bmatrix}(4\tau)
    + 2E_2 \begin{bmatrix}
      -1 \\ q
    \end{bmatrix}(4\tau)
    \nonumber\\
  & \ - E_1 \begin{bmatrix}
      -1 \\ q
    \end{bmatrix}(4\tau)^2
    + 2 E_1 \begin{bmatrix}
    -1 \\ q
  \end{bmatrix}(4\tau)E_1 \begin{bmatrix}
    + 1 \\ q
  \end{bmatrix}(4\tau) \bigg) \\
  = & \ 1 + 2 q + 6 q^2 + 20 q^3 + 50 q^4 + 120 q^5 + 280 q^6 + 616 q^7 + 
 1303 q^8 + \cdots. \nonumber
\end{align}
It is annihilated by a 14-th order monic modular differential operator,
{\small\begin{align}
  & D_q^{(14)} -\frac{41356535  E_4}{13471}D_q^{(12)} -\frac{1651331220 E_6}{13471}D_q^{(11)}  + \frac{11772659875 E_4^2}{13471}D_q^{(10)}  \nonumber\\
  & \ +\frac{1018752772625 E_6 E_4}{13471} D_q^{(9)} 
  +\frac{625\left(8603926907 E_4^3+11155004154 E_6^2\right)}{13471} D_q^{(8)}  \nonumber\\
  & \ +\frac{100111961381250 E_6 E_4^2}{13471} D_q^{(7)}
  -\frac{1250  \left(1174872118745 E_4^3-4752943356784 E_6^2\right) E_4}{13471}D_q^{(6)} \nonumber\\
  & \ -\frac{13125  E_6 \left(125333104655 E_4^3-2372060191536 E_6^2\right)}{13471}D_q^{(5)} \nonumber\\
  & \ +\frac{62500  \left(1660587751085 E_4^3-6515126570848 E_6^2\right) E_4^2}{13471} D_q^{(4)}\\
  & \ -\frac{109375  E_6 \left(18868960484060 E_4^3+63405268916423 E_6^2\right) E_4}{13471}D_q^{(3)} \nonumber\\
  & \ -\frac{15625 \left(815283970908600 E_4^6+5873692279036750 E_6^2 E_4^3+2077585779126753 (E_6)^4\right)}{13471} D_q^{(2)} \nonumber\\
  & \ -\frac{1640625 E_6 \left(283955673730300 E_4^3+654661271404749 E_6^2\right) E_4^2}{13471} D_q^{(1)} \nonumber\\
  & \ -\frac{7109375 \text{cch} E_4 \left(47764261202400 E_4^6+726944797649700 E_6^2 E_4^3+416991195316829 E_6^4\right)}{13471} \ . \nonumber
\end{align}}
The unflavored MLDE has 11 rational indicial roots,
\begin{align}
  \alpha = - \frac{1}{6}, \quad \left[\frac{1}{12}\right]_2, \quad \left[\frac{1}{3}\right]_2, \quad \frac{7}{12}, \quad \left[\frac{5}{6}\right]_2, \quad
  \left[\frac{13}{12}\right]_2, \quad \frac{25}{12} \ ,
\end{align}
where the subscripts denote the multiplicity of the roots, with additional 3 more irrational roots. The eleven rational roots implies the existence of an 11-th order unflavored MLDE at higher modular weight.

As we saw earlier, the integrand $\mathcal{Z}(\mathfrak{a})$ has several residues $R_{j}$, but only two of them are independent. They form one combination which admits smooth unflavoring limit, giving 
\begin{align}
  \operatorname{ch}_1 = & \ - \frac{\eta(4\tau)^4}{\sqrt{q}\eta(\tau)^2 \vartheta_1(\tau|4\tau)^4} \bigg(
    1 + 4 E_1 \begin{bmatrix}
      1 \\ q
    \end{bmatrix}(4 \tau)\bigg) \nonumber\\
  = & \ q^{\frac{1}{12}} + 10 q^{\frac{13}{12}} + 51 q^{\frac{25}{12}} + 190 q^{\frac{37}{12}} + 601 q^{\frac{49}{12}} + 1702 q^{\frac{61}{12}} + 4422 q^{\frac{73}{12}} + \cdots \ .
\end{align}
Again, this is another solution to the unflavored MLDE corresponding to the second smallest indicial root $\frac{1}{12}$.

\subsection{\texorpdfstring{Exact Formula for the $D_p^b(\mathfrak{sl}_N,[Y])$ Theory}{}}

In this subsection we briefly discuss the Schur index of the more general $D_p^b(\mathfrak{sl}_N,[Y])$ theory, where $[Y]$ is a generic regular puncture. Recall from \eqref{general duality} that a generic $D_p^b(\mathfrak{sl}_N,[Y])$ theory is dual to a type I Argyres--Douglas theory together with a set of free hypermultiplets. For later convenience, we record here the duality relation explicitly:
\begin{align}
	D_p^b\left(\mathfrak{sl}_N,[Y]\right)\cong & \ D_p\left(\mathfrak{sl}_{(N-b)p+b},[(p-1)^{N-b},Y]\right) \nonumber \\
	& \ \qquad \otimes (N-b)\sum_{i=p}^N(i-p+1)\ell_i \text{ free hypermultiplets }.
\end{align}
The right-hand side includes free hypermultiplets originating from the breaking of flavor symmetry after choosing the general regular puncture $[Y]$. In the special case of the full puncture $[Y] = [1^N]$, the duality simplifies to
\begin{equation}\label{sduality}
	D_p^b\left(\mathfrak{sl}_N\right)\cong D_p\left(\mathfrak{sl}_{(N-b)p+b},[(p-1)^{N-b},1^N]\right).
\end{equation}
Here the right-hand side is a type I Argyres--Douglas theory $D_{p'}(\mathfrak{sl}(N'), [r^m, 1^{N' - rm}])$, where the relevant data are given by
\begin{equation}
	b' = N' =(N-b)p+b,
	\quad k'=p-b,
	\quad r=(p-1),
	\quad m=N-b.
\end{equation}

We would like to compute the Schur index of $D_p^b(\mathfrak{sl}_N, [Y])$. A convenient approach exploits the TQFT nature of the Schur index \cite{Gadde:2011ik,Gadde:2011uv,Gadde:2009kb,Gaiotto:2012xa}. From the TQFT perspective, the Schur index is expressed as a 2d $q$-deformed Yang-Mills correlator
\begin{equation}\label{eq:schurindexTQFT}
	\mathcal{I}_{D_p^b(\mathfrak{sl}_N, [Y])} (q, \boldsymbol{z})=\sum_{\lambda}\psi_{\lambda}^{p, b, N}(q)\psi_{\lambda}^{Y}(\boldsymbol{z}) \ ,
\end{equation}
where $\psi^{p, b, N}_\lambda(q)$ denotes the wavefunction of the irregular puncture, and the wavefunction associated with a generic regular puncture is given by
\begin{equation}
	\psi_{\lambda}^Y(\boldsymbol{a})=\mathrm{PE}\left[\sum_j\frac{q^{j+1}}{1-q}\mathrm{Tr}_{R_j}(\boldsymbol{a})\right]\chi_{\lambda}(\boldsymbol{a}q^{\Lambda_Y}).
\end{equation}
Specializing to $[Y] = [1^N]$, using (\ref{sduality}), the left hand side of (\ref{eq:schurindexTQFT}) is known by analyzing the type I Schur index on the right, and one can solve the wavefunction $\psi^{p, b, N}_\lambda(q)$ of the irregular puncture. Then we can combine $\psi^{p,b,N}_\lambda(q)$ with the known wavefunction of the chosen regular puncture to deduce the Schur index of $D_p^b(\mathfrak{sl}_N,[Y])$. 

For simplicity, we focus on the case with $\gcd(p,b) = 1$, although the idea certainly extends beyond this case. In this setting, the theory $D_p\left(\mathfrak{sl}_{(N-b)p+b},[(p-1)^{N-b},1^N]\right)$ contains no mass parameters in the irregular puncture and is completely determined by the formula
\eqref{schurrm}. The general Schur index of $D_p^b(\mathfrak{sl}_N)$ takes the form
\begin{equation}\label{eq:schurindexTQFT-full-puncture}
	\mathcal{I}_{D_p^b(\mathfrak{sl}_N)}(q, \boldsymbol{z}) = \sum_\lambda\psi_\lambda^\mathrm{irregular}{(q)}\psi_\lambda^{\text{full}}(\boldsymbol{z}) \ ,
\end{equation} 
where 
\begin{equation}
	\psi_\lambda^{\text{full}}(\boldsymbol{z})=\mathrm{PE}\left[\frac{q}{1-q}\chi_{\mathrm{adj}}(\boldsymbol{z})\right]\chi_{\lambda}(\boldsymbol{z}).
\end{equation}
Denote the Schur index of $D_p\left(\mathfrak{sl}_{(N-b)p+b},[(p-1)^{N-b},1^N]\right)$ by $\mathcal{I}(q,\boldsymbol{z})$. Then the irregular puncture contributes
\begin{equation}
	\psi_\lambda^{p,b,N}{(q)} = \oint[d\boldsymbol{z}]\mathrm{PE}\left[-\frac{q}{1-q}\chi_\mathrm{adj}(\boldsymbol{z})\right]\mathcal{I}(q,\boldsymbol{z})\chi_{\lambda}(\boldsymbol{z})\ .
\end{equation}

Substituting $\psi^{p,b,N}_\lambda(q)$ into (\ref{eq:schurindexTQFT-full-puncture}), and using the identity  
\begin{equation}
	\sum_{\lambda}\chi_{\lambda}(\boldsymbol{z})\,\chi_{\lambda}(\boldsymbol{z}^{\prime})
	= \frac{1}{\Delta_{J}(\boldsymbol{z})}\,\delta(\boldsymbol{z}-\boldsymbol{z}^{\prime}),
\end{equation}
with respect to the Haar measure $\Delta_{J}(\boldsymbol{z})$, we obtain the wave function associated with a generic regular puncture specified by the Young diagram $Y$:  
\begin{equation}
	\psi^Y_\lambda(\mathbf{a}) = \mathrm{PE}\left[\sum_j\frac{q^{j+1}}{1-q}\mathrm{Tr}_{R_j}(\boldsymbol{a})\right]\chi_{\lambda}(\boldsymbol{a}q^{\Lambda_Y}) \ .
\end{equation}

We find the $\mathcal{I}_{D_p^b(\mathfrak{sl}_N, [Y])}$ to be simply
\begin{equation}
	\mathcal{I}_{D_p^b(\mathfrak{sl}_N, [Y])} = \mathrm{PE}\left[\sum_j\frac{q^{j+1}}{1-q}\mathrm{Tr}_{R_j}(\boldsymbol{a})\right]\mathrm{PE}\left[-\frac{q}{1-q}\chi_\mathrm{adj}(\boldsymbol{a}q^{\Lambda_Y})\right]\mathcal{I}(q,\boldsymbol{a}q^{\Lambda_Y}).
\end{equation}
Recalling $\boldsymbol{a}q^{\Lambda_Y}$ solves
\begin{equation}
	\chi_{\text{adj}}(\boldsymbol{a}q^{\Lambda_Y})=\sum_j\chi_{\mathcal{R}_j}(\mathbf{a})\chi_{V_j}^{\mathfrak{su}(2)},
\end{equation} 
the Schur index simplifies to
\begin{equation}\label{general Y}
	\begin{aligned}
		&\mathrm{PE}\left[\sum_j\frac{q^{j+1}}{1-q}\mathrm{Tr}_{R_j}(\boldsymbol{a})\right]\mathrm{PE}\left[-\sum_j\frac{q}{1-q}\mathrm{Tr}_{R_j}(\boldsymbol{a})\sum_{m=-j}^jq^m\right]\mathcal{I}(q,\boldsymbol{a}q^{\Lambda_Y})\\
		=& \mathrm{PE}\left[-\sum_j\frac{q}{1-q}\mathrm{Tr}_{R_j}(\boldsymbol{a})\sum_{m=-j}^{j-1}q^m\right]\mathcal{I}(q,\boldsymbol{a}q^{\Lambda_Y}).
	\end{aligned}
\end{equation}

As an explicit example, let us consider the theory $D_3^5(\mathfrak{sl}_6, [3,3])$, with $k = -2$, $b = 5$, $p = 3$, $N = 6$. To begin, we first study the theory $D_3^5(\mathfrak{sl}_6)$ with a full puncture. The duality relation implies
\begin{equation}\label{genexample1}
	D_3^5(\mathfrak{sl}_6)\cong D_3(\mathfrak{sl}_8,[2,1^6]) \ ,
\end{equation}
from which we deduce the the Schur index of $D_3^5(\mathfrak{sl}_6)$ is given by 
\begin{equation}
	\operatorname{PE}\left[\frac{\big(q^{3/2}-q^{5/2}\big) \big(a^{-1}\chi_{\square}^{SU(6)}+a \chi_{\bar{\square}}^{SU(6)}\big)+\big(q-q^3\big) \chi_{\text{adj}}^{SU(6)}-q^3+q}{(1-q) \big(1-q^3\big)}\right] \ .
\end{equation}
Now we partically close the full puncture to $Y=[3,3]$, corresponding to the flavor symmetry $SU(2)$, by working out the representation $R_j$ in the decomposition of adjoint of $\mathfrak{sl}_6$. From the character decomposition, we know the fugacity should take the form
\begin{equation}
	\{aq,a,aq^{-1},bq,b,bq^{-1}\},~~ab=1\ .
\end{equation} 
for the Young diagram $Y=[3,3].$ The character decompose as 
\begin{equation}
	\chi_{1}(a)+\chi_1(q)(1+\chi_{1}^f(a))+\chi_2(q)(1+\chi_{1}^f(a)),
\end{equation}
we denote the character of the spin-$j$ representation of the flavor symmetry $SU(2)$ by $\chi_{i}^f$. We read off
\begin{equation}
  \operatorname{tr}_{\mathcal{R}_{j = 1,2}} (\boldsymbol{a}) = (1 + \chi_1^f(a))\ .
\end{equation}
Hence we can conclude that the prefactor is
\begin{equation}
	\mathrm{PE}\left[-\frac{q}{1-q}(1+\chi_{1}(a))\sum_{j=1}^2\sum_{m=-j}^{j-1}q^m\right].
\end{equation}
Combining these two contributions, we obtain the Schur index of  $D_3^5(\mathfrak{sl}_6,[3,3])$.

Now we proceed to check the consistency of our formula \eqref{general Y}. Recall that the general duality involves $(N-b)\sum_{i=p}^{N}(i-p+1)l_{i}$ free hypermultiplets. If we choose the regular puncture $Y$ such that the only non-zero $l_{i}$ are those with $i<p$, then the duality directly implies the equivalence
\begin{equation}
	D_p^b\left(\mathfrak{sl}_N,[Y]\right)\cong D_p\left(\mathfrak{sl}_{(N-b)p+b},[(p-1)^{N-b},Y]\right),
\end{equation}
which allows us to test the consistency of the formula.

As an example, let us again start from \eqref{genexample1}, but now consider the regular puncture
\begin{equation}
	[Y]=[2^3].
\end{equation}
In this case, the duality predicts
\begin{equation}
	D_3^5(\mathfrak{sl}_6,[2^3])\cong D_3(\mathfrak{sl}_8,[2^4]).
\end{equation}
On the right-hand side, the regular puncture $[2^4]$ induces the following decomposition under $\mathfrak{su}(2) \times \mathfrak{su}(4)$
\begin{equation}
	\chi_{\text{adj}}^{SU(4)}+\chi_1(1+\chi_{\text{adj}}^{SU(4)}).
\end{equation}
Hence the Schur index of $D_3(\mathfrak{sl}_8, [2^4])$ takes the form
\begin{equation}
	\operatorname{PE}\left[\frac{q(1+q)\chi_{\text{adj}}^{SU(4)}}{1-q^3}\right]
\end{equation}
On the left-hand side, our formula prescribes the substitution of variables as
\begin{equation}
	a q^{1/2},aq^{-1/2},bq^{1/2},bq^{-1/2},cq^{1/2},cq^{-1/2},~abc=1.
\end{equation}
The decomposition under $\mathfrak{su}(2)\times\mathfrak{su}(3)$ is then
\begin{equation}
	\chi_{\text{adj}}^{SU(3)}+\chi_1(1+\chi_{\text{adj}}^{SU(3)}).
\end{equation}
The prefactor contributes
\begin{equation}
	\mathrm{PE}\left[-\frac{q}{1-q}(1+\chi_{\text{adj}}^{SU(3)})(1+q^{-1})\right].
\end{equation}
Collecting all contributions, the Schur index of $D_3^5(\mathfrak{sl}_6,[2^3])$ becomes
\begin{equation}
	\operatorname{PE}\left[\frac{q (q+1) \left(a b d^2 \left(a^2 b^2+a+b\right)+d \left(a^2 b^2+a+b\right) (a b (a+b)+1)+a b (a b (a+b)+1)\right)}{a^2 b^2 d \left(1-q^3\right)}\right].
\end{equation} 
After a redefinition of the variables $a,b,d,$ this expression reproduces the Schur index of $D_3(\mathfrak{sl}_8,[2^4]).$

\section{\texorpdfstring{$D_{N-4}(\mathfrak{sl}(N), [N-4, 4])$}{}}\label{sec:DN-4}

With the simplest examples worked out in the previous section, we now turn to the series $D_{N-4}(\mathfrak{sl}(N), [N-4, 4])$ where $N = 4n + 2 = 2(2n+1)\ge 18$. This series admit the following gauge theory description,
\begin{equation}
	\begin{tikzpicture}[baseline=(current bounding box.center)]
		\node (left) {$D_{\frac{N}{2}-2}(\mathfrak{sl}(\frac{N}{2} -4),[\frac{N}{2}-6,1,1])$};
		\node (mid) [right=of left, xshift=0.5cm] {$\mathfrak{su}(2)$};
		\node (right) [right=of mid, xshift=0.5cm] {$D_{\frac{N}{2}-2}(\mathfrak{sl}(\frac{N}{2}),[\frac{N}{2}-2,1,1])$};
		\draw[-{Stealth[scale=1.2]}] (mid) -- (right);
		\draw[-{Stealth[scale=1.2]}] (mid) -- (left); 
		\node (box) [draw, rectangle, below=0.5cm of mid] {$0$};
		\draw (mid) -- (box);
	\end{tikzpicture}
\end{equation}
As discussed in \cite{Beem:2023ofp,Xie:2019yds}, this theory is dual to the $D_{N-4}(\mathfrak{sl}_4, [4])$, which include the example $D_{10}(\mathfrak{sl}_4, [4])$ discussed above when we specialize $N = 14$. In the notation of \cite{Cecotti:2010fi}, the theory is is also known as $(A_3, A_{N - 9})$. The theory receives $U(1)$ flavor symmmetry from the irregular puncture. The associated VOA is given by the logarithmic $\mathcal{B}(\frac{N-4}{4})_{\mathfrak{sl}_4}$ \cite{Creutzig:2017qyf}. In the following, we stick to the notation $D_{N-4}(\mathfrak{sl}(N), [N-4, 4])$.

When $N = 4n + 2$, the central charges of $D_{N - 4}(\mathfrak{sl}(N), [N-4, 4])$ satisfy
\begin{equation}
  a_\text{4d} - c_\text{4d} = - \frac{1}{12} + \frac{1}{4(N-4)}, \qquad N = 4n + 2 \ge 18\ ,
\end{equation}
which is analogous to the well-known $a_\text{4d} = c_\text{4d}$ relation. As a side note, for $N = 4n$, there is a simpler relation
\begin{equation}
  a_\text{4d} - c_\text{4d} = \frac{1}{8}, \qquad N = 4n \ge 20 \ .
\end{equation}
This $N = 4n$ series admit an $SU(3)$ gauge description, whose Schur index can also be computed using our integration formula. However, for computational simplicity, we will only focus on the $N = 4n + 2$ series in this paper.

\subsection{The flavored Schur index}

The Schur index of the $D_{\frac{N}{2} - 2}(\mathfrak{sl}(\frac{N}{2} -4),[\frac{N}{2}-6,1,1])$ and $D_{\frac{N}{2} - 2}(\mathfrak{sl}(\frac{N}{2}),[\frac{N}{2}-2,1,1])$ theories can be computed using \eqref{schurrm},
\begin{align}
  & \ \mathcal{I}_{D_{\frac{N}{2} - 2}(\mathfrak{sl}(\frac{N}{2} -4),[\frac{N}{2}-6,1,1])} \nonumber \\
  = & \ \operatorname{PE}\bigg[
    \frac{1}{(1-q)(1-q^{\frac{N}{2} - 2})}\Big(
      (q - q^{\frac{N}{2} - 2}) \, \chi_{\text{adj}}^{SU(2)}(a) + \sum_{i = 1}^{\frac{N}{2} - 6} (q^i - q^{\frac{N}{2} - 2 - i + 1})
    \Big)
  \bigg]\nonumber\\
  & \ \times \operatorname{PE} \bigg[
    \frac{q^{\frac{1}{2}(\frac{N}{2} - 6 + 1)}
    - q^{\frac{N}{2} - 2 - \frac{1}{2}(\frac{N}{2} - 6 - 1)}
    }{(1 - q)(1 - q^{\frac{N}{2} - 2})}(b_1 + b_1^{-1})(a + a^{-1})
  \bigg] \ ,
\end{align}
and 
\begin{align}
  & \ \mathcal{I}_{D_{\frac{N}{2} - 2}(\mathfrak{sl}(\frac{N}{2}),[\frac{N}{2}-2,1,1])} \nonumber\\
  = & \ \operatorname{PE}\bigg[
    \frac{1}{(1-q)(1-q^{\frac{N}{2} - 2})}\Big(
      (q - q^{\frac{N}{2} - 2}) \, \chi_{\text{adj}}^{SU(2)}(a) + \sum_{i = 1}^{\frac{N}{2} - 2} (q^i - q^{\frac{N}{2} - 2 - i + 1})
    \Big)
  \bigg]\nonumber\\
  & \ \times \operatorname{PE} \bigg[
    \frac{q^{\frac{1}{2}(\frac{N}{2} - 2 + 1)}
    - q^{\frac{N}{2} - 2 - \frac{1}{2}(\frac{N}{2} - 2 - 1)}
    }{(1 - q)(1 - q^{\frac{N}{2} - 2})}(b_2 + b_2^{-1})(a + a^{-1})
  \bigg] \nonumber \\
  = & \ \operatorname{PE} \bigg[
    \frac{(q - q^{\frac{N}{2} - 2})}{(1-q)(1- q^{\frac{N}{2} - 2})} \chi^{SU(2)}_\text{adj}(a)
  \bigg] 
  = \mathcal{I}_{D_{\frac{N}{2} - 2}(\mathfrak{sl}(2), [1,1])}
\end{align}
where we have identified the two $SU(2)$ flavor fugacities from the two theories. The last equality manifests the duality $D_{N-4}(\mathfrak{sl}(N), [N-4, 4]) = D_{N-4}(\mathfrak{sl}_4, [4])$. The plethystic exponential can be simplified using
\begin{equation}
  \frac{1}{(1 - q)(1 - q^{\frac{N}{2} - 2})}\sum_{i = 1}^{\frac{N}{2} - 6}(q^i - q^{\frac{N}{2} - 2 - i + 1}) = \frac{(1 + q + q^2 + q^3)}{(1 - q^{\frac{N}{2} - 2})} \sum_{i = 1}^{\frac{N}{2} - 6}q^i \ .
\end{equation}
Then the Schur index of the full theory is given by the integral
\begin{equation}
  \mathcal{I} = \oint \frac{da}{2\pi i a} \mathcal{I}_{\text{VM}}(\mathfrak{a}) \mathcal{I}_{D_{\frac{N}{2} - 2}(\mathfrak{sl}(\frac{N}{2} -4),[\frac{N}{2}-6,1,1])}(\mathfrak{a},\mathfrak{b}_1) \mathcal{I}_{D_{\frac{N}{2} - 2}(\mathfrak{sl}(\frac{N}{2}),[\frac{N}{2}-2,1,1])}(\mathfrak{a})\ ,
\end{equation}
where the integrand reads
\begin{align}
  \mathcal{Z}(\mathfrak{a})
  = & \ \frac{q^{- \frac{7}{24}(N-4)} \eta((\frac{N}{2} - 2)\tau)^8}{2}
  \prod_{i = 1}^{\frac{N}{2}-6} \prod_{j = 0}^3 \frac{1}{(q^{i+j}, q^{\frac{N}{2} - 2})} \\ 
  & \ \times \frac{\vartheta_1(2 \mathfrak{a} + (\frac{N}{2} - 2)\tau | (\frac{N}{2} - 2) \tau)^2}{
    \prod_{\alpha = \pm}\prod_{\ell = -1, 1, 3,5}
    \vartheta_1(\mathfrak{a} + \alpha \mathfrak{b}_1 + (\ell - \frac{N}{2}) \frac{\tau}{2}| (\frac{N}{2} - 2)\tau)
  } \nonumber \ .
\end{align}
The relevant poles are simple and imaginary, located at
\begin{equation}
  \alpha \mathfrak{b}_1 - (\ell - \frac{N}{2}) \frac{\tau}{2}, \qquad \alpha = \pm, \ell = -1, 1, 3, 5 \ ,
\end{equation}
with residues
\begin{align}\label{eq:residue-D_N-4}
  R_{\alpha, \ell} = & \ \frac{q^{- \frac{7}{24}(N-4)} \eta((\frac{N}{2} - 2)\tau)^8}{2}
  \prod_{i = 1}^{\frac{N}{2}-6} \prod_{j = 0}^3 \frac{1}{(q^{i+j}, q^{\frac{N}{2} - 2})} \nonumber \\
  & \ \times \frac{\vartheta_1((N-2-\ell)\tau + 2\alpha \mathfrak{b}_1 | (\frac{N}{2} - 2)\tau)^2}{
    \prod_{\alpha' = \pm}\prod_{\ell' = -1, 1, 3,5}
    \vartheta_1( (\alpha' + \alpha) \mathfrak{b}_1 + (\ell' - \ell) \frac{\tau}{2}| (\frac{N}{2} - 2)\tau)
  } \ .
\end{align}
As the result of the residue computation, any $\vartheta_1(0|(\frac{N}{2} - 2)\tau)$ in the denominator should be replaced by $\frac{1}{2\pi i}\vartheta_1'(0|(\frac{N}{2} - 2)\tau) = -i \eta((\frac{N}{2} - 2)\tau)^3$. The Schur index is then given by
\begin{align}
  \mathcal{I}_{D_{N-4}(\mathfrak{sl}(N), [N-4,4])}
  = \sum_{\alpha = \pm}\sum_{\ell = -1,1,3,5}R_{\alpha, \ell} E_1 \begin{bmatrix}
    -1 \\ b_1^\alpha q^{\frac{1}{2}(\frac{N}{2} - \ell)}q^{\frac{1}{2}}
  \end{bmatrix}\Big((\frac{N}{2} -2)\tau\Big) \ .
\end{align}

The integrand $\mathcal{Z}(\mathfrak{a})$ can also be rewritten as follows, using the shift property of the Jacobi theta functions,
\begin{align}
  \mathcal{Z}(\mathfrak{a})
  = & \ \frac{q^{- \frac{7}{24}(N-4) + \frac{1}{4}} }{2 \prod_{i = 1}^{\frac{N}{2} - 6}\prod_{j = 1}^4 (q^{i + j}, q^{\frac{N}{2} - 2})} \frac{\eta((\frac{N}{2} - 2)\tau)^8\vartheta_1(2 \mathfrak{a}|(\frac{N}{2} - 2)\tau)^2}{
    \prod_{\alpha = \pm} \prod_{\ell = 1}^4 \vartheta_4(\pm \mathfrak{a} + \mathfrak{m}_\ell| (\frac{N}{2} - 2)\tau)
  } \ ,
\end{align}
where the parameters $\mathfrak{m}_i$ are defined as
\begin{equation}
  \mathfrak{m}_\ell = \mathfrak{b}_1 - \frac{3\tau}{2} + (\ell - 1) \tau, \qquad \ell = 1, \dots, 4 \ .
\end{equation}
We notice that the second factor is nothing but the integrand that computes the $SU(2)$ theory with four flavors, whose flavor parameters $\mathfrak{m}_i$ are set to the above special values and the modular parameter $\tau$ is rescaled to $\tilde \tau$. Concretely,
\begin{equation}
  \mathcal{I}_{D_{N-4}(\mathfrak{sl}(N), [N-4,4])}(q, b_1)
  = \frac{q^{- \frac{7}{24}(N-4) + \frac{1}{4}} }{2 \prod_{i = 1}^{\frac{N}{2} - 6}\prod_{j = 1}^4 (q^{i + j}, q^{\frac{N}{2} - 2})} \mathcal{I}_{SU(2) ~ \text{SQCD}}(q^{\frac{N}{2} - 2}, m_\ell) \ ,
\end{equation}
where
\begin{equation}
  m_\ell = b_1 q^{- \frac{3}{2}} q^{(\ell - 1)} \ .
\end{equation}

\subsection{The unflavored limit}

It is straightforward to compute the unflavored limit. We use the following identity to simplify the expressions,
\begin{equation}
  (-i)^n q^{\frac{1}{12}n(n+1)}\prod_{\ell = 1}^{\lfloor \frac{2n + 1}{2} \rfloor} \vartheta_1(\ell \tau | (2n+1)\tau) = \eta(\tau) \eta((2n+1)\tau)^{n-1} \ , \quad n = 1, 2, \cdots, \ ,
\end{equation}
\begin{align}
  \prod_{i = 1}^{\frac{N}{2} - 6}\prod_{\ell = 0}^{3}(q^{i + \ell}, q^{\frac{N}{2} - 2}) = & \ - \frac{
    \prod_{\ell = 1}^{\lfloor \frac{N}{4} - 1 \rfloor} \vartheta_1(\ell \tau|(\frac{N}{2} - 2)\tau)^{\min(4, \ell)}}{
    q^{-\frac{1}{4}(2N - 29)}
    q^{- \frac{1}{24}(\frac{N}{2} - 2)M}
    \eta((\frac{N}{2} - 2)\tau)^M
  }\ ,  \nonumber\\
  M \coloneqq & \ \sum_{\ell = 1}^{\lfloor \frac{N}{4} - 1 \rfloor} \min(\ell, 4) \ .
\end{align}
To list a few explicit examples with low $N$, we have
\begin{align}
  \mathcal{I}_{D_{14}(\mathfrak{sl}_4, [4])}(q)
  = & \ \frac{iq^{-\frac{13}{3}}\eta(7\tau)^2}{\eta(\tau)^3 \vartheta_1(2\tau|7\tau)}
  \bigg( E_1 \begin{bmatrix} 1 \\ q \end{bmatrix}(7\tau) - 2 E_1 \begin{bmatrix} 1 \\ q^2 \end{bmatrix}(7\tau) + E_1 \begin{bmatrix} 1 \\ q^3 \end{bmatrix}(7\tau) \nonumber \\
  & + 2 E_2(7\tau) + 2 E_2 \begin{bmatrix} 1 \\ q^2 \end{bmatrix}(7\tau) + 2 E_2 \begin{bmatrix} 1 \\ q^3 \end{bmatrix}(7\tau) \nonumber\\
  & + E_1 \begin{bmatrix} 1 \\ q^3 \end{bmatrix}(7\tau)^2 - E_1 \begin{bmatrix} 1 \\ q^2 \end{bmatrix}(7\tau)^2 \nonumber \\
  & + 4 E_1 \begin{bmatrix} 1 \\ q \end{bmatrix}(7\tau) E_1 \begin{bmatrix} 1 \\ q^3 \end{bmatrix}(7\tau) - 2 E_1 \begin{bmatrix} 1 \\ q \end{bmatrix}(7\tau) E_1 \begin{bmatrix} 1 \\ q^2 \end{bmatrix}(7\tau)
  \bigg) \ ,
\end{align}
\begin{align}
  \mathcal{I}_{D_{18}(\mathfrak{sl}_4,[4])}(q)
  = & \ \frac{ 
     \vartheta_1(\tau|9\tau)
     \vartheta_1(3\tau|9\tau)}{
     q^{\frac{35}{6}}\eta(\tau)^4
    } \bigg(
      - E_1 \begin{bmatrix} 1 \\ q \end{bmatrix}(9\tau)
      + 2 E_1 \begin{bmatrix} 1 \\ q^2 \end{bmatrix}(9\tau)
      - E_1 \begin{bmatrix} 1 \\ q^3 \end{bmatrix}(9\tau) \nonumber\\
    & - 2 E_2(9\tau)
    - 2 E_2 \begin{bmatrix} 1 \\ q^3 \end{bmatrix}(9\tau)
    - 2 E_2 \begin{bmatrix} 1 \\ q^4 \end{bmatrix}(9\tau)
    \nonumber\\
    & - 3 E_1 \begin{bmatrix} 1 \\ q^3 \end{bmatrix}(9\tau)^2
     - E_1 \begin{bmatrix} 1 \\ q^4 \end{bmatrix}(9\tau)^2
     \nonumber\\
    & + 2 E_1 \begin{bmatrix} 1 \\ q \end{bmatrix}(9\tau) E_1 \begin{bmatrix} 1 \\ q^3 \end{bmatrix}(9\tau)
    - 4 E_1 \begin{bmatrix} 1 \\ q \end{bmatrix}(9\tau) E_1 \begin{bmatrix} 1 \\ q^4 \end{bmatrix}(9\tau) 
     \nonumber\\
    & \ + 2 E_1 \begin{bmatrix} 1 \\ q^2 \end{bmatrix}(9\tau) E_1 \begin{bmatrix} 1 \\ q^3 \end{bmatrix}(9\tau)
    + 2 E_1 \begin{bmatrix} 1 \\ q^2 \end{bmatrix}(9\tau) E_1 \begin{bmatrix} 1 \\ q^4 \end{bmatrix}(9\tau)
    \bigg) \ ,
\end{align}
\begin{align}
  \mathcal{I}_{D_{22}(\mathfrak{sl}_4, [4])}(q)
  = & \ \frac{\vartheta_1(\tau|11\tau) \vartheta_1(3\tau|11\tau)}{q^{\frac{25}{3}}\eta(\tau)^4}\bigg(
    - E_1 \begin{bmatrix} 1 \\ q \end{bmatrix}(11\tau)
    + 2 E_1 \begin{bmatrix} 1 \\ q^2 \end{bmatrix}(11\tau)
    - E_1 \begin{bmatrix} 1 \\ q^3 \end{bmatrix}(11\tau) \nonumber\\
    & - 2 E_2(11\tau)
    - 2 E_2 \begin{bmatrix} 1 \\ q^4 \end{bmatrix}(11\tau) 
    - 2 E_2 \begin{bmatrix} 1 \\ q^5 \end{bmatrix}(11\tau)\nonumber \\
    & - E_1 \begin{bmatrix} 1 \\ q^4 \end{bmatrix}(11\tau)^2
    - E_1 \begin{bmatrix} 1 \\ q^5 \end{bmatrix}(11\tau)^2
    \nonumber \\
    & + 2 E_1 \begin{bmatrix} 1 \\ q \end{bmatrix}(11\tau) E_1 \begin{bmatrix} 1 \\ q^4 \end{bmatrix}(11\tau)
    - 4 E_1 \begin{bmatrix} 1 \\ q \end{bmatrix}(11\tau) E_1 \begin{bmatrix} 1 \\ q^5 \end{bmatrix}(11\tau) \nonumber \\
    & \ + 2 E_1 \begin{bmatrix} 1 \\ q^2 \end{bmatrix}(11\tau) E_1 \begin{bmatrix} 1 \\ q^4 \end{bmatrix}(11\tau)
    + 2 E_1 \begin{bmatrix} 1 \\ q^2 \end{bmatrix}(11\tau) E_1 \begin{bmatrix} 1 \\ q^5 \end{bmatrix}(11\tau) \nonumber \\
    & - 2 E_1 \begin{bmatrix} 1 \\ q^3 \end{bmatrix}(11\tau) E_1 \begin{bmatrix} 1 \\ q^4 \end{bmatrix}(11\tau) 
  \bigg) \ .
\end{align}
\begin{align}
  \mathcal{I}_{D_{26}(\mathfrak{sl}_4, [4])}(q) = & \ \frac{\vartheta_1(\tau|13\tau) \vartheta_1(3 \tau|13\tau)}{q^{\frac{65}{6}}\eta(\tau)^4} \bigg(
    - E_1 \begin{bmatrix} 1 \\ q \end{bmatrix}(13\tau) + 2 E_1 \begin{bmatrix} 1 \\ q^2 \end{bmatrix}(13\tau) - E_1 \begin{bmatrix} 1 \\ q^3 \end{bmatrix}(13\tau)  \nonumber\\
    & - 2 E_2(13\tau) - 2 E_2 \begin{bmatrix} 1 \\ q^5 \end{bmatrix}(13\tau) - 2 E_2 \begin{bmatrix} 1 \\ q^6 \end{bmatrix}(13\tau)  \nonumber\\
    & - E_1 \begin{bmatrix} 1 \\ q^5 \end{bmatrix}(13\tau)^2 - E_1 \begin{bmatrix} 1 \\ q^6 \end{bmatrix}(13\tau)^2  \nonumber\\
    & + 2 E_1 \begin{bmatrix} 1 \\ q \end{bmatrix}(13\tau) E_1 \begin{bmatrix} 1 \\ q^5 \end{bmatrix}(13\tau) - 4 E_1 \begin{bmatrix} 1 \\ q \end{bmatrix}(13\tau) E_1 \begin{bmatrix} 1 \\ q^6 \end{bmatrix}(13\tau)  \nonumber\\
    &  + 2 E_1 \begin{bmatrix} 1 \\ q^2 \end{bmatrix}(13\tau) E_1 \begin{bmatrix} 1 \\ q^5 \end{bmatrix}(13\tau) 
    + 2 E_1 \begin{bmatrix} 1 \\ q^2 \end{bmatrix}(13\tau) E_1 \begin{bmatrix} 1 \\ q^6 \end{bmatrix}(13\tau) \nonumber\\
    & \ - 2 E_1 \begin{bmatrix} 1 \\ q^3 \end{bmatrix}(13\tau) E_1 \begin{bmatrix} 1 \\ q^5 \end{bmatrix}(13\tau) 
    \bigg) \ ,
\end{align}
\begin{align}
  \mathcal{I}_{D_{30}(\mathfrak{sl}_4, [4])}(q)
   & = \frac{\vartheta_1(\tau|15\tau) \vartheta_1(3\tau|15\tau)}{q^{40/3}\eta(\tau)^4}\bigg(
    - E_1 \begin{bmatrix} 1 \\ q \end{bmatrix}(15\tau)
    + 2 E_1 \begin{bmatrix} 1 \\ q^2 \end{bmatrix}(15\tau)
    - E_1 \begin{bmatrix} 1 \\ q^3 \end{bmatrix}(15\tau)
     \nonumber\\
    &  - 2 E_2(15\tau)
    - 2 E_2 \begin{bmatrix} 1 \\ q^6 \end{bmatrix}(15\tau) 
    - 2 E_2 \begin{bmatrix} 1 \\ q^7 \end{bmatrix}(15\tau)\nonumber\\
    & - E_1 \begin{bmatrix} 1 \\ q^6 \end{bmatrix}(15\tau)^2
    - E_1 \begin{bmatrix} 1 \\ q^7 \end{bmatrix}(15\tau)^2 \nonumber\\
    & + 2 E_1 \begin{bmatrix} 1 \\ q \end{bmatrix}(15\tau) E_1 \begin{bmatrix} 1 \\ q^6 \end{bmatrix}(15\tau)
    - 4 E_1 \begin{bmatrix} 1 \\ q \end{bmatrix}(15\tau) E_1 \begin{bmatrix} 1 \\ q^7 \end{bmatrix}(15\tau) \nonumber \\
    & \ + 2 E_1 \begin{bmatrix} 1 \\ q^2 \end{bmatrix}(15\tau) E_1 \begin{bmatrix} 1 \\ q^6 \end{bmatrix}(15\tau)
    + 2 E_1 \begin{bmatrix} 1 \\ q^2 \end{bmatrix}(15\tau) E_1 \begin{bmatrix} 1 \\ q^7 \end{bmatrix}(15\tau) \nonumber\\
    & - 2 E_1 \begin{bmatrix} 1 \\ q^3 \end{bmatrix}(15\tau) E_1 \begin{bmatrix} 1 \\ q^6 \end{bmatrix}(15\tau)
  \bigg) \ .
\end{align}
From the above results, the closed form unflavored Schur index can be easily read off, with $\tilde \tau \coloneqq \frac{N-4}{2}\tau$,
\begin{align}\label{eq:unflavoredSchurIndex-series-1}
  & \ \mathcal{I}_{D_{N - 4}(\mathfrak{sl}(N), [N-4, 4])}(q)
  = \mathcal{I}_{D_{N - 4}(\mathfrak{sl}_4, [4])}(q) \nonumber \\
  = & \ \frac{\vartheta_1(\tau|\frac{N - 4}{2}\tau) \vartheta_1(3\tau|\frac{N-4}{2}\tau)}{q^{\frac{10N-190}{24}}\eta(\tau)^4}\bigg(
    - E_1 \begin{bmatrix} 1 \\ q \end{bmatrix}(\tilde \tau)
    + 2 E_1 \begin{bmatrix} 1 \\ q^2 \end{bmatrix}(\tilde \tau)
    - E_1 \begin{bmatrix} 1 \\ q^3 \end{bmatrix}(\tilde \tau)\nonumber\\
    & \qquad\qquad  - 2 E_2(\tilde \tau)
    - 2 E_2 \begin{bmatrix} 1 \\ q^{\frac{N-10}{4}} \end{bmatrix}(\tilde \tau) 
    - 2 E_2 \begin{bmatrix} 1 \\ q^{\frac{N-6}{4}} \end{bmatrix}(\tilde \tau)\nonumber\\
    & \qquad\qquad - E_1 \begin{bmatrix} 1 \\ q^{\frac{N-10}{4}} \end{bmatrix}(\tilde \tau)^2
    - E_1 \begin{bmatrix} 1 \\ q^{\frac{N-6}{4}} \end{bmatrix}(\tilde \tau)^2 \nonumber\\
    & \qquad\qquad + 2 E_1 \begin{bmatrix} 1 \\ q \end{bmatrix}(\tilde \tau) E_1 \begin{bmatrix} 1 \\ q^{\frac{N-10}{4}} \end{bmatrix}(\tilde \tau)
    - 4 E_1 \begin{bmatrix} 1 \\ q \end{bmatrix}(\tilde \tau) E_1 \begin{bmatrix} 1 \\ q^{\frac{N-6}{4}} \end{bmatrix}(\tilde \tau) \nonumber \\
    & \qquad\qquad \ + 2 E_1 \begin{bmatrix} 1 \\ q^2 \end{bmatrix}(\tilde \tau) E_1 \begin{bmatrix} 1 \\ q^{\frac{N-10}{4}} \end{bmatrix}(\tilde \tau)
    + 2 E_1 \begin{bmatrix} 1 \\ q^2 \end{bmatrix}(\tilde \tau) E_1 \begin{bmatrix} 1 \\ q^{\frac{N-6}{4}} \end{bmatrix}(\tilde \tau) \nonumber\\
    & \qquad\qquad - 2 E_1 \begin{bmatrix} 1 \\ q^3 \end{bmatrix}(\tilde \tau) E_1 \begin{bmatrix} 1 \\ q^{\frac{N-10}{4}} \end{bmatrix}(\tilde \tau)
  \bigg) \ .
\end{align}

\subsection{The large \texorpdfstring{$N$}{N} limit}

Let us series expand the Schur index for $N = 18, 22, 26, 30, 34$, where we drop the $q^{ - \frac{c}{24}}$ factor,
\begin{align}
  & \ \mathcal{I}_{D_{14}(\mathfrak{sl}(18), [14,4])}(b)
  =  1
  +q +3 q^2+6 q^3+13 q^4
  + (b^2+\frac{1}{b^2}+23) q^5 \nonumber\\
  & \ \qquad +(2 b^2+\frac{2}{b^2}+45) q^6
  +(78 + 6 b^2+\frac{6 }{b^2}) q^{7}  + (141 + 12 b^2+\frac{12}{b^2}) q^8\nonumber\\
  & \ \qquad +(239 + 27 b^2+\frac{27}{b^2}) q^9 + (409 + b^4+52 b^2+\frac{52}{b^2}+\frac{1}{b^4})q^{10} + \cdots \ , \\
  & \mathcal{I}_{D_{18}(\mathfrak{sl}(22), [18,4])}(b)
  = 1 +q +3 q^2+6 q^3+13 q^4 + 23 q^5 + 45 q^6
  + (78 + b^2+\frac{1}{b^2})q^{7} \nonumber\\
  & \qquad + (141 + 2 b^2+\frac{2}{b^2})q^{8} + (239 + 6/b^2 + 6 b^2)q^9 + (409 + 12 b^2+\frac{12}{b^2})q^{10} \nonumber\\
  & \qquad + (27 b^2+\frac{27}{b^2}+674)q^{11} + \cdots \ ,  \nonumber\\
  & \ \mathcal{I}_{D_{22}(\mathfrak{sl}(26), [22,4])}(b) = 1 + q + 3 q^2 + 6 q^3 + 13 q^4 + 23 q^5 + 45 q^6 + 78 q^7 + 141 q^8 \nonumber\\
  & \qquad + (239 + 1/b^2 + b^2)q^9 + (409 + \frac{2}{b^2} + 2 b^2)q^{10} + (674 + \frac{6}{b^2} + 6 b^2)q^{11}  \nonumber\\
  & \qquad + (12 b_1^2+\frac{12}{b_1^2}+1116)q^{12} + \cdots \ .
\end{align}
From these results, we observe a clear large-$N$ limit,
\begin{align}
  \lim_{N \to \infty} \mathcal{I}_{D_{N - 4}(\mathfrak{sl}(N), [N-4, 4])}(b) 
  = 1 + q + 3q^2 + 6q^3 + 13 q^4 + 23q^5 + 45q^6 + 78 q^7  \nonumber \\
   + 141 q^8 + 239 q^9 + 409 q^{10} + 674 q^{11} + 1116 q^{12} + \cdots \ ,
\end{align}
where the explicit $q$-series arise from the following analytic function,
\begin{align}\label{eq:large-N-series-1}
  \frac{(1 - q)^3 (1 - q^2)^2 (1 - q^3)}{(q;q)^4} \ .
\end{align}
To show this large-$N$ behavior, first note that the large-$N$ limit is independent of the flavor fugacity $b$: whenever a term in the $q$-series develops independence of $b$, it stabilizes and assumes the fixed value in the large-$N$ limit. Since $b$ drops out, this limit can be seen by analyzing the closed-form unflavored Schur index (\ref{eq:unflavoredSchurIndex-series-1}). First of all the denominator in (\ref{eq:large-N-series-1}) corresponds to the $\eta(\tau)^{-4}$ in (\ref{eq:unflavoredSchurIndex-series-1}). As $N \to +\infty$, the numerator in (\ref{eq:unflavoredSchurIndex-series-1}) contributes only
\begin{align}
  \vartheta_1(\tau|&  \tilde \tau)\vartheta_1(3\tau|\tilde \tau) \nonumber \\
  & \ \sim q^{2 \frac{N-4}{2 \times 8}}q^{1/2}q^{3/2}(1 - q^{\frac{N-4}{2}})^2 
  (1 - q q^{\frac{N-4}{2}})
  (1 - q^{-1})
  (1 - q^3 q^{\frac{N-4}{2}})
  (1 - q^{-3}) \nonumber \\
  & \ \sim q^{\frac{N}{8} - \frac{5}{2}} - q^{\frac{N}{8} - \frac{3}{2}}
  - q^{\frac{N}{8} + \frac{1}{2}} + q^{\frac{N}{8} + \frac{3}{2}} + O(q^{\frac{5N}{8}}) \ .
\end{align}
The combination of Eisenstein is trickier, but by inspection, it approximates to
\begin{align}
  \sim q^{\frac{N-4}{3} - 3}(- 1 + 2q + q^2 - 4 q^3 + q^4 + 2 q^5 - q^6 ) + O(q^{N-10}) \ .
\end{align}
Putting all pieces together including a suitable overall $q$ factors, we see that
\begin{equation}
  \mathcal{I}_{D_{N - 4}(\mathfrak{sl}(N), [N-4,4])}(q)
  \xrightarrow{N \to +\infty} \frac{(1 - q)^3 (1 - q^2)^2 (1 - q^3)}{(q;q)^4} \ .
\end{equation}
Performing plethystic logarithm, we obtain the single letter index
\begin{equation}
  i_{D_{\infty-4}(\mathfrak{sl}(\infty), [\infty-4,4])}(q) = \frac{q + q^2 + q^3 + q^4}{1 - q}\ .
\end{equation}

At closer inspection, we observe that at large finite $N$,
\begin{align}
  & \ (q,q)^4\mathcal{I}_{D_{N - 4}(\mathfrak{sl}(N), [N-4,4])}(q) \nonumber \\
  = & \ (1 - q)^3 (1 - q^2)^2 (1 - q^3)
  + 2q^{\frac{N}{2} - 4} \Big[(1-q)(1-q^2)(1-q^3)\Big]^2(1 + q) \nonumber \\
  & \ + q^{2(\frac{N}{2} - 4)}(1 - q)^2 (1 - q^2)^2 (1 - q^3) (1 - q^5) (2 - 5 q + 3 q^2 - 5 q^3 +
  2 q^4)  \nonumber \\
  & \ + q^{3(\frac{N}{2} - 4)}(1 - q)^3 (\cdots) \ .
\end{align}
In terms of ratio,
\begin{align}
  & \ \frac{\mathcal{I}_{D_{N-4}(\mathfrak{sl}(N), [N-4,4])}}{
    \mathcal{I}_{D_{\infty-4}(\mathfrak{sl}(\infty), [\infty-4,4])}} \nonumber \\
    = & \ 1 + 2q^{\frac{N}{2} - 4}(1 + q^2)(1 + q + q^2)
    + q^{2(\frac{N}{2} - 4)}(\sum_{i=1}^{4}q^i) (2 - 5q + 3q^2 - 5q^3 + 2q^4) \nonumber\\
    & \ + q^{3(\frac{N}{2} - 4)} 2 (\sum_{i = 1}^{4}q^i) (1 - 2q + q^2 - 3q^3 + 2q^4 - 3q^5 + q^6 - 2q^7 + q^8)  + \cdots
\end{align}
This structure is very much similar to the giant graviton expansion of $\mathcal{N} = 4$ $U(N)$ theory discussed in \cite{Gaiotto:2021xce,Beccaria:2024szi}, and it would be interesting to explore the holography interpretation of the large-$N$ expansion.

\subsection{Wilson line index}

Using the $SU(2)$ gauge theory description of $D_{N - 4}(\mathfrak{sl}(N),[N-4,4])$, we include a spin-$j$ Wilson line coupled to the $SU(2)$ gauge group. The Wilson line index is given by the integral
\begin{equation}
  \mathcal{I}_j = \oint \frac{da}{2\pi i a} \chi_j(a) \mathcal{Z}(\mathfrak{a}) \ ,
\end{equation}
where $\chi_j(a) = \sum_{m = -j}^{+j} a^{2m} $ is the $SU(2)$ character. When $j \in \mathbb{Z}$, the $m = 0$ term integrates to the original index $\mathcal{I}_{D_{N - 4}(\mathfrak{sl}(N), [N-4,4])}$.

The flavored Wilson line index can be easily computed using the integration formula,
\begin{align}
  \mathcal{I}_j = & \ \delta_{j \in \mathbb{Z}}\mathcal{I}_{D_{N - 4}(\mathfrak{sl}(N), [N-4,4])} - \sum_{\substack{m = -j \\ m\ne 0}}^{+j}\sum_{\alpha=\pm} \sum_{\ell} R_{\alpha, \ell} \frac{(b^\alpha q^{-\frac{1}{2}(\ell - \frac{\tau}{2})})^{2m}q^{- \frac{2m}{2}(\frac{N}{2} - 2)}} {q^{\frac{2m}{2}(\frac{N}{2} - 2)}-q^{-\frac{2m}{2}(\frac{N}{2} - 2)}} \ ,
\end{align}
where the residues $R_{\alpha, \ell}$ are given in (\ref{eq:residue-D_N-4}).

It is straightforward to compute the unflavored limit. For example when $N = 14$
\begin{align}
  \mathcal{I}_j = & \ \delta_{j \in \mathbb{Z}} \mathcal{I}_{D_{10}(\mathfrak{sl}(14), [10,4])}(q) \nonumber\\
  & \ - \frac{\eta(\tilde \tau)^2}{\vartheta_1(\tau|\tilde \tau)^2 \vartheta_1(2 \tau|\tilde \tau)^2} \frac{\Lambda_{14}}{q^{m(\frac{N}{2} - 2)} - q^{-m(\frac{N}{2} - 2)}} \sum_{\substack{m = -j \\ m \ne 0}}^{+j} \bigg( \frac{
    2m(-1-q^{2m} q^{4m} - q^{6m})
  }{
      2q^{\frac{18m+16}{6}}
  }\nonumber\\
  & \ \qquad\qquad\qquad + \frac{
     +2 (q^{6m} - 1)
     + 2(-1 + 2q^{2m} - 2 q^{4m} + q^{6m}) E_1 \big[\substack{1 \\ q}\big](\tilde \tau)
  }{
      2q^{\frac{18m+16}{6}}
  }\nonumber\\
  & \ \qquad\qquad\qquad\frac{+ 2(-2 - q^{2m} + q^{4m} + 2q^{6m})E_1 \big[\substack{1\\q^2}\big](\tilde \tau)}{2q^{\frac{18m+16}{6}}}\bigg) \ .
\end{align}
For $N = 18$,
\begin{align}
  \mathcal{I}_j = & \ \delta_{j \in \mathbb{Z}} \mathcal{I}_{D_{10}(\mathfrak{sl}(14), [10,4])}(q) \nonumber\\
  & \ + \frac{\eta(\tilde \tau)^2}{\vartheta_1(\tau|\tilde \tau)^2 \vartheta_1(2 \tau|\tilde \tau)^2} \frac{\Lambda_{18}}{q^{m(\frac{N}{2} - 2)} - q^{-m(\frac{N}{2} - 2)}} \sum_{\substack{m = -j \\ m \ne 0}}^{+j} \bigg( \frac{
    -2m(1+q^{2m} + q^{4m} + q^{6m})
  }{
      2q^{\frac{18m+23}{6}}
  } \nonumber\\
  & \ \qquad\qquad\qquad + \frac{
     2(-1 + 2q^{2m} - 2 q^{4m} + q^{6m}) E_1 \big[\substack{1 \\ q}\big](\tilde \tau)
  }{
      2q^{\frac{18m+23}{6}}
  }\\
  & \ \qquad\qquad\qquad+\frac{ 2(-1 + q^{2m}) (1+q^{2m})^2E_1 \big[\substack{1\\q^2}\big](\tilde \tau) + 2 (1 - q^{6m})E_1 \Big[\substack{1\\q^3}\Big](\tilde \tau)}{2q^{\frac{18m+23}{6}}}\bigg) \ . \nonumber
\end{align}
For general $N = 4n + 2 \ge 18$ and $j$
\begin{align}
  \mathcal{I}_j = & \ \delta_{j \in \mathbb{Z}} \mathcal{I}_{D_{10}(\mathfrak{sl}(14), [10,4])}(q) \nonumber\\
  & \ + \frac{\eta(\tilde \tau)^2}{\vartheta_1(\tau|\tilde \tau)^2 \vartheta_1(2 \tau|\tilde \tau)^2} \frac{\Lambda_{N}}{q^{m(\frac{N}{2} - 2)} - q^{-m(\frac{N}{2} - 2)}} \sum_{\substack{m = -j \\ m \ne 0}}^{+j} \bigg( \frac{
    -2m(1+q^{2m} + q^{4m} + q^{6m})
  }{
      2q^{\frac{1}{12}(N+36m+38)}
  } \nonumber\\
  & \ \qquad\qquad\qquad + \frac{
     2(-1 + 2q^{2m} - 2 q^{4m} + q^{6m}) E_1 \big[\substack{1 \\ q}\big](\tilde \tau)
  }{
      2q^{\frac{1}{12}(N+36m+38)}
  }\\
  & \ \qquad\qquad\qquad+\frac{ 2(-1 + q^{2m}) (1+q^{2m})^2E_1 \big[\substack{1\\q^2}\big](\tilde \tau) + 2 (1 - q^{6m})E_1 \Big[\substack{1\\q^3}\Big](\tilde \tau)}{2q^{\frac{1}{12}(N+36m+38)}}\bigg) \ . \nonumber
\end{align}

Now we can compute the multi-point functions of Wilson line index, and work out the large-$N$ limit,
\begin{align}
  \langle W_j W_j \rangle \xrightarrow{N \to +\infty} & \ q^{\frac{1}{4}} \frac{(1-q)^3(1-q^2)^2(1-q^3)}{(q,q)^4} \\
  \langle W_{j} W_{j} W_{j} \rangle \bigg|_{j \in \mathbb{Z} + \frac{1}{2}}  \xrightarrow{N\to +\infty} & \ 4 q^{\frac{N - 9}{4}} \frac{(1-q)^2 (1-q^2)^2 (1-q^3)(1-q^4)}{(q,q)^4} \ ,\\
  \langle W_{j} W_{j} W_{j} \rangle\bigg|_{j \in \mathbb{Z}}  \xrightarrow{N\to +\infty} & \ q^{\frac{1}{4}} \frac{(1-q)^3(1-q^2)^2(1-q^3)}{(q,q)^4} \ , \\
  \langle W_j W_j W_j W_j \rangle \xrightarrow{N \to +\infty} & \ q^{\frac{1}{4}}(2j+1) \frac{(1-q)^3(1-q^2)^2(1-q^3)}{(q,q)^4} \\
  \langle W_jW_jW_jW_jW_j \rangle\bigg|_{j \in \mathbb{Z}} \xrightarrow{N \to +\infty} & \ q^{\frac{N - 9}{4}} \frac{(1-q)^2 (1-q^2)^2 (1-q^3)(1-q^4)}{(q,q)^4}\\
  \langle W_jW_jW_jW_jW_j \rangle\bigg|_{j \in \mathbb{Z} + \frac{1}{2}} \xrightarrow{N \to +\infty} & \ q^{\frac{1}{4}} \frac{5j^2 + 5j + 2}{2} \frac{(1-q)^3(1-q^2)^2(1-q^3)}{(q,q)^4} \ .
\end{align}
Note that the $\langle W_j W_j \rangle$, $\langle (W_j)^3|_{j \in \mathbb{Z}} \rangle$, $\langle (W_j)^4\rangle$, $\langle (W_j)^5|_{j \in \mathbb{Z} + \frac{1}{2}} \rangle$ are all proportional to the large-$N$ Schur index up to a simple $q$ factor,
\begin{equation}
  \mathcal{I}_{D_{\infty-4}(\mathfrak{sl}(\infty), [\infty-4,4])} = \frac{(1 - q)^3 (1 - q^2)^2 (1 - q^3)}{(q;q)^4}  \ .
\end{equation}

\section{\texorpdfstring{$D_{N - 2}(\mathfrak{sl}(N), [N-3, 3])$}{}}\label{sec:DN-2}

In this section we analyze the type-I series $D_{N - 2}(\mathfrak{sl}(N), [N-3, 3])$ with even $N \ge 12$. Alternatively, this series coincide with the type-II series $D_{N - 2}^{b = 2}(\mathfrak{sl}_3, [3])$. They admit the following gauge theory description,
\begin{equation}
	\begin{tikzpicture}[baseline=(current bounding box.center)]
		\node (left) {$D_{\frac{N}{2} - 1}(\mathfrak{sl}(\frac{N}{2} - 2),[\frac{N}{2} - 4,1,1])$};
		\node (mid) [right=of left, xshift=0.3cm] {$\mathfrak{su}(2)$};
		\node (right) [right=of mid, xshift=0.3cm] {$D_{\frac{N}{2} - 1}(\mathfrak{sl}(\frac{N}{2}),[\frac{N}{2} - 2, 1,1])$};
		\draw[-{Stealth[scale=1.2]}] (mid) -- (right);
		\draw[-{Stealth[scale=1.2]}] (mid) -- (left); 
	\end{tikzpicture}
\end{equation}
The 4d central charges $a_\text{4d}, c_\text{4d}$ are given by the formula
\begin{equation}
	a_{4d}=\frac{202-101N+12N^2}{12(N-2)},~c_{4d}=\frac{50-25N+3N^2}{3(N-2)} \ .
\end{equation}
Note that these central charges satisfy a property reminiscent to the $a_\text{4d} = c_\text{4d}$ relation valid for $\mathcal{N} = 4$ theories and the special famlies of $\widehat{\Gamma}(G)$ in \cite{Kang:2021lic},
\begin{equation}
  a_\text{4d}  = c_\text{4d} - \frac{1}{12} \ .
\end{equation}
This central charge relation is also similar to that of the $D_{N -4}(\mathfrak{sl}(N), [N-4,4])$, without the $O(1/N)$ tail.

\subsection{The flavored Schur index}

The Schur index for the left and right nodes in the gauge theory description reads
\begin{align}
  & \ \mathcal{I}_{D_{\frac{N}{2} - 1}(\mathfrak{sl}(\frac{N}{2} - 2),[\frac{N}{2} - 4,1,1])}\\
  = & \ \operatorname{PE}\bigg[
    \frac{1}{(1 - q)(1 - q^{\frac{N}{2} - 1})}\Big(
      (q - q^{\frac{N}{2} - 1}) \, \chi_{\text{adj}}^{SU(2)}(a) + \sum_{i = 1}^{\frac{N}{2} - 4} (q^i - q^{\frac{N}{2} - 1 - i + 1})
  \bigg] \nonumber\\
  & \ \times \operatorname{PE}\bigg[
    \frac{(q^{\frac{1}{2}(\frac{N}{2} - 4 + 1)} - q^{\frac{N}{2} - 1 - \frac{1}{2}(\frac{N}{2} - 4 - 1)} )}{(1 - q)(1 - q^{\frac{N}{2} - 1})}(b_1 + b_1^{-1})(a + a^{-1})
  \bigg]
\end{align}
\begin{align}
  & \ \mathcal{I}_{D_{\frac{N}{2} - 1}(\mathfrak{sl}(\frac{N}{2}),[\frac{N}{2} - 2, 1,1])} \nonumber\\
  = & \ \operatorname{PE}\bigg[
    \frac{1}{(1 - q)(1 - q^{\frac{N}{2} - 1})} \Big(
      (q - q^{\frac{N}{2} - 1}) \, \chi_{\text{adj}}^{SU(2)}(a) + \sum_{i = 1}^{\frac{N}{2} - 2} (q^i - q^{\frac{N}{2} - 1 - i + 1})
    \Big)
  \bigg] \nonumber\\
  & \times \operatorname{PE}\bigg[
    \frac{(q^{\frac{1}{2}(\frac{N}{2} - 2 + 1)} - q^{\frac{N}{2} - 1 - \frac{1}{2}(\frac{N}{2} - 2 - 1)} )}{(1 - q)(1 - q^{\frac{N}{2} - 1})} 
      (b_2 + b_2^{-1})(a + a^{-1}) 
  \bigg] \ .
\end{align}

After carefully simplifying the plethystic exponential, we find that  the full Schur index of $D_{\frac{N}{2} - 1}(\mathfrak{sl}(N), [N-3, 3])$ is given by the following integral expression:
\begin{align}
  \mathcal{I}_{D_{\frac{N}{2} - 1}(\mathfrak{sl}(N), [N-3, 3])}
  = \oint \frac{da}{2\pi i a} & \ \frac{
      q^{-\frac{1}{6}(N-2)(N-3)}\eta(\frac{N}{2}-1)^{10}
      \prod_{i = 1}^{\frac{N}{2} - 2}(q^i, q^{\frac{N}{2} - 1})
    }{
      2\eta(\tau)^2
      \prod_{i = 0}^{\frac{N}{2} - 5}\prod_{j = 0}^2 (q^{i + j + 1}, q^{\frac{N}{2} - 1})
      \prod_{i = 0}^{\frac{N}{2} - 3}(q^{1 + i}, q^{\frac{N}{2} - 1})
    } \nonumber\\
  & \ \qquad\qquad\qquad \times  \frac{
    \vartheta_1(2 \mathfrak{a}| (\frac{N}{2} -1)\tau)}{
    \prod_{\alpha = \pm} \vartheta_4(\alpha\mathfrak{a} + \mathfrak{m}_\ell | (\frac{N}{2} -1)\tau)
    } \ ,
\end{align}
where $\mathfrak{m}_\ell$ are given by
\begin{equation}
  \mathfrak{m}_{\ell} = \mathfrak{b}_{1,2}, \qquad \pm \mathfrak{b}_{1} \pm \tau\ . 
\end{equation}
For brevity we often denote the overall factor of $q$-Pochhammer symbols
\begin{align}
  \Lambda_N
  = & \ \prod_{i = 1}^{\frac{N}{2}-2}(q^i, q^{\frac{N}{2} - 1})
  \prod_{i = 0}^{\frac{N}{2} - 5}\prod_{j = 0}^2 \frac{1}{(q^{i+j+1}, q^{\frac{N}{2} - 1})}
  \prod_{i = 0}^{\frac{N}{2} - 3} \frac{1}{(q^{1+i}, q^{\frac{N}{2} - 1})}\nonumber \\
  = & \ \frac{1}{\prod_{\ell = 2}^{\frac{N}{2} - 3}(q^\ell, q^{\frac{N}{2} - 1})
  \prod_{\ell = 3}^{\frac{N}{2} - 4}(q^\ell, q^{\frac{N}{2} - 1})} \ .
\end{align}

Similar to the theories $D_{N - 4}(\mathfrak{sl}(N), [N - 4, 4])$, the Schur index is also closely related to that of the $SU(2)$ SQCD with four flavors,
\begin{align}
  & \ \mathcal{I}_{D_{N - 2}(\mathfrak{sl}(N), [N - 2, 2])}(q, b) \nonumber \\
  = & \ \frac{
      q^{-\frac{1}{6}(N-2)(N-3)}\eta(\frac{N}{2}-1)^{2}
      \prod_{i = 1}^{\frac{N}{2} - 2}(q^i, q^{\frac{N}{2} - 1})
    }{
      \eta(\tau)^2
      \prod_{i = 0}^{\frac{N}{2} - 5}\prod_{j = 0}^2 (q^{i + j + 1}, q^{\frac{N}{2} - 1})
      \prod_{i = 0}^{\frac{N}{2} - 3}(q^{1 + i}, q^{\frac{N}{2} - 1})
    }\mathcal{I}_{SU(2) ~ \text{SQCD}}(q^{\frac{N}{2} - 1}, m_\ell) \ ,
\end{align}
where the four flavor fugacities on the right take the special values
\begin{equation}
  m_\ell = b_1, \ b_2, \ b_1 q^{-1}, \ b_1 q \ .
\end{equation}

The integrand is elliptic with respect to $(\frac{N}{2} - 1)\tau$. It has eight imaginary simple poles,
\begin{align}
  & \pm \mathfrak{b}_1 + \ell\tau + \frac{1}{2}\big(\frac{N}{2} - 1\big)\tau, \qquad \ell = -1,0,1, \ ,\\
  & \pm \mathfrak{b}_2 + \frac{1}{2}(\frac{N}{2} - 1)\tau \ .
\end{align}
The residues are given by, with $\tilde \tau = (\frac{N}{2} - 1)\tau$,
\begin{align}\label{eq:residues-D_N-2-1}
  R_{\alpha \mathfrak{b}_1 + \ell\tau + \frac{1}{2}(\frac{N}{2} - 1)\tau}
  = & \ \Lambda_N \frac{i q^{-\frac{N-3}{3}}\eta(\tilde\tau)^7}{2\eta(\tau)^2\vartheta_1(\tau|\tilde \tau)}\\
  & \ \times \frac{\alpha(-1)^{\ell - 1}\vartheta_1(2\ell \tau + 2 \mathfrak{b}_1|\tilde \tau)}{
    \vartheta_1((|\ell| + 1)\tau|\tilde \tau)
    \prod_\pm \vartheta_1(\alpha\ell \tau + \mathfrak{b}_1 \pm \mathfrak{b}_2)
    \prod_{\substack{\ell' \ne -\ell\\ \in \{0,-1,1\}}}\vartheta_1(2\mathfrak{b}_1 + \alpha\ell'\tau)
  } \ , \nonumber\\
  R_{\alpha \mathfrak{b}_2 + \frac{1}{2}(\frac{N}{2} - 1) \tau}
  = & \ - \Lambda_N\alpha\frac{i q^{- \frac{N-3}{3}} \eta(\tilde \tau)^{10}\vartheta_1(2 \mathfrak{b}_2|\tilde \tau)}{
    2\eta(\tau)^2 \prod_{\ell = -1}^1 \prod_\pm \vartheta_1(\mathfrak{b}_1 \pm \mathfrak{b}_2 + \ell\tau|\tilde \tau)\label{eq:residues-D_N-2-2}
  } \ ,
\end{align}
Note that these eight residues sum to zero, and only four are linear independent.

Using these residues, the Schur index then reads
\begin{align}
  \mathcal{I}_{D_{N - 2}(\mathfrak{sl}(N), [N - 2, 2])}(q, b)
  = & \ \sum_{\ell = -1,0,1}\sum_{\alpha = \pm}R_{\alpha \mathfrak{b}_1 + \ell \tau + (\frac{N}{2} - 1)\tau} E_1 \begin{bmatrix}
    -1 \\ b_1^\alpha q^\ell + q^{\frac{1}{2}(\frac{N}{2} - 1)}q^{-\frac{1}{2}}
  \end{bmatrix}(\tilde \tau) \nonumber\\
  & \ + \sum_{\alpha = \pm} R_{\alpha \mathfrak{b}_2 + \frac{1}{2}(\frac{N}{2} - 1) \tau} E_1 \begin{bmatrix}
    -1 \\ b_2^\alpha q^{\frac{1}{2}(\frac{N}{2} - 1)}q^{-\frac{1}{2}}
  \end{bmatrix}(\tilde \tau) \ .
\end{align}

\subsection{The large-\texorpdfstring{$N$}{N} unflavored Schur index}

To analyze the unflavored Schur index, we consider two distinct subcases, $N = 4n$, and $N = 4n + 2$, $n = 3, 4, \cdots$, which have different unflavored Schur index closed-form formula, but sharing the same large-$N$ behavior. First we list a few flavored Schur index.
\begin{align}
  \mathcal{I}_{D_{10}(\mathfrak{sl}(12), [9,3])}
  = & \ 1+2 q+6 q^2+14 q^3  +(32 + b_1^2+b_1 b_2+b_1 b_2^{-1}+ b_i \leftrightarrow b_i^{-1}) q^4  \nonumber\\
  & \ + (66 + 3 b_1^2+3 b_1 b_2+3 b_1 b_2^{-1} + b_i \leftrightarrow b_i^{-1})q^5 + \cdots\\
  \mathcal{I}_{D_{14}(\mathfrak{sl}(16), [13,3])}
  = & \ 1 +2 q+6 q^2+14 q^3+32 q^4+66 q^5  \nonumber\\
  & \ +(136 + b_1^2+b_1 b_2 + b_1 b_2^{-1} + b_i \leftrightarrow b_i^{-1}) q^6  \nonumber\\
  & \ + (262 + 3b_1^2 + 3 b_1 b_2 + 3 b_1 b_2^{-1} + b_i \leftrightarrow b_i^{-1}) q^7 + \cdots\\
  \mathcal{I}_{D_{22}(\mathfrak{sl}(24), [21,3])}
  = & \ 1+2 q+6 q^2+14 q^3+32 q^4+66 q^5+136 q^6+262 q^7+499 q^8+916 q^9 \nonumber\\
  & \ +(1654 +  b_1^2+b_1 b_2 + b_1 b_2^{-1} + b_i \leftrightarrow b_i^{-1}) q^{10} \nonumber\\
  & \ + (2912 + 3 b_1^2 + 3 b_1 b_2 + 3 b_1 b_2^{-1} + b_i \leftrightarrow b_i^{-1}) q^{11}
\end{align}

\begin{align}
  \mathcal{I}_{D_{10}(\mathfrak{sl}(12), [9,3])}
  = & \ 1 + q + 6q^2 + 14q^3 + \big(32 + b_1^2 + b_1 b_2 + \frac{b_1}{b_2} + b_i \leftrightarrow b_i^{-1}\big)q^4\nonumber\\
  & \ + (66 + 3b_1^2 + 3b_1 b_2 + 3b_1b_2^{-1} + b_i \leftrightarrow b_i^{-1})q^5
  + \cdots \ ,\\
  \mathcal{I}_{D_{14}(\mathfrak{sl}(16), [13,3])}
  = & \ 1 + 2q + 6q^2 + 14 q^3 + 32 q^4 + 66q^5 \nonumber\\
  & + (136 + b_1^2 + b_1 b_2 + b_1 b_2^{-1} + b_i \leftrightarrow b_i^{-1})q^6 \nonumber\\
  & + (262 + 3 b_1^2 + 3 b_1 b_2 + 3 b_1 b_2^{-1} + b_i \leftrightarrow b_i^{-1})q^7 + \cdots \ ,\\
  \mathcal{I}_{D_{18}(\mathfrak{sl}(20), [17,3])}
  = & \ 1 + 2q + 6q^2 + 14 q^3 + 32 q^4 + 66q^5 + 136 q^6 + 262 q^7 \nonumber\\
  & + (499 + b_1^2 + b_1 b_2 + b_1 b_2^{-1} + b_i \leftrightarrow b_i^{-1})q^8\nonumber\\
  & + (916 + 3 b_1^2 + 3 b_1 b_2 + 3 b_1 b_2^{-1} + b_i \leftrightarrow b_i^{-1})q^9 + \cdots \ .
\end{align}

We first consider the unflavored Schur index for $N = 4n$. The above explicit $q$-series suggest the index to have a large-$N$ limit that is independent of the flavor fugacities $b_i$. Let us now work out what the limiting $q$-series is by computing the unflavored Schur index.

Direct computation shows that the unflavored Schur index for $N=4n$ takes the form with $\tilde \tau \coloneqq (\frac{N}{2} - 1)\tau$, where the $q$ factor normalizes the series to start with one,
\begin{align}\label{eq:unflavoredSchurIndex-series-2-4n}
  & \ \mathcal{I}_{D_{N - 2} (\mathfrak{sl}(N), [N-3, 3])}(q)
  =  - \Lambda_N \frac{q^{- \frac{N}{3} + \frac{3}{4}}\eta(\tilde \tau)^4}{2 \eta(\tau)^2 \vartheta_1(\tau|\tilde \tau)} \bigg( \nonumber\\
  & \ + 1 + 4 E_1\begin{bmatrix}
      1 \\ q
    \end{bmatrix}(\tilde \tau)
    -2 E_1\begin{bmatrix}
      1 \\ q^2
    \end{bmatrix}(\tilde \tau)
    + 4 E_1 \begin{bmatrix}
      1 \\ q^{\frac{1}{2}(\frac{N}{2} - 1)}
    \end{bmatrix}(\tilde \tau) \nonumber\\
  & \ -6 E_2(\tilde \tau)
  - 4 E_2 \begin{bmatrix}
      1 \\ q^{\frac{1}{2}(\frac{N}{2} - 3)}
    \end{bmatrix}(\tilde \tau)
  - 8 E_2 \begin{bmatrix}
    1 \\ q^{\frac{1}{2}(\frac{N}{2} - 1)}
  \end{bmatrix}(\tilde \tau) 
  -2 E_1 \begin{bmatrix}
    1 \\ q^{\frac{1}{2}(\frac{N}{2} - 3)}
  \end{bmatrix}(\tilde \tau)^2\nonumber\\
  & \ 
   + 8 E_1 \begin{bmatrix}
    1 \\ q
  \end{bmatrix}(\tilde \tau)E_1 \begin{bmatrix}
    1 \\ q^{\frac{1}{2}(\frac{N}{2} - 3)}
  \end{bmatrix}(\tilde \tau)
  - 4 E_1 \begin{bmatrix}
    1 \\ q^2
  \end{bmatrix}(\tilde \tau)E_1 \begin{bmatrix}
    1 \\ q^{\frac{1}{2}(\frac{N}{2} - 3)}
  \end{bmatrix}\Big(\tilde \tau\Big) \bigg) \ ,
\end{align}
where $\Lambda_N$ is define above,
\begin{equation}
  \Lambda_N = \frac{1}{\prod_{\ell = 2}^{\frac{N}{2} - 3}(q^\ell, q^{\frac{N}{2} - 1})
  \prod_{\ell = 3}^{\frac{N}{2} - 4}(q^\ell, q^{\frac{N}{2} - 1})} \ .
\end{equation}
In deriving this close form, we also make use of the identity
\begin{align}
  \prod_{\ell = 1}^n \vartheta_1(\ell \tau | 2n\tau) = & \ i^n q^{- \frac{1}{24}(n^2 + 4n + 1)} \eta(2n\tau)^{n - 1}(q^n, q^{2n}) \ , \\
  (q^n, q^{2n})^2 = & \ \frac{-i q^{\frac{n}{3}}\vartheta_1(n\tau|2n\tau)}{\eta(2n\tau)} \ .
\end{align}

Without the $\Lambda_N$ factor, we observe that the remaining pieces of $\mathcal{I}_{D_{N - 2}(\mathfrak{sl}(N), [N-3,3])}(q)$ has a large-$N$ limit simply given by
\begin{align}
  \Lambda_N^{-1}\mathcal{I}_{D_{N - 2}(\mathfrak{sl}(N), [N-3,3])}(q)
  \xrightarrow{N \to +\infty} \frac{1}{(q,q)^2} = 1 + 2q + 5q^2 + 10q^3 + 20q^4 + \cdots.
\end{align}
For finite $N$, the correction sets in at the $(\frac{N}{2}-3)$-th order,
\begin{equation}
  \Lambda_N^{-1}\mathcal{I}_{D_{N - 2}(\mathfrak{sl}(N), [N-3,3])}(q)
  = \frac{1}{(q,q)^2} + \mathcal{O}(q^{\frac{N}{2} - 3}) \ .
\end{equation}
The $\Lambda_N$ factor itself admits a simple large-$N$ limit,
\begin{equation}
  \Lambda_N^{-1} = \frac{(1-q)^2(1-q^2)}{(q;q)^2} + O(q^{\frac{N}{2} - 3}) \ ,
\end{equation}
which gives rise to the large-$N$ limit
\begin{equation}
  \mathcal{I}_{D_{N - 2}(\mathfrak{sl}(N), [N-3,3])}(q)
  = \frac{(1-q)^2(1-q^2)}{(q;q)^4} + q^{\frac{N}{2} - 3}\mathcal{I}^{(N)}_1(q)
\end{equation}
In fact, at large finite $N = 4n$, the correction term itself is also a nested sum of stabilizing series,
\begin{align}
  \mathcal{I}^{(N)}_1(q) = 1 + 10 q + 29q^2 + 74 q^3 + 157 q^4 + \cdots + q^{\frac{N}{2} - 3}\mathcal{I}^{(N)}_2(q)\\
  \mathcal{I}^{(N)}_n(q) = \mathcal{I}^{(\infty)}_n(q) + q^{\frac{N}{2} - 3}\mathcal{I}^{(N)}_{n+1}(q) \ , \quad n = 2, 3, \cdots
\end{align}
This is similar to the discussions in the previous section and the giant graviton expansion \cite{Gaiotto:2021xce,Beccaria:2024szi}.

Next we comment on the case with $N = 4n + 2$. The unflavored Schur index takes the following form, with $\tilde \tau \coloneqq (\frac{N}{2} - 1)\tau$,
\begin{align}\label{eq:unflavoredSchurIndex-series-2-4n+2}
  & \ \mathcal{I}_{D_{N - 2}(\mathfrak{sl}(N), [N-3, 3])}(q)
  = \ \Lambda_{N} \frac{q^{- \frac{N}{3} + \frac{3}{4}}\eta(\tilde \tau)^4}{4\eta(\tau)^2 \vartheta_1(\tau, q^{14})^4}\Bigg( \nonumber\\
  & \ -1 - 8 E_1 \begin{bmatrix}
    1 \\ q
  \end{bmatrix}(\tilde \tau)
  + 4 E_1 \begin{bmatrix}
    1 \\ q^2
  \end{bmatrix}(\tilde \tau)
  + 2 E_1 \begin{bmatrix}
    1 \\ q^{\frac{1}{2}(\frac{N}{2} - 3)}
  \end{bmatrix}(\tilde \tau)
  - 6 E_1 \begin{bmatrix}
    1 \\ q^{\frac{1}{2}(\frac{N}{2} - 1)}
  \end{bmatrix}(\tilde \tau) \nonumber\\
  & \ + 12 E_2(\tilde \tau)
   + 8 E_2 \begin{bmatrix}
    1 \\ q^{\frac{1}{2}(\frac{N}{2} - 3)}
  \end{bmatrix}(\tilde \tau)
  + 16 E_2 \begin{bmatrix}
    1 \\ q^{\frac{1}{2}(\frac{N}{2} - 1)}
  \end{bmatrix}(\tilde \tau)
  + 4 E_1 \begin{bmatrix}
    1 \\ q^{\frac{1}{2}(\frac{N}{2} - 3)}
  \end{bmatrix}(\tilde \tau)^2 \nonumber\\
  & \ - 16 E_1 \begin{bmatrix}
    1 \\ q
  \end{bmatrix}(\tilde \tau)E_1 \begin{bmatrix}
    1 \\ q^{\frac{1}{2}(\frac{N}{2} - 3)}
  \end{bmatrix}(\tilde \tau)
  + 8 E_1 \begin{bmatrix}
    1 \\ q^2
  \end{bmatrix}(\tilde \tau)E_1 \begin{bmatrix}
    1 \\ q^{\frac{1}{2}(\frac{N}{2} - 3)}
  \end{bmatrix}(\tilde \tau) \nonumber\\
  & \ 
  + 4 E_1 \begin{bmatrix}
    1 \\ q^{\frac{1}{2}(\frac{N}{2} - 3)}
  \end{bmatrix}(\tilde \tau)E_1 \begin{bmatrix}
    1 \\ q^{\frac{1}{2}(\frac{N}{2} - 1)}
  \end{bmatrix}(\tilde \tau)
  \Bigg) \ .
\end{align}
Here the overall $q$ factor normalizes the series to start with $1$.

Although this $\mathcal{I}_{D_{N - 2}(\mathfrak{sl}(N), [N-3,3])}$ looks rather different from that of the $N = 4n$ case, the large-$N$ behavior is actually the same. Plugging in $N = 4n + 2$ into \eqref{eq:unflavoredSchurIndex-series-2-4n}, the resulting $q$-series differs from the correct answer using (\ref{eq:unflavoredSchurIndex-series-2-4n+2}) precisely starting from order $q^{\frac{N}{2} - 3}$.

To summarize, for both $N = 4n, 4n +2$, we have found the large-$N$ behavior
\begin{align}
  \mathcal{I}_{D_{N - 2}(\mathfrak{sl}(N), [N-3,3])}(q) = \frac{(1-q)^2(1-q^2)}{(q;q)^4} + q^{\frac{N}{2} - 3}(1 + \cdots) \ .
\end{align}
Its plethystic logarithm gives a single particle index
\begin{equation}
  i_{D_{\infty-2}(\mathfrak{sl}(\infty), [\infty-3,3])}(q) = \frac{2q + q^2 + q^3}{1 - q} \ .
\end{equation}

\subsection{Wilson line index}

Using the $SU(2)$ gauge theory description, we may include a spin-$j$ Wilson line coupled to the $SU(2)$ gauge group. The Wilson line index is given by the integral
\begin{equation}
  \mathcal{I}_j = \oint \frac{da}{2\pi i a} \chi_j(a) \mathcal{Z}(\mathfrak{a}) \ ,
\end{equation}
where $\chi_j(a) = \sum_{m = -j}^{+j} a^{2m} $ is the $SU(2)$ character. When $j \in \mathbb{Z}$, the $m = 0$ term integrates to the original index $\mathcal{I}_{D_{N - 2}(\mathfrak{sl}(N), [N-3,3])}$.

The Wilson line index $\mathcal{I}_j$ can be easily computed using the integration formula,
\begin{align}
  \mathcal{I}_j = & \ \delta_{j \in \mathbb{Z}}\mathcal{I}_{D_{N - 2}(\mathfrak{sl}(N), [N-3,3])}(q, b) \nonumber \\
  & \ + \sum_{\substack{m = - j \\ m\ne 0}}^{+j} \sum_{\alpha} \sum_{\ell}  R_{\alpha \mathfrak{b}_1 + \ell \tau + \frac{1}{2}(\frac{N}{2} - 1) \tau} \frac{(b_1^\alpha q^\ell q^{\frac{1}{2}(\frac{N}{2} - 1)})^{2m}\tilde q^{- m}}{\tilde q^{m} - \tilde q^{-m}} \nonumber \\
  & \ + \sum_{\alpha = \pm} R_{\alpha \mathfrak{b}_2 + \frac{1}{2}(\frac{N}{2} - 1)\tau} \frac{(b_2^\alpha q^{\frac{1}{2}(\frac{N}{2} - 1)})^{2m} \tilde q^{-m}}{q^m - q^{-m}} \ ,
\end{align}
where the residues are computed in (\ref{eq:residues-D_N-2-1}) and (\ref{eq:residues-D_N-2-2}).

The unflavoring limit of the Wilson line index is straightforward. For example, when $j = \frac{1}{2}$, the unflavored Schur index is, with $\tilde\tau = (\frac{N}{2} - 1)\tau$,
\begin{align}
  \mathcal{I}_{\frac{1}{2}} = & \ - \Lambda_N
  \frac{\eta(\tilde\tau)^4}{\eta(\tau)^2 \vartheta_1(\tau|\tilde\tau)}
  \frac{1}{q^{\frac{N}{2}}}
  \frac{1}{q^{\frac{1}{2}(\frac{N}{2} - 1)} - q^{-\frac{1}{2}(\frac{N}{2} - 1)}} \\
  & \ \bigg(-1 -4 q - q^2 - 4 E_1\begin{bmatrix}
    1 \\ q
  \end{bmatrix}(\tilde \tau)
  + 4q^2 E_1 \begin{bmatrix}
    1 \\ q
  \end{bmatrix}(\tilde \tau)
  + 2 E_1 \begin{bmatrix}
    1 \\ q^2
  \end{bmatrix}(\tilde \tau)
  - 2q^2 E_1 \begin{bmatrix}
    1 \\ q^2
  \end{bmatrix}(\tilde \tau)
  \bigg) \ , \nonumber
\end{align}
and when $j = 1$,
\begin{align}
  \mathcal{I}_1 = & \ \mathcal{I}_{D_{N - 2}(\mathfrak{sl}(N), [N-3,3])}(q)
  + 2 \Lambda_N \frac{\eta(\tilde \tau)^4}{\eta(\tau)^2 \vartheta_1(\tau|\tilde \tau)^4} \frac{1}{q^{\frac{N}{3} + 1}} \frac{1}{q^{\frac{N}{2} - 1} - q^{-\frac{N}{2} + 1}}  \\ 
  & \ \bigg(
    1 + 4q^2 + q^4 + 2 E_1 \begin{bmatrix}
      1 \\ q
    \end{bmatrix}(\tilde \tau)
    - 2q^4 E_1 \begin{bmatrix}
      1 \\ q
    \end{bmatrix}(\tilde \tau)
    - E_1 \begin{bmatrix}
      1 \\ q^2
    \end{bmatrix}(\tilde \tau)
    + q^4 E_1 \begin{bmatrix}
      1 \\ q^2
    \end{bmatrix}(\tilde \tau)
  \bigg) \ . \nonumber
\end{align}
For general $j$ and $N$, the unflavored Wilson line index is given by
\begin{align}
  \mathcal{I}_j = & \ \delta_{j \in \mathbb{Z}}\mathcal{I}_{D_{N - 2}(\mathfrak{sl}(N), [N-3,3])}(q)\\
  & \ - \sum_{\substack{m = -j\\m \ne 0}}^{+j}
  \frac{\Lambda_N\eta(\tilde \tau)^4}{\eta(\tau)^2 \vartheta_1(\tau|\tilde \tau)^4}
   \frac{m(1 + 4q^{2m} + q^{4m}) - (q^{4m} - 1)(2 E_1\big[\substack{1\\q}\big](\tilde \tau) - E_1\big[\substack{1\\q^2}\big](\tilde \tau))}{q^{\frac{N}{3} + 2m - 1}(q^{m(\frac{N}{2} - 1)} - q^{-m(\frac{N}{2} - 1)})} \ . \nonumber
\end{align}

Let us consider the basic building block of these Wilson line index (without $\Lambda_N$),
\begin{equation}
  \mathcal{I}(m) \coloneqq \frac{\eta(\tilde \tau)^4}{\eta(\tau)^2 \vartheta_1(\tau|\tilde \tau)^4}
   \frac{m(1 + 4q^{2m} + q^{4m}) - (q^{4m} - 1)(2 E_1\big[\substack{1\\q}\big](\tilde \tau) - E_1\big[\substack{1\\q^2}\big](\tilde \tau))}{q^{\frac{N}{3} + 2m - 1}(q^{m(\frac{N}{2} - 1)} - q^{-m(\frac{N}{2} - 1)})} \ .
\end{equation}
Using plethystic logarithm, we make the following observations of the large-$N$ behavior,
\begin{align}
  \mathcal{I}(m = \frac{1}{2}) \xrightarrow{N \to \infty} & \ q^{\frac{N-6}{4}}\frac{2(1 + q)^2}{(q,q)^2}, \qquad
  & \mathcal{I}(1) \xrightarrow{N \to \infty} & \ \frac{1}{(q,q)^2} \ .
\end{align}
Similarly, we can compute the multi-point functions of the spin-$j$ Wilson lines and their large-$N$ limits,
\begin{align}
  \mathcal{I}_{\frac{1}{2}} \xrightarrow{N \to +\infty} & \ \frac{1}{(q,q)^2}, \quad
  & \langle W_j W_j \rangle = \sum_{\ell = 0}^{2j} \mathcal{I}_\ell \xrightarrow{N \to \infty} & \ \frac{1}{(q,q)^2} \ ,\\
  \langle W_j W_j W_j \rangle|_{j \in \mathbb{Z}} \xrightarrow{N \to + \infty} & \ \frac{1}{(q,q)^2}, \qquad
  & \langle W_j W_j W_j \rangle|_{j \in \mathbb{Z}+\frac{1}{2}} \xrightarrow{N \to +\infty} & \ q^{\frac{N-6}{4}} \frac{4(1 + q)^2}{(q,q)^2} \ .
\end{align}
where the large-$N$ limit of the two-point function is independent of $j$.

\section{Minahan-Nemeschansky theories}\label{sec:MN}

\subsection{Integration formula}

In this section, we will compute the partially flavored Schur index for the Minahan-Nemeschansky (MN) theories with $E_6, E_7$ flavor symmetry, exploiting the generalized partition formula proposed in \cite{Deb:2025ypl}. The formula inevitably involves contour integrals of elliptic functions with higher order poles, which is beyond the scope of the integration formula in \cite{Pan:2021mrw}. We will first generalize the integration formula to higher order poles. In this subsection we first generalized the formula to accommodate higher order poles, and then apply it to compute the Schur index of the MN theories in the next subsection.

Consider an elliptic function of $\mathfrak{a}$ with double periodicity, 
\begin{equation}
  \mathcal{Z}(\mathfrak{a} + n\tau) = \mathcal{Z}(\mathfrak{a} + 1) = \mathcal{Z}(\mathfrak{a}) \ .
\end{equation}
Here for generality we assume a periodicity $\tilde \tau \coloneqq n\tau$ in the variable $\mathfrak{a}$ with some positive integer $n$ instead of $\tau$; we also denote $\tilde q \coloneqq q^n$. Moreover, the integrand may have $n_j$-th order poles $\mathfrak{a}_j$ inside the fundamental parallelogram bounded by the vertices $0, \tilde\tau, 1, \tilde\tau + 1$. Following the notation in \cite{Pan:2021mrw}, we call $\mathfrak{a}_j$ a real pole when $\mathfrak{a}_j \in [0, 1)$, and imaginary if otherwise.

The integrand $\mathcal{Z}(\mathfrak{a})$ can be expanded using Weierstrass $\zeta$ functions,
\begin{equation}
  \mathcal{Z}(\mathfrak{a}) = C(\tilde\tau) + \frac{1}{2\pi i}\sum_{j}\sum_{m = 1}^{+\infty} R_{j,m} \zeta^{(m - 1)}(\mathfrak{a} - \mathfrak{a}_j|\tilde \tau) \ ,
\end{equation}
where $C(\tilde\tau)$ is a function independent of the variable $\mathfrak{a}$, and we define the generalized residues $R_{j,m}$,
\begin{equation}
  R_{j, m} \coloneqq \operatorname{Res}_{\mathfrak{a} = \mathfrak{a}_j} \bigg[2\pi i \frac{(-1)^{m - 1}}{(m - 1)!} (\mathfrak{a} - \mathfrak{a}_j)^{m - 1} \mathcal{Z}(\mathfrak{a})\bigg] \ .
\end{equation}
Note that $R_{j,m} = 0$ when $m$ is larger than the order $n_j$ of the pole $\mathfrak{a}_j$, hence the above is a finite sum over $\zeta$ functions. Also note that
\begin{equation}
  \sum_j R_{j, m} = 0 \qquad \text{when $m$ is odd} \ .
\end{equation}

The Weierstrass functions $\zeta(\mathfrak{a}|\tilde \tau)$ can be expanded in Fourier series,
\begin{align}
  \zeta(\mathfrak{a} - \mathfrak{a}_j|\tilde \tau) 
  = & \ - 4\pi^2 (\mathfrak{a} - \mathfrak{a}_j)E_2(\tilde\tau)
  \mp \pi i
  + \pi \sum_{n}' \frac{1}{\sin n \pi \tilde\tau} \tilde q^{\mp \frac{n}{2}} e^{2\pi i n (\mathfrak{a} - \mathfrak{a}_j)} \\
  \zeta^{(1)}(\mathfrak{a} - \mathfrak{a}_j|\tilde \tau) 
  = & \ - 4\pi^2E_2(\tilde\tau)
  + \pi \sum_{n}' \frac{2\pi i n}{\sin n \pi \tilde \tau} \tilde q^{\mp \frac{n}{2}} e^{2\pi i n (\mathfrak{a} - \mathfrak{a}_j)} \\ 
  \zeta^{(m - 1)}(\mathfrak{a} - \mathfrak{a}_j|\tilde \tau) 
  = & \ 
  + \pi \sum_{n}' \frac{(2\pi i n)^{m - 1}}{\sin n \pi \tilde \tau} \tilde q^{\mp \frac{n}{2}} e^{2\pi i n (\mathfrak{a} - \mathfrak{a}_j)} , \qquad m \ge 3 \ ,
\end{align}
where the $\mp$ corresponds to real/imaginary poles. The Fourier series can then be used to derive the integration formula, giving
\begin{align}
  \oint \frac{da}{2\pi i a} \mathcal{Z}(\mathfrak{a})
  = C(\tilde\tau) + \sum_{\text{real/img}~ \mathfrak{a}_j} R_{j, 1} \Big(\mp \frac{1}{2}  - 2\pi i\mathfrak{a}_j E_2(\tilde\tau) \Big)
  + 2\pi i E_2(\tilde\tau) \sum_{\text{real/img}~\mathfrak{a}_j} R_{j,2}  \ . \nonumber
\end{align}
We can compute $C(\tilde\tau)$ from any reference point $\mathfrak{a}_0$,
\begin{equation}
  C(\tilde\tau) = \mathcal{Z}(\mathfrak{a}_0) - \frac{1}{2\pi i}\sum_{j}\sum_{m \ge 1} R_{j, m} \zeta^{(m - 1)}(\mathfrak{a} - \mathfrak{a}_j| \tilde \tau)\bigg|_{\mathfrak{a} = \mathfrak{a}_0} \ .
\end{equation}
It is often convenient to choose $\mathfrak{a}_0$ such that $\mathcal{Z}(\mathfrak{a}_0) = 0$ to get rid of the first term.

Analogous to \cite{Pan:2021mrw}, we may rewrite the results in terms of Eisenstein series. The Weierstrass $\zeta$ function is related to the Eisenstein series by
\begin{align}
  \zeta(\mathfrak{z}| \tilde\tau) \pm \pi i + 4\pi^2 \mathfrak{z}E_2(\tilde\tau) =  2\pi iE_1 \begin{bmatrix}
    -1 \\ zq^{\mp1/2}
  \end{bmatrix}(\tilde\tau) \ ,
\end{align}
which hepls simplify the final result,
\begin{align}\label{integration-formula-1}
  \oint \frac{da}{2\pi i a}& \  \mathcal{Z}(\mathfrak{a})
  = \mathcal{Z}(\mathfrak{a}_0)
  + \sum_j R_{j, 1} E_1 \begin{bmatrix}
    -1 \\ a_j/a_0 q^{\pm \frac{1}{2}}
  \end{bmatrix}(\tilde\tau) \\
  & \ - \frac{1}{2\pi i} \sum_j R_{j, 2} \bigg(\zeta'(\mathfrak{a}_0 - \mathfrak{a}_j| \tilde \tau) + 4\pi^2 E_2(\tilde\tau)\bigg)
  - \frac{1}{2\pi i} \sum_{j}\sum_{m > 2} R_{j, m} \zeta^{(m - 1)}(\mathfrak{a}_0 - \mathfrak{a}_j| \tilde \tau)\ . \nonumber
\end{align}
The $\zeta'(\mathfrak{z}|\tau)$ in the second line can be rewritten in Eisenstein series,
\begin{equation}
  \zeta'(\mathfrak{z}|\tau) = 4\pi^2 \bigg(
    E_1 \begin{bmatrix}
      1 \\ z
    \end{bmatrix}(\tau)^2
    + 2 E_2 \begin{bmatrix}
      1 \\ z
    \end{bmatrix}(\tau)
  \bigg) \ .
\end{equation}
The subsequent derivatives $\zeta^{(m - 1)}(\mathfrak{z}|\tau)$ can also be reorganized in terms of Eisenstein series, but we shall not spell out here.

Alternatively, one can stick to the Weierstrass $\zeta$ function expression, and write
\begin{align}
  \oint \frac{da}{2\pi i a}\mathcal{Z}(\mathfrak{a})
  = & \ \mathcal{Z}(\mathfrak{a}_0)
  - \frac{1}{2\pi i} \sum_{j} \sum_{m = 1}^{n_j}  R_{j,m} \zeta^{(m - 1)}(\mathfrak{a}_0 - \mathfrak{a}_j|\tilde \tau) \nonumber\\
  & \ + \frac{1}{2\pi i} \sum_{j}R_{j,1}(\pm \pi i) - 2\pi i E_2(\tilde \tau)
  \sum_{j} R_{j,1} \mathfrak{a}_j + 2\pi i E_2(\tilde \tau) \sum_{j} R_{j,2} \ .
\end{align}
Using the definition of standard residue, the generalized residue can be written as
\begin{equation}
  R_{j,m} = 2\pi i \frac{(-1)^{m - 1}}{(m - 1)!} \frac{1}{(n_j-m)!} \frac{d^{n_j - m}}{da^{n_j - m}} \bigg|_{\mathfrak{a} = \mathfrak{a}_j} \Big[(\mathfrak{a} - \mathfrak{a}_j)^{n_j} \mathcal{Z}(\mathfrak{a})\Big] \ .
\end{equation}
Also note that
\begin{equation}
  \zeta^{(m - 1)}(\mathfrak{a}_0 - \mathfrak{a}_j|\tilde \tau) = (-1)^{m -1} \frac{d^{m - 1}}{d\mathfrak{a}^{m - 1}} \zeta(\mathfrak{a}_0 - \mathfrak{a}|\tilde \tau)
\end{equation}
Putting these ingredients together, the integral can be written simply as combinations of derivatives of the integrand $\mathcal{Z}(\mathfrak{a})$,
\begin{align}
  \oint \frac{da}{2\pi i a}\mathcal{Z}(\mathfrak{a})
  = & \ \mathcal{Z}(\mathfrak{a}_0)
  - \sum_{j} \frac{1}{(n_j - 1)!} \frac{d^{n_j - 1}}{d\mathfrak{a}^{n_j - 1}} \bigg|_{\mathfrak{a} = \mathfrak{a}_j} \Big[(\mathfrak{a} - \mathfrak{a}_j)^{n_j} \mathcal{Z}(\mathfrak{a})\zeta(\mathfrak{a}_0 - \mathfrak{a}|\tilde \tau)\Big] \nonumber\\
  & \ - \frac{1}{2} \sum_{j}(\pm)R_{j,1} - 2\pi i E_2(\tilde \tau)
  \sum_{j} R_{j,1} \mathfrak{a}_j + 2\pi i E_2(\tilde \tau) \sum_{j} R_{j,2} \ .
\end{align}
Here $\pm \pi i = + \pi i$ for a real pole, and $-\pi i$ for an imaginary pole.

Finally, in the presence of a monomial, the integration of Fourier series is fairly simple, giving
\begin{equation}\label{integration-formula-2}
  \oint \frac{da}{2\pi i a}a^k f(\mathfrak{a})
  = \sum_{j} \sum_{m \ge 1}R_{j,m} \frac{1}{\tilde q^{k/2} - \tilde q^{-k/2}} (2\pi i k)^{m - 1}\tilde q^{\pm \frac{k}{2}} a_j^k \ .
\end{equation}
This formula can be used to compute the Wilson line index.

\subsection{Partially flavored \texorpdfstring{$E_6, E_7$}{E6, E7} index}

Minahan-Nemeschansky theories are a class of $\mathcal{N}=2$ SCFTs with exceptional flavor symmetry $E_{6,7,8}$ \cite{Minahan:1996fg}, which can be realized as class-$\mathcal{S}$ theories with regular punctures. The associated VOAs are given by the $\widehat{\mathfrak{e}}_6, \widehat{\mathfrak{e}}_7, \widehat{\mathfrak{e}}_8$ entries of Deligne-Cvitanovi\'c exceptional series at non-admissible level $k = -\frac{h^\vee}{6} - 1$ \cite{Beem:2013sza, Beem:2014rza}, where $h^\vee$ is the dual Coxeter number of the Lie algebra. The Schur index of these theories have been extensively studied in the physics literature \cite{Gadde:2010te,Eguchi:1986sb,Razamat:2012uv}. For example,  in \cite{Razamat:2012uv} it is shown that the $E_6$ Schur index can be written as a sum of $SU(3)$ SQCD Schur index, exploiting the $S$-duality relation and inversion formula \cite{Gadde:2010te,2004math.....11044S}. In \cite{Pan:2021mrw} the flavored Schur index of the $SU(3)$ SQCD is computed using the integration formula, which produces the $E_6$ Schur index in closed form. The unflavored Schur index of $E_{6,7,8}$ satisfy second order unflavored modular differential equations, whose solution can be explicitly written in terms of quasi-modular forms \cite{Arakawa:2016hkg,Arakawa:2015jya}.

The associated VOAs are non-admissible quasi-lisse, whose irreducible highest weight modules were known in \cite{Arakawa:2015jya}. Alternatively, the spectrum can be derived using flavor modular differential equations and quasi-modular bootstrap approach \cite{Pan:2023jjw,Pan:2024dod}. In this approach, the fully flavored characters are completely constrained by a set of flavored modular differential equations coming from special null states in the VOA. Conversely, once the flavored MLDEs are known, any solution must be some linear combination of the irreducible characters. Although the vacuum character of $E_6$ theory is known in closed form, the closed-form of the non-vacuum characters are less studied. In \cite{Eager:2019zrc} a combination of $(E_6)_{-3}$ flavored characters is proposed in closed form using the pure spinor formalism. 

In this section we continue to explore the flavored Schur index of the $E_{6,7}$ theory and non-vacuum characters of the associated VOAs by generalizing the result in \cite{Deb:2025ypl}. The unflavored Schur index of $E_{6, 7, 7+\frac{1}{2}, 8}$ theory can be computed by the contour integral (up to an overall numerical factor)
\begin{equation}
  \mathcal{I}_\alpha = \frac{\eta(\tau)^{-2(\alpha - 1)}}{B(2\alpha -1, \alpha)} \oint \frac{da}{2\pi i a} \mathcal{Z}(\mathfrak{a}, \mathfrak{b})^\alpha \bigg|_{\mathfrak{b}_i = 0} \ ,
\end{equation}
where
\begin{equation}
  \mathcal{Z}(\mathfrak{a}, \mathfrak{b}) = + \frac{1}{2}\vartheta_1(2 \mathfrak{a})^2 \prod_{i = 1}^4 \frac{\eta(\tau)^2}{\vartheta_4( \mathfrak{a} + \mathfrak{b}_i) \vartheta_4(- \mathfrak{a} + \mathfrak{b}_i)} \ ,
\end{equation}
and $B(\alpha, \beta) \coloneqq \frac{\alpha!}{\beta!(\alpha - \beta)!}$ denotes the Binomial coefficient. The dictionary is as follows,
\begin{center}
  \begin{tabular}{c|c|c|c|c}
  $\alpha$ & 2 & 3 & 4 & 5 \\
  \hline
  Theory & $E_6$ & $E_7$ & $E_{7+\frac{1}{2}}$ & $E_8$
\end{tabular}
\end{center}

We begin with the closed-form unflavored index. Setting all the flavor fugacities $\mathfrak{b}_i = 0$, the integrand reads
\begin{equation}
  \mathcal{Z}(\mathfrak{a}) = \frac{1}{2} \eta(\tau)^{8\alpha} \frac{\vartheta_1(2 \mathfrak{a})^{2\alpha}}{\vartheta_4(\mathfrak{a})^{8\alpha}} \ ,
\end{equation}
which is elliptic with periodicity $\tau$ and has only one $n = 8\alpha$-order imaginary pole $\mathfrak{a}_1 = \frac{\tau}{2}$. The closed form of Schur index is given in terms of the residues
\begin{equation}
  R_{1, m} = 2\pi i \frac{(-1)^{m - 1}}{(m - 1)!} \frac{1}{(n - m)!} \frac{d^{n - m}}{d\mathfrak{a}^{n - m}} \bigg|_{\mathfrak{a} = \frac{\tau}{2}} \Big[(\mathfrak{a} - \frac{\tau}{2})^{n} \mathcal{Z}(\mathfrak{a})\Big] \ .
\end{equation}
Since there is only one pole, the residue $R_{1,m = 1}$ must vanish. With this simplification, we can write down the unflavored Schur index compactly as
\begin{equation}
  \mathcal{I}_\alpha(q) = \frac{\eta(\tau)^{2 - 2\alpha}}{B(2\alpha -1, \alpha)}\bigg[ \frac{1}{(n-1)!} \frac{d^{n - 1}}{d\mathfrak{a}^{n - 1}} \bigg|_{\mathfrak{a} = \frac{\tau}{2}} \Big[(\mathfrak{a} - \frac{\tau}{2})^{n} \mathcal{Z}(\mathfrak{a})\zeta(\mathfrak{a}| \tau)\Big]
  - 2\pi i E_2( \tau) R_{1,2}\bigg] \ .
\end{equation}

Now we turn on one flavor fugacity in the above computation, obtaining the partially flavored Schur index using the generalized partition function. The starting point is the fully flavored partition function
\begin{equation}\label{eq:Ialpha}
  \mathcal{I}_{\alpha}(\mathfrak{b}) = \eta(\tau)^{2(\alpha - 1)}\oint \frac{da}{2\pi i a} \mathcal{Z}(\mathfrak{a}, \mathfrak{b})^\alpha \ .
\end{equation}

Consider the simplest case of $\alpha = 2$, corresponding to the $E_6$ case. The integrand is periodic with respect to $\tau$, but it now has eight imaginary double poles,
\begin{equation}
  \mathfrak{a} = \pm \mathfrak{b}_i + \frac{\tau}{2}, \qquad i = 1, 2, 3, 4 \ .
\end{equation}
The presence of double poles requires the use of the integration formula (\ref{integration-formula-1}). The relevant residues are
\begin{align}
  \operatorname{Res}_{\pm \mathfrak{b}_i + \frac{\tau}{2}, 1} = \frac{\pm \eta(\tau)^8\vartheta_1(2 \mathfrak{b}_1)^2}{
    \prod_{j \ne i}\vartheta_1(\mathfrak{b}_i - \mathfrak{b}_j)^2
    \vartheta_1(\mathfrak{b}_i + \mathfrak{b}_j)^2
  }
  \bigg(
    - E_1 \begin{bmatrix}
      1 \\ b_i^2
    \end{bmatrix}
    + \frac{1}{3} \sum_{\alpha = \pm}  \sum_{j \ne i} E_1 \begin{bmatrix}
      1 \\ b_i b_j^\alpha
    \end{bmatrix} 
  \bigg) \ ,
\end{align}
and
\begin{align}
  \operatorname{Res}_{\pm \mathfrak{b}_i + \frac{\tau}{2}, 2}
  =  \frac{\pm \eta(\tau)^8\vartheta_1(2 \mathfrak{b}_1)^2}{
    12\pi i \prod_{j \ne i}\vartheta_1(\mathfrak{b}_i - \mathfrak{b}_j)^2
    \vartheta_1(\mathfrak{b}_i + \mathfrak{b}_j)^2
  }\ .
\end{align}
From this we have
\begin{align}
  \mathcal{I}_{\alpha = 2}(\mathfrak{b})
  = \sum_{i = 1}^{4}\sum_{\pm} \Bigg[ R_{\pm\mathfrak{b}_i + \frac{\tau}{2}, 1} E_1 \begin{bmatrix}
    -1 \\ b_i^\pm
  \end{bmatrix}
  - \frac{1}{2\pi i}R_{\pm\mathfrak{b}_i + \frac{\tau}{2}, 2} \Big(\zeta'(\mp \mathfrak{b}_i - \frac{\tau}{2}) + 4\pi^2 E_2(\tau) \Big)\Bigg] \ .
\end{align}
Now we take the special limit
\begin{equation}
  \mathfrak{b}_{1} \to 3\mathfrak{b}, \qquad \mathfrak{b}_{2,3,4} \to \mathfrak{b}\ ,
\end{equation}
and we find that the the result can be series expanded as
{\small\begin{align}
  = & \ q^{13/12} + \frac{ (8 b^8+16 b^6+30 b^4+16 b^2+8) q^{25/12}}{b^4}\\
  & \ +\frac{\left(35 b^{16}+112 b^{14}+310 b^{12}+480 b^{10}+635 b^8+480 b^6+310 b^4+112 b^2+35\right) q^{37/12}}{b^8} \nonumber \\
  & \ +\frac{2 \left(56 b^{24}+224 b^{22}+749 b^{20}+1632 b^{18}+3030 b^{16}+4184 b^{14}+4885 b^{12}+4184 b^{10}\right) q^{49/12}}{b^{12}} + \cdots  \ .\nonumber 
\end{align}
}
Each coefficient of $q^{\#}$ is a palindromic polynomial of $b$ with integer coefficients, and is in fact a character of $E_6$ representation where the fugacities are set to a special limit. For example,
\begin{equation}
  30 + 8b^{-4} + 16b^{-2} + 16 b^2 + 8 b^4
  = \chi^{E_6}_{\mathbf{78}}(b)\bigg|_{b_{1,2,4} = b, b_3 = b^2, b_5,b_6 = 1} \ ,
\end{equation}
\begin{align}
  & \ 635 + 35b^{-8} + 112b^{-6} + 310b^{-4} + 480b^{-2} + 480 b^2 + 310 b^4 + 
  112 b^6 + 35 b^8 \nonumber \\
  = & \ \chi^{E_6}_{\mathbf{2430}}(b) + \chi^{E_6}_{\mathbf{78}}(b) + \chi^{E_6}_{\mathbf{1}}(b)\bigg|_{b_{1,2,4} = b, b_3 = b^2,b_5,b_6 = 1} \ .
\end{align}
The integral $\mathcal{I}_{\alpha = 2}(\mathfrak{b}_{1} \to 3\mathfrak{b}, \mathfrak{b}_{2,3,4} \to \mathfrak{b})$ thus gives the partially flavored Schur index of $E_6$ theory
\begin{align}
  \mathcal{I}_{\alpha = 2}(\mathfrak{b}_{1} \to 3\mathfrak{b}, \mathfrak{b}_{2,3,4} \to \mathfrak{b}) = & \ \mathcal{I}^{E_6}(b)\bigg|_{b_{1,2,4} = b, b_3 = b^2,b_5,b_6 = 1} \ .
\end{align}
Further sending $b \to 1$ gives the Schur index in terms of Eisenstein series,
\begin{align}
  \mathcal{I}^{E_6}(q) = \frac{32}{77\eta(\tau)^{22}} \bigg(
    & \ 
    760 E_4(\tau) E_2
    \big[\substack{-1 \\
    1}\big]^4+1232 E_6(\tau) E_2
    \big[\substack{-1 \nonumber \\
    1}\big]^3\\ 
    & \ -1725 E_4(\tau)^2 E_2
    \big[\substack{-1 \\
    1}\big]^2-5320 E_4(\tau) E_6(\tau) E_2
    \big[\substack{-1 \\
    1}\big]
    \nonumber\\
    & \ -80 E_2
    \big[\substack{-1 \\
    1}\big]^6+1125 E_4(\tau)^3-1890 E_2(\tau) E_4(\tau) E_6(\tau)
  \bigg) \ .
\end{align}

The partially flavored Schur index satisfies the partially flavored modular differential equations,
\begin{align}
  \bigg[D_q^{(1)} - \frac{1}{48}D_b^2 + \Big(- \frac{1}{6}E_1 \begin{bmatrix}
    1 \\ b^2
  \end{bmatrix}
  - \frac{2}{3}E_1 \begin{bmatrix}
    1 \\ b^4
  \end{bmatrix}
  \Big) D_b
  + \Big(4E_2\begin{bmatrix}
    1 \\ b^2
  \end{bmatrix}
  + 8 E_2 \begin{bmatrix}
    1 \\ b^4
  \end{bmatrix}\Big) \bigg]\mathcal{I} = 0 \ ,
\end{align}
\begin{align}
  \bigg[D_q^{(2)} + \bigg(4E_3 \begin{bmatrix}
    1 \\ b^2
  \end{bmatrix}
  + 4 E_3 \begin{bmatrix}
    1 \\ b^4
  \end{bmatrix}\bigg) D_b
  - \biggl(283E_4(\tau) + 288E_4\begin{bmatrix}
    1 \\ b^2
  \end{bmatrix} + 144E_4 \begin{bmatrix}
    1 \\ b^4
  \end{bmatrix}\biggr)\bigg] \mathcal{I} = 0 \ ,
\end{align}
resulting from two null states in the chiral algebra $(\widehat{E}_6)_{-3}$ by applying Zhu's recursion formula \cite{zhu1996modular,Gaberdiel:2008pr,Beem:2017ooy,Pan:2023jjw},
\begin{align}
  |\mathcal{N}_\text{Sug}\rangle = & \ \bigg(L_{-2} - \frac{1}{2(k + h^\vee)} \sum_{a,b} K_{ab} J^a_{-1}J^b_{-1}\bigg)|0\rangle \ , \\
  |\mathcal{N}_T\rangle = & \ \bigg(L_{-2}^2 - \frac{8 + c}{6(k+h^\vee)} K_{ab}J^a_{-3}J^b_{-1} + \frac{2 + c}{8(k + h^\vee)}K_{ab}J^a_{-2}J^b_{-2} \bigg)|0\rangle \ .
\end{align}

When $\alpha = 1$, the full flavored integral gives the Schur index $\mathcal{I}$ of $SU(2)$ SQCD. The associated VOA of the theory is $\widehat{\mathfrak{so}}(8)_{-2}$, whose irreducible highest weight modules are classified in \cite{Arakawa:2015jya}, with the finite highest weight given by $0, -2 \omega_1, - \omega_2, - 2 \omega_3, -2\omega_4$, where $0$ corresponds to the vacuum module. The flavored characters of these modules are integral linear combination of the residue $R_{j,1}$ at the simple poles $\mathfrak{a}_j = \mathfrak{b}_j + \frac{\tau}{2}$ of $\mathcal{Z}(\mathfrak{a}, \mathfrak{b})$ \cite{2023arXiv230409681L}
\begin{align}
  \operatorname{ch}_{-2\omega_1} = &\ \mathcal{I} - 2R_{1,1}\ ,\\
  \operatorname{ch}_{-\omega_2} = &\ -2 \mathcal{I} + 2 R_{1,1} + 2R_{2,1}\ ,\\
  \operatorname{ch}_{-2\omega_3} = &\ \mathcal{I} - R_{1,1} - R_{2,1} - R_{3,1} - R_{4,1}\ ,\\
  \operatorname{ch}_{-2\omega_3} = &\ \mathcal{I} - R_{1,1} - R_{2,1} - R_{3,1} + R_{4,1}\ .
\end{align}
All these characters satisfy the same set of flavored modular differential equations derived from the null states $|\mathcal{N}\rangle_\text{Sug}, |\mathcal{N}_T\rangle$ in the VOA $\widehat{SO}(8)_{-2}$. From the physics point of view, this is the manifestation of the relation between Wilson line operators in four dimension and non-vacuum modules in the VOA.

For $\alpha = 2$, a priori we would not expect that relation between the residue of $\mathcal{Z}(\mathfrak{a}, \mathfrak{b})^\alpha$ and the characters of the VOA $(E_6)_{-3}$, since the integral (\ref{eq:Ialpha}) does not follow from a Lagrangian description, and there is no natural way to couple the non-Lagrangian theory to a Wilson line operator. However, it turns out that we can extract additional solution to the above partially flavored MLDEs. This solution is given precisely by the residue
\begin{align}
  & \ R_{3 \mathfrak{b} + \frac{\tau}{2}, m = 1} \mathcal{Z}(\mathfrak{a}, \mathfrak{b})\bigg|_{\mathfrak{b}_{1} \to 3\mathfrak{b}, \mathfrak{b}_{2,3,4} \to \mathfrak{b}} = \frac{\eta(\tau)^8 \vartheta_1(6 \mathfrak{b})^2}{\vartheta_1(2 \mathfrak{b})^6 \vartheta_1(4 \mathfrak{b})^2} \bigg(
    E_1 \begin{bmatrix}
      1 \\ b^2
    \end{bmatrix}
    + E_1 \begin{bmatrix}
      1 \\ b^4
    \end{bmatrix}
    - E_1 \begin{bmatrix}
      1 \\ b^6
    \end{bmatrix}
  \bigg) \nonumber\\
  = & \ -\frac{b^{12} \left(b^4+b^2+1\right) \left(b^8+3 b^6+6 b^4+3 b^2+1\right)}{2 \left(b^2-1\right)^{11} \left(b^2+1\right)^7}q^{-\frac{11}{12}} \nonumber\\
  & \ + \frac{b^{10} \left(4 b^{16}+27 b^{14}+60 b^{12}+90 b^{10}+100 b^8+90 b^6+60 b^4+27 b^2+4\right)}{\left(b^2-1\right)^{11} \left(b^2+1\right)^7} q^{\frac{1}{12}} \ .
\end{align}
Note that at the limit $\mathfrak{b}_1 \to 3 \mathfrak{b}, \mathfrak{b}_{2,3,4} \to \mathfrak{b}$, there is only one independent residue given by the above expression. The higher residues $R_{j, m > 1}$ do not give rise to further solutions to the modular differential equations, nor do their linear combinations.

Next we turn to $\alpha = 3$, corresponding to the $E_{7}$ case. In this case we consider the following contour integral
\begin{equation}
  \mathcal{I}_{\alpha = 3}(\mathfrak{b}) = \eta(\tau)^{-4} \oint \frac{da}{2\pi i a} \mathcal{Z}(\mathfrak{a}, \mathfrak{b})^3 \bigg|_{\mathfrak{b}_i = \mathfrak{b}} \ .
\end{equation}
In the integral, we have taken the required partially flavored limit. The integrand therefore has two 12-order poles
\begin{equation}
  \mathfrak{a} = \pm \mathfrak{b} + \frac{\tau}{2} \ .
\end{equation}
It is straightforward but tedious to compute the generalized residues $R_{\pm \mathfrak{b} + \frac{\tau}{2}, m}$ with $m = 1, 2, \cdots, 12$. Here we list a few simplest results,
\begin{align}
  \operatorname{Res}_{\mathfrak{b} + \frac{\tau}{2}, 12} = & \ - \frac{1}{40874803200\pi^{11}\eta(\tau)^{36}\vartheta_1(2 \mathfrak{b})^{6}} \ ,\\
  \operatorname{Res}_{\mathfrak{b} + \frac{\tau}{2}, 10} = & \ - \frac{i}{\pi^9\eta(\tau)^{16} \vartheta_1(2 \mathfrak{b}_1)^6} \bigg(\frac{1}{30965760}E_1 \begin{bmatrix}
    1 \\ b^2
  \end{bmatrix}^2
  + \frac{1}{15482880}E_2\begin{bmatrix}
    1 \\ b^2
  \end{bmatrix}\bigg)\ , \\
  \operatorname{Res}_{\mathfrak{b} + \frac{\tau}{2}, 9} = & \ \frac{1}{860160 \pi^8 \eta (\tau)^{16} \vartheta_1(2 \mathfrak{b})^6}\bigg(
    E_1 \begin{bmatrix}
      1 \\ b^2
    \end{bmatrix}^3
    + 3 E_1 \begin{bmatrix}
      1 \\ b^2
    \end{bmatrix}E_2 \begin{bmatrix}
      1 \\ b^2
    \end{bmatrix}
    + 3 E_3 \begin{bmatrix}
      1 \\ b^2
    \end{bmatrix}
  \bigg)\\
  \operatorname{Res}_{\mathfrak{b} + \frac{\tau}{2}, 8} = & \ \frac{1}{215040 \pi^7 \eta (\tau)^{16} \vartheta_1(2 \mathfrak{b})^6}\bigg(
    6E_4(\tau)
    + E_1 \begin{bmatrix}
      1 \\ b^2
    \end{bmatrix}^4
    + 4 E_1 \begin{bmatrix}
      1 \\ b^2
    \end{bmatrix}^2 E_2 \begin{bmatrix}
      1 \\ b^2
    \end{bmatrix} \nonumber\\
    & \ - 10 E_2 \begin{bmatrix}
      1 \\ b^2
    \end{bmatrix}^2
    + 28 E_1 \begin{bmatrix}
      1 \\ b^2
    \end{bmatrix}E_3 \begin{bmatrix}
      1 \\ b^2
    \end{bmatrix}
    + 28 E_4 \begin{bmatrix}
      1 \\ b^2
    \end{bmatrix}
  \bigg) \ , \\
  \operatorname{Res}_{\mathfrak{b} + \frac{\tau}{2}, 7} = & \ \frac{1}{1280\pi^6\eta(\tau)^{16}\vartheta_1(2 \mathfrak{b})^6}\bigg(
    E_1 \begin{bmatrix}
      1 \\ b^2
    \end{bmatrix}^5
    + 5 E_1 \begin{bmatrix}
      1 \\ b^2
    \end{bmatrix}^3 E_2 \begin{bmatrix}
      1 \\ b^2
    \end{bmatrix}
    + 7 E_1 \begin{bmatrix}
      1 \\ b^2
    \end{bmatrix}E_2 \begin{bmatrix}
      1 \\ b^2
    \end{bmatrix}^2 \nonumber\\
    & \ + E_1 \begin{bmatrix}
      1 \\ b^2
    \end{bmatrix}^2 E_3 \begin{bmatrix}
      1 \\ b^2
    \end{bmatrix}
    + 7 E_2 \begin{bmatrix}
      1 \\ b^2
    \end{bmatrix}E_3 \begin{bmatrix}
      1 \\ b^2
    \end{bmatrix}
    - 5 E_1 \begin{bmatrix}
      1 \\ b^2
    \end{bmatrix}E_4 \begin{bmatrix}
      1 \\ b^2
    \end{bmatrix}
    - 5 E_5 \begin{bmatrix}
      1 \\ b^2
    \end{bmatrix}
  \bigg)\ . \nonumber
\end{align}

In the end, we obtain a long expression for the partially flavored Schur index of $E_7$ theory, which series expands as
\begin{align}
  \mathcal{I}_{\alpha = 3}(\mathfrak{b}) = & \ q^{19/12} + (79 + 27b^{-2} + 27 b^2)q^{31/12} \nonumber \\
  & \ + (3239 + 351b^4 + 351b^{-4} + 1782 b^2 + 1782 b^{-2})q^{43/12} + \cdots \ .
\end{align}
The coefficients
\begin{align}
  79 + 27b^{-2} + 27 b^2 = & \ \chi^{E_7}_{\mathbf{133}}(b)\bigg|_{b \to b^*} \ ,\\
  3239 + 351b^4 + 351b^{-4} + 1782 b^2 + 1782 b^{-2} = & \ \chi^{E_7}_{\mathbf{7371}}(b) + \chi^{E_7}_{\mathbf{133}}(b) + \chi^{E_7}_{\mathbf{1}}\bigg|_{b \to b^*} \ ,
\end{align}
where $b \to b^*$ represents the special limit
\begin{equation}
  b_1 \to b^2, b_2 \to b^2, b_3 \to b^2, b_4 \to b, b_5 \to 1, b_6 \to b, b_7 \to b \ .
\end{equation}
Accordingly, we have computed the partially flavored Schur index of $E_7$ theory in terms of the generalized partition function,
\begin{equation}
  \mathcal{I}_{\alpha = 3}(\mathfrak{b}) = \mathcal{I}^{E_7}(b)\bigg|_{b \to b^*} \ .
\end{equation}
The partially flavored Schur index of $E_7$ theory also satisfies the following partially flavored modular differential equations,
\begin{align}
  0 = & \ \bigg[D_q^{(1)} - \frac{1}{24}D_b^2 - \frac{3}{2}E_1 \begin{bmatrix}
    1\\b^2
  \end{bmatrix}D_b + \bigg(E_2 + 18 E_2 \begin{bmatrix}
    1 \\ b^2
  \end{bmatrix}\bigg)\bigg] \mathcal{I}^{E_7}_{b \to b^*} \\
  0 = & \ \bigg[D_q^{(2)} + 18 E_3 \begin{bmatrix}
    1 \\ b^2
  \end{bmatrix} D_b
  - \bigg(967 E_4(\tau) + 648 E_4 \begin{bmatrix}
    1 \\ b^2
  \end{bmatrix}\bigg)\bigg] \mathcal{I}^{E_7}_{b \to b^*} \ .
\end{align}
Similar to the $\alpha = 2$ case, there is another solution to the above modular differential equations, which is given precisely by the residue
\begin{equation}
  \operatorname{Res}_{\mathfrak{b} + \frac{\tau}{2}, 1} \mathcal{Z}(\mathfrak{a}, \mathfrak{b})^3\bigg|_{\mathfrak{b}_i = \mathfrak{b}} \ .
\end{equation}
The explicit expresion of this residue is straightforward to compute, although it is too long to be presented here. We record its series expansion here,
{\small\begin{align}
  & \ \operatorname{Res}_{\mathfrak{a} \mathfrak{b} + \frac{\tau}{2}, 1}\eta(\tau)^{-4}\mathcal{Z}(\mathfrak{a}, \mathfrak{b})^3
  = \frac{b^{12}(1 + b^2)(1 + 9 b^2 + 19 b^4 + 9 b^6 + b^8)}{(1 - b^2)^{17}}q^{- \frac{17}{12}} \nonumber \\
  & \ + \frac{79 b^{12} + 439b^{14} + 808 b^{16} + 808 b^{18} + 439 b^{20} + 79b^{22}}{(1 - b^2)^{17}} q^{- \frac{5}{12}}\\
  & \ + \frac{+27 b^{24}+2780 b^{22}+8333 b^{20}+13391 b^{18}+13391 b^{16}+8333 b^{14}+2780 b^{12}+27 b^{10}}{(1 - b^2)^{17}} q^{\frac{7}{12}} + \cdots \ .\nonumber
\end{align}}

\section*{Acknowledgments}
The authors would like to thank Anirudh Deb, Yinan Wang, Wenbin Yan for useful discussions. The work of Y.P. is supported by the National Natural Science Foundation of China (NSFC) under Grant No. 11905301.
The work of P.Y. is supported by the National Natural Science Foundation of China, Grant No. 12447142, No. 12175004, No. 12422503.

\appendix
\section{Special functions and notation \label{app:special-functions}}

In the following and in the main text, we use the normal and fraktur font to denote fugacities in characters and Schur indices,
\begin{equation}
  a = e^{2\pi i \mathfrak{a}}, \quad
  b = e^{2\pi i \mathfrak{b}}, \quad
  \tilde b = e^{2\pi i \tilde{\mathfrak{b}}}, \quad
  \cdots, \quad
  z = e^{2\pi i \mathfrak{z}} \ ,
\end{equation}
The only exception is $q = e^{2\pi i \tau}$, which is the standard notation.

\subsection{Dedekind eta function}

The Dedekind eta function $\eta(\tau)$ is defined as
\begin{equation}
  \eta(\tau) \coloneqq q^{\frac{1}{24}} \prod_{n = 1}^{+\infty}(1 - q^n) \ .
\end{equation}
Under modular transformations,
\begin{equation}
  \eta(\tau + 1) = e^{\frac{\pi i}{12}}\eta(\tau), \qquad
  \eta( - \frac{1}{\tau}) = \sqrt{-i\tau}\eta(\tau) \ .
\end{equation}

\subsection{Weierstrass \texorpdfstring{$\zeta$}{Zeta} function}

The Weierstrass $\zeta$ function is defined as
\begin{equation}
  \zeta(\mathfrak{z}|\tau)
	= \frac{1}{\mathfrak{z}} + \sum_{\substack{(m,n) \in \mathbb{Z}^2 \\ (m,n)\ne (0,0)}}^{\prime} \bigg[\frac{1}{\mathfrak{z} - m - n \tau} + \frac{1}{m + n \tau} + \frac{\mathfrak{z}}{
	^2} (m+n\tau)\bigg] \ .
\end{equation}
The $\zeta$ function is an odd function, $\zeta(\mathfrak{z}|\tau) + \zeta(- \mathfrak{z}|\tau) = 0$. It is non-elliptic, but satisfies the quasi-periodic property
\begin{equation}
	\zeta(\mathfrak{z} + 1|\tau) = \zeta(\mathfrak{z}|\tau) + 2 \eta_1(\tau), \quad
	\zeta(\mathfrak{z} + \tau|\tau) = \zeta(\mathfrak{z}|\tau) + 2 \eta_2(\tau) \ ,
\end{equation}
where
\begin{equation}
	\eta_1(\tau) = -2\pi^2 E_2(\tau), \quad
	\eta_2(\tau) = \tau \eta_1(\tau) - \pi i \ .
\end{equation}
For our purpose, $\zeta(\mathfrak{z}|\tau)$ and its derivatives form a basis of elliptic functions, which allows contour integration to be carried out analytically.

\subsection{Jacobi theta functions}

The standard Jacobi theta functions are defined using the $q$-Pochhammer symbol $(z;q) \coloneqq \prod_{k = 0}^{+\infty}(1 - zq)$,
\begin{align}
  \vartheta_1(\mathfrak{z}|\tau) = & \ i q^{\frac{1}{8}} z^{-\frac{1}{2}}(q;q)(z;q)(z^{-1}q;q) 
	= - i q^{\frac{1}{8}} z^{\frac{1}{2}}(q;q)(zq;q)(z^{-1};q) \ , \\
  \vartheta_2(\mathfrak{z}|\tau) = & \ q^{\frac{1}{8}}z^{-\frac{1}{2}}(q;q)(-z;q)(- z^{-1}q;q) = q^{\frac{1}{8}}z^{\frac{1}{2}}(q;q) (- zq;q) (- z^{-1};q) \ , \\
  \vartheta_3(\mathfrak{z}|\tau)= & \ (q;q)(-zq^{1/2};q)(- z^{-1}q^{1/2};q) \ , \\
  \vartheta_4(\mathfrak{z}|\tau)= & \ (q;q)(zq^{1/2};q)(z^{-1}q^{1/2};q) \ .
\end{align}
We will often omit the $\tau$ from the notation, and in particular,
\begin{equation}
  \vartheta_i(0) = \vartheta_i(\mathfrak{z} = 0|\tau)\ , \qquad
  \vartheta_i^{(k)}(0) = \partial_{\mathfrak{z}}^n \Big|_{\mathfrak{z} = 0} \vartheta_i(\mathfrak{z}|\tau) \ .
\end{equation}
These functions can be rewritten in infinite sum,
\begin{align}
	\vartheta_1(\mathfrak{z}|\tau) \coloneqq & \ -i \sum_{r \in \mathbb{Z} + \frac{1}{2}} (-1)^{r-\frac{1}{2}} e^{2\pi i r \mathfrak{z}} q^{\frac{r^2}{2}} ,
	& \vartheta_2(\mathfrak{z}|\tau) \coloneqq & \sum_{r \in \mathbb{Z} + \frac{1}{2}} e^{2\pi i r \mathfrak{z}} q^{\frac{r^2}{2}} \ ,\\
	\vartheta_3(\mathfrak{z}|\tau) \coloneqq & \ \sum_{n \in \mathbb{Z}} e^{2\pi i n \mathfrak{z}} q^{\frac{n^2}{2}},
	& \vartheta_4(\mathfrak{z}|\tau) \coloneqq & \sum_{n \in \mathbb{Z}} (-1)^n e^{2\pi i n \mathfrak{z}} q^{\frac{n^2}{2}} \ .
\end{align}
The Jacobi theta functions are quasi-periodic functions of $\mathfrak{z}$, for example,
\begin{align}
	\vartheta_1(\mathfrak{z} + m \tau + n) = (-1)^{m + n} e^{-2\pi i m \mathfrak{z}} q^{ - \frac{1}{2}m^2}\vartheta_1(\mathfrak{z})\ .
\end{align}
When $\mathfrak{z}$ is shifted by half-periods, the four Jacobi theta functions transform into each other. We summarize the shift property in the following diagram,
\begin{center}
	\includegraphics[height=100pt]{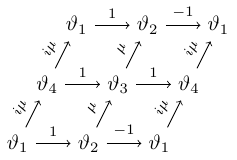}
\end{center}
where $\mu = e^{- \pi i \mathfrak{z}} e^{- \frac{\pi i}{4}}$, and $f \xrightarrow{a} g$ means
\begin{align}
	\text{either}\qquad  f\left(\mathfrak{z} + \frac{1}{2}\right) = a g(\mathfrak{z}) \qquad \text{or} \qquad
	f\left(\mathfrak{z} + \frac{\tau}{2}\right) = a g(\mathfrak{z}) \ ,
\end{align}
depending on whether the arrow is horizontal or (slanted) vertical respectively.

In the main text we often encounter products of theta functions with fractional shifts, which can be simplified using the identities
\begin{align}
	\prod_{\ell = 0}^{n-1} \vartheta_1(\mathfrak{z} - \ell \tau|n\tau)
	= & \ (-iz^{1/2})^{n - 1} q^{-\frac{(n-1)(2n - 1)}{12}} \vartheta_1(\mathfrak{z})
	\frac{\eta(n\tau)^n}{\eta(\tau)} \ , \\
	\prod_{\ell = 1}^n \vartheta_1(\ell \tau| 2n \tau)
	= & \ i^n q^{- \frac{1}{24}(2n^2 +4n + 1)}\eta(2n)^{n-1}\eta(\tau)(q^n, q^{2n}) \ , \\
	\prod_{\ell = 1}^n \vartheta_1(\ell \tau |(2n+1)\tau) = & \ i^n\eta(\tau)\eta((2n+1)\tau)^{n-1} \ .
\end{align}

The modularity of $\vartheta_i(\mathfrak{z} | \tau)$ is well-known. Under the $S$ and $T$ transformations, which act on the nome and flavor fugacity as $(\frac{\mathfrak{z}}{\tau}, - \frac{1}{\tau})\xleftarrow{~~S~~}(\mathfrak{z}, \tau) \xrightarrow{~~T~~} (\mathfrak{z}, \tau + 1)$,
\begin{center}
  \begin{tikzpicture}

    \node(m1) at (0,0) {$\vartheta_1$};
    \node(m2) at (0,-1) {$\vartheta_2$};
    \node(m3) at (0,-2) {$\vartheta_3$};
    \node(m4) at (0,-3) {$\vartheta_4$};

    \node(l1) at (-2,0) {$-i\alpha\vartheta_1$};
    \node(l2) at (-2,-1) {$\alpha\vartheta_2$};
    \node(l3) at (-2,-2) {$\alpha\vartheta_3$};
    \node(l4) at (-2,-3) {$\alpha\vartheta_4$};

    \node(r1) at (1.5,0) {$e^{\frac{\pi i}{4}}\vartheta_1$};
    \node(r2) at (1.5,-1) {$e^{\frac{\pi i}{4}}\vartheta_2$};
    \node(r3) at (1.5,-2) {$\vartheta_3$};
    \node(r4) at (1.5,-3) {$\vartheta_4$};

    \draw[->=stealth] (m1)--node[above]{$S$}(l1);
    \draw[->=stealth] (m2)--(l4);
    \draw[->=stealth] (m3)--node[above]{$S$}(l3);
    \draw[->=stealth] (m4)--(l2);
    \draw[->=stealth] (m1)--node[above]{$T$}(r1);
    \draw[->=stealth] (m2)--node[above]{$T$}(r2);
    \draw[->=stealth] (m3)--node[above]{$T$}(r4);
    \draw[->=stealth] (m4)--(r3);
  \end{tikzpicture}
\end{center}
Here we define $\alpha \coloneqq \sqrt{-i \tau}e^{\frac{\pi i}{\tau} \mathfrak{z}^2}$. These transformation can be used to deduce the modular transformation of the Eisenstein series that we will now review.

\subsection{Eisenstein series}
The Eisenstein series\footnote{In the literature these functions are often called \emph{twisted} Eisenstein series. In this paper we will suppress the word twisted.} depend on two characteristics $\phi, \theta$, and are defined by the infinite sum,
\begin{align}
	E_{k \ge 1}\left[\begin{matrix}
		\phi \\ \theta
	\end{matrix}\right] \coloneqq & \ - \frac{B_k(\lambda)}{k!} \\
	& \ + \frac{1}{(k-1)!}\sum_{r \ge 0}' \frac{(r + \lambda)^{k - 1}\theta^{-1} q^{r + \lambda}}{1 - \theta^{-1}q^{r + \lambda}}
	+ \frac{(-1)^k}{(k-1)!}\sum_{r \ge 1} \frac{(r - \lambda)^{k - 1}\theta q^{r - \lambda}}{1 - \theta q^{r - \lambda}} \ , \nonumber
\end{align}
where $\phi \coloneqq e^{2\pi i \lambda}$ with $0 \le \lambda < 1$, $B_k(x)$ denotes the $k$-th Bernoulli polynomial, and the prime in the sum indicates that the $r = 0$ should be omitted when $\phi = \theta = 1$. To make certain formula more compact, we also define
\begin{align}
	E_0\left[\begin{matrix}
		\phi \\ \theta
	\end{matrix}\right] = -1 \ .
\end{align}
We have suppressed the $\tau$-dependence in the notation. However, when the argument is rescaled to $n\tau$, we will write it explicitly, like
\begin{equation}
	E_2 \begin{bmatrix}
		-1 \\ z
	\end{bmatrix}(3\tau) \ , \qquad
	E_3 \begin{bmatrix}
		1 \\ q^{1/4} b
	\end{bmatrix}(4\tau) \ , \qquad
	\text{etc.} \ .
\end{equation}

Eisenstein series with even $k = 2n$ are related to the usual Eisenstein series $E_{2n}(\tau)$ by sending the $\theta, \phi \to 1$. When $k$ is odd, $\theta = \phi = 1$ is a vanishing limit, except for the case $k = 1$ where it is singular,
\begin{align}
	E_{2n}\left[\begin{matrix}
		+1 \\ +1
	\end{matrix}\right] = E_{2n} \ , \qquad E_1\left[\begin{matrix}
		+ 1 \\ z
	\end{matrix}\right] = \frac{1}{2\pi i }\frac{\vartheta'_1(\mathfrak{z})}{\vartheta_1(\mathfrak{z})}, \qquad
	E_{2n + 1 \ge 3}\left[\begin{matrix}
		+1 \\ +1
	\end{matrix}\right] = 0 \ .
\end{align}
In fact, all $E_k\big[\substack{\pm 1 \\ z}\big]$ are regular as $z \to 1$, except for $E_1\big[{\substack{1 \\ z}}\big]$ which has a simple pole. We often omit $(\tau)$ in $E_n(\tau)$, unless the argument is $n\tau$, where we will write $E_k(n\tau)$.

A closely related property is the symmetry of the Eisenstein series
\begin{align}\label{Eisenstein-symmetry}
	E_k\left[\begin{matrix}
	  \pm 1 \\ z^{-1}
	\end{matrix}\right] = (-1)^k E_k\left[\begin{matrix}
	  \pm 1 \\ z
	\end{matrix}\right] \ .
\end{align}

Often it is useful to rewrite the Eisenstein series in terms of the Jacobi theta functions, 
\begin{align}\label{EisensteinToTheta}
	E_k\left[\begin{matrix}
		+ 1 \\ z
	\end{matrix}\right] = \sum_{\ell = 0}^{\lfloor k/2 \rfloor}  \frac{(-1)^{k + 1}}{(k - 2\ell)!}\left(\frac{1}{2\pi i}\right)^{k - 2\ell} \mathbb{E}_{2\ell} \frac{\vartheta_1^{(k - 2\ell)}(\mathfrak{z})}{\vartheta_1(\mathfrak{z})} \ ,
\end{align}
Here $\vartheta_i^{(k)}(\mathfrak{z})$ is the $k$-th derivative in $\mathfrak{z}$. The conversion from $E_k\left[\substack{- 1 \\ \pm z}\right]$ can be obtained by replacing $\vartheta_1$ with $\vartheta_{2,3,4}$ appropriately. An immediate consequence of this relation is that the Eisenstein series satisfy the following shift property
\begin{align}\label{eq:Eisenstein-shift}
	E_k\left[\begin{matrix}
		\pm 1\\ z q^{\frac{n}{2}}
	\end{matrix}\right]
	=
	\sum_{\ell = 0}^{k} \left(\frac{n}{2}\right)^\ell \frac{1}{\ell !}
	E_{k - \ell}\left[\begin{matrix}
		(-1)^n(\pm 1) \\ z
	\end{matrix}\right] \ .
\end{align}

Using the known modular property of $\vartheta_i$, one can deduce that under the $S: (\tau, \mathfrak{z}) \to (- \frac{1}{\tau}, \frac{\mathfrak{z}}{\tau})$, the Eisenstein series transform according to the following set of formula,
\begin{align}\label{Eisenstein-S-transformation}
  E_n \begin{bmatrix}
    +1 \\ +z
  \end{bmatrix} \xrightarrow{S} \ &
	E_n \begin{bmatrix}
    +1 \\ e^{\frac{2\pi i \mathfrak{z}}{\tau}}
  \end{bmatrix}(- \frac{1}{\tau}) = 
  \left(\frac{1}{2\pi i}\right)^n \sum_{\ell = 0}^k \frac{(- \log z)^{n - \ell} (\log q)^\ell}{(n - \ell)!} E_\ell \begin{bmatrix}
    +1 \\ z
\end{bmatrix}\ ,\\
  E_n \begin{bmatrix}
    -1 \\ +z
  \end{bmatrix} \xrightarrow{S} \ &
	E_n \begin{bmatrix}
    -1 \\ e^{\frac{2\pi i \mathfrak{z}}{\tau}}
  \end{bmatrix}(- \frac{1}{\tau})
	=
  \left(\frac{1}{2\pi i}\right)^n \sum_{\ell = 0}^k \frac{(- \log z)^{n - \ell} (\log q)^\ell}{(n - \ell)!} E_\ell \begin{bmatrix}
    +1 \\ -z
\end{bmatrix}\ ,\\
  E_n \begin{bmatrix}
    1 \\ -z
  \end{bmatrix} \xrightarrow{S} \ &
	E_n \begin{bmatrix}
    1 \\ -e^{\frac{2\pi i \mathfrak{z}}{\tau}}
  \end{bmatrix}(- \frac{1}{\tau})
	=
  \left(\frac{1}{2\pi i}\right)^n \sum_{\ell = 0}^k \frac{(- \log z)^{n - \ell} (\log q)^\ell}{(n - \ell)!} E_\ell \begin{bmatrix}
    -1 \\ z
\end{bmatrix}\ ,\\
  E_n \begin{bmatrix}
    -1 \\ -z
  \end{bmatrix} \xrightarrow{S} \ &
	E_n \begin{bmatrix}
    -1 \\ -e^{\frac{2\pi i \mathfrak{z}}{\tau}}
  \end{bmatrix}(- \frac{1}{\tau})
	=
  \left(\frac{1}{2\pi i}\right)^n \sum_{\ell = 0}^k \frac{(- \log z)^{n - \ell} (\log q)^\ell}{(n - \ell)!} E_\ell \begin{bmatrix}
    -1 \\ -z
\end{bmatrix}\ .
\end{align}
Under the $T$-transformation, the Eisenstein series transform according to the following rules,
\begin{align}\label{Eisenstein-T-transformation}
  E_n \begin{bmatrix}
    + 1 \\ + z
  \end{bmatrix} \xrightarrow{T}& \ E_n \begin{bmatrix}
    + 1 \\ + z
  \end{bmatrix}, & 
  E_n \begin{bmatrix}
    - 1 \\ + z
  \end{bmatrix} \xrightarrow{T}& \
  E_n \begin{bmatrix}
    - 1 \\ - z
  \end{bmatrix} \\
  E_n \begin{bmatrix}
    + 1 \\ - z
  \end{bmatrix} \xrightarrow{T}& \ E_n \begin{bmatrix}
    + 1 \\ - z
  \end{bmatrix}, & 
  E_n \begin{bmatrix}
    - 1 \\ - z
  \end{bmatrix} \xrightarrow{T}& \ 
  E_n \begin{bmatrix}
    - 1 \\ + z
  \end{bmatrix} \ .
\end{align}
For reference, we give a few explicit examples of modular transformation of Eisenstein series. Consider the special case $E_2(\tau) = E_2 \big[\substack{+1\\+1}\big] = \lim_{z \to 1}E_2 \big[\substack{+1\\z}\big]$. Under the $S$-transformation $(\tau, \mathfrak{z}) \to (- \frac{1}{\tau}, \frac{\mathfrak{b}}{\tau})$,
\begin{equation}
	E_2 \begin{bmatrix}
		1 \\ e^{2\pi i \mathfrak{z}}
	\end{bmatrix}(\tau) \to - \frac{\mathfrak{z}^2}{2} - \mathfrak{z}\tau E_1 \begin{bmatrix}
		1 \\ e^{2\pi i \mathfrak{z}}
	\end{bmatrix}
	+ \tau^2 E_2 \begin{bmatrix}
		1 \\ e^{2\pi i \mathfrak{z}}
	\end{bmatrix} \ .
\end{equation}
When taking the limit $\mathfrak{z} \to 0$, the second term does not vanishe due the fact that $E_1 \big[\substack{1 \\ z}\big]$ has a simple pole at $z = 1$,
\begin{equation}
	\lim_{\mathfrak{z} \to 0}\mathfrak{z} E_1 \begin{bmatrix}
		1 \\ e^{2\pi i \mathfrak{z}}
	\end{bmatrix} = - \frac{i}{2\pi} \ .
\end{equation}
As a result, we recover the standard modular transformation of $E_2(\tau)$,
\begin{equation}
	E_2(\tau) \xrightarrow{S} \tau^2 E_2(\tau) + \frac{i}{2\pi}\tau \ .
\end{equation}
Besides $E_2 \big[\substack{1 \\ 1}\big]$, no other $E_{k > 2}(\tau)$ has this extra $\tau$-term, since $\lim_{\mathfrak{z} \to 0}\mathfrak{z}^{k - 1} E_1\big[\substack{1 \\ z}\big] = 0$.

As a second example, we consider two different ways of computing the modular transformation of $E_2 \big[\substack{+1 \\ -1}\big]$. First, according to the above rules,
\begin{align}
	E_2 \begin{bmatrix}
		+ 1 \\ -1
	\end{bmatrix}(\tau) \xrightarrow{S}
	\left(\frac{1}{2\pi i}\right)^2 \sum_{\ell = 0}^k \frac{(- \log z)^{2 - \ell} (\log q)^\ell}{(2 - \ell)!} E_\ell \begin{bmatrix}
    -1 \\ z
	\end{bmatrix}(\tau)\Bigg|_{z \to 1} = \tau^2 E_2 \begin{bmatrix}
		-1 \\ + 1
	\end{bmatrix} \ ,
\end{align}
Altenatively we can treat $-1 = e^{\frac{2 \pi i \mathfrak{z}}{\tau}}$ with $\mathfrak{z} \coloneqq \frac{\tau}{2}$, $z \coloneqq e^{2\pi i \mathfrak{z}} = q^{\frac{1}{2}}$,
\begin{align}
	E_2 \begin{bmatrix}
		+ 1 \\ -1
	\end{bmatrix}(\tau) \xrightarrow{S}
	E_2 \begin{bmatrix}
		+ 1 \\ e^{ \frac{ \pi i \tau}{\tau}}
	\end{bmatrix}(- \frac{1}{\tau})
	= & \ \left({\frac{1}{2\pi i}}\right)^2 \sum_{\ell = 0}^2 \frac{(- \pi i \tau)^{2 - \ell} (2\pi i \tau)^\ell}{(2 - \ell)!} E_\ell \begin{bmatrix}
		-1 \\ q^{+\frac{1}{2}}
	\end{bmatrix}(\tau) \nonumber\\
	= & \ \tau^2 E_2 \begin{bmatrix}
		-1 \\ + 1
	\end{bmatrix}(\tau) \ .
\end{align}
In going to the second line, we make use of the shift property (\ref{eq:Eisenstein-shift}). Combining the $S, T$ transformation, we obtain
\begin{align}
  E_n \begin{bmatrix}
    -1 \\ z
  \end{bmatrix} \xrightarrow{STS}
  \left(\frac{1}{2\pi i}\right)^n\left[\bigg(\sum_{k \ge 0}\frac{1}{k!}(- \log z)^k y^k\bigg)
  \bigg(\sum_{\ell \ge 0}(\log q - 2\pi i)^\ell y^\ell E_\ell \begin{bmatrix}
    -1 \\ +z
  \end{bmatrix}\bigg)\right]_n\ . \nonumber
\end{align}

There are some useful identities involving Eisenstein series and their products. For example, using the following elegant identity, we can trade the argument $p\tau$ with $\tau$,
\begin{equation}
	E_n \begin{bmatrix}
		\pm 1 \\ b^p
	\end{bmatrix}(p \tau)
	= \frac{1}{p} \sum_{\ell = 0}^{p - 1} E_n \begin{bmatrix}
		\pm 1 \\ b e^{2\pi i \ell/p}
	\end{bmatrix} (\tau) \ , \qquad p = 1, 2, 3, \cdots \ .
\end{equation}

\bibliographystyle{utphys2}

\bibliography{ref}

\end{document}